\documentclass[final,leqno,onefignum,onetabnum]{siamltexmm}
\pdfoutput=0

\usepackage[numbers,sort&compress]{natbib}

\usepackage[letterpaper,centering]{geometry}
\usepackage{subfigure}
\usepackage{amsmath}
\usepackage{amssymb}
\usepackage{amsfonts, bm}
\usepackage{multirow}

\usepackage[linesnumbered,ruled,vlined]{algorithm2e}

\usepackage{soul}
\usepackage{color}

\usepackage[textsize=tiny]{todonotes}

\usepackage{pgf}
\usepackage{pgfplots}
\usepgfplotslibrary{groupplots}
\usetikzlibrary{plotmarks}
\usetikzlibrary{arrows,automata}
\usetikzlibrary{calc}

\title{A multiscale strategy for Bayesian inference using transport 
maps\thanks{This research was supported in part by the US Department of Energy 
Office of Science Graduate Fellowship Program (DOE SCGF), made possible in part 
by  the American Recovery and Reinvestment Act of 2009, administered by 
ORISE-ORAU under contract DE-AC05-06OR23100. This research is also supported by the US Department of Energy, Office of Advanced Scientific Computing Research (ASCR), under grant number DE-SC0009297.}}

\author{Matthew Parno\footnotemark[2] \and Tarek Moselhy\footnotemark[3] \and Youssef Marzouk\footnotemark[3]}

\newcommand{\EditsText}[1]{#1}
\newcommand{\EditsMath}[1]{#1}
  
\newcommand{\real}{\mathbb{R}}

\newcommand{\Ex}[1]{\mathbb{E}_{#1}}
\newcommand{\cov}{\mathbb{C}\mathrm{ov}}

\newcommand{\Var}[1]{{\mathbb{V}\mathrm{ar}}_{#1}}

\newcommand{\eqd}{\buildrel i.d. \over =}
\newcommand{\setf}[1]{\mathcal{#1}}

\newcommand{\trv}{x}
\newcommand{\rrv}{r}
\newcommand{\pd}{d}

\newcommand{\rd}{p}     

\newcommand{\basis}{\psi} 

\newcommand{\crv}{\gamma} 
\newcommand{\frv}{\theta} 
\newcommand{\drv}{y} 
\newcommand{\jrv}{x} 

\newcommand{\finc}{n} 

\newcommand{\tmeas}{\mu}
\newcommand{\rmeas}{\nu}

\newcommand{\emap}{T} 
\newcommand{\demap}{{\nabla T}}

\newcommand{\efmap}{S} 
\newcommand{\ifmap}{\widetilde{S}} 

\newcommand{\imap}{{\widetilde{T}}} 

\newcommand{\espace}{{\mathcal{T}}}
\newcommand{\ispace}{{\widetilde{\mathcal{T}}}}

\newcommand{\approxeqd}{\buildrel i.d. \over \approx}

\begin{document}

\maketitle

\newcommand{\slugmaster}{}
\renewcommand{\thefootnote}{\fnsymbol{footnote}}

\footnotetext[2]{U. S. Army Cold Regions Research and Engineering Laboratory, Hanover, NH 03755, USA. \texttt{matthew.d.parno@usace.army.mil}}

\footnotetext[3]{Department of Aeronautics and Astronautics, Massachusetts 
Institute of Technology, Cambridge, MA 02139, USA. \texttt{\{tmoselhy,ymarz\}@mit.edu}}

\renewcommand{\thefootnote}{\arabic{footnote}}

\begin{abstract}
In many inverse problems, model parameters cannot be precisely determined from observational data. Bayesian inference provides a mechanism for capturing the resulting parameter uncertainty, but typically at a high computational cost.  This work introduces a multiscale decomposition that exploits conditional independence across scales, when present in certain classes of inverse problems, to decouple Bayesian inference into two stages: (1) a computationally tractable coarse-scale inference problem; and (2) a mapping of the low-dimensional coarse-scale posterior distribution into the original high-dimensional parameter space. This decomposition relies on a characterization of the non-Gaussian joint distribution of coarse- and fine-scale quantities via optimal transport maps.  
We demonstrate our approach on a sequence of inverse problems arising in subsurface flow, using the multiscale finite element method to discretize the steady state pressure equation. We compare the multiscale strategy with full-dimensional Markov chain Monte Carlo on a problem of moderate dimension (100 parameters) and then use it to infer a conductivity field described by over 10\,000 parameters.

\end{abstract}

\begin{keywords} Bayesian inference, inverse problems, multiscale
  modeling, multiscale finite element method, optimal transportation,
  Markov chain Monte Carlo
\end{keywords}

\section{Introduction}\label{sec:intro}

Mathematical models often contain parameters that must be estimated from observational data, before the models can be used for prediction or design. Deterministic approaches have long been applied to such inverse problems (e.g., \cite{Doherty1994,Vogel2002,Aster2011,Epanomeritakis2008,Bui-Thanh2012a}).  \EditsText{Yet observations can seldom constrain model parameters precisely, and the resulting inverse problems are thus ill-posed.} In this context, statistical approaches---e.g., the Bayesian approach \cite{TarantolaBook,KaipioBook,Stuart2010}---provide a rigorous framework for simultaneously characterizing model parameters \textit{and} their uncertainties \cite{Biegler2011}. \EditsText{This characterization is}
%
particularly crucial for applications requiring quantified uncertainties in model predictions (e.g., \cite{Wagner1989,Freeze1990,EPA2005,Vesselinov2013}).  

Ill-posed inverse problems often result from the combination of high-dimensional parameters and indirect observations, especially when the observations smooth or integrate the parameter field of interest. In this setting, the map from parameters to observations is called the \textit{forward} model. Smoothing forward models are ubiquitous in science and engineering applications, ranging from flow through porous media to tomography and remote sensing.
A significant body of research has sought to develop and analyze \textit{multiscale} methods for evaluating such forward models \cite{msfem,Hughes98,Arbogast00,He06,Weinan07,Jenny03,Jenny06,Aarnes08,Juanes08}. Many of these methods rely on the idea that a finite dimensional coarse-scale representation can include the impact of fine-scale structure in the parameters without resolving the problem to that level. For example, in the context of partial differential equations (PDEs), fine-scale spatial variation in a coefficient or initial condition can be captured by homogenized coefficients, coarse-scale basis functions, or other correction terms to the variational statement of the problem, enabling the PDE solution to be accurately approximated at reduced cost. 
A host of theoretical developments and numerical approaches \cite{Hughes_Sangalli_2007,Pavliotis_Stuart_2008,Jagular2014} have yielded multiscale solution strategies for linear and nonlinear PDEs, ODEs, and other systems \cite{Miller_Tadmor_2009}. 

Implicit in the success of a multiscale strategy for solving the forward model is a notion of \textit{conditional independence}: observations of the solution can be directly predicted by some coarse-scale quantities; conditioned on these coarse-scale quantities, the observations and the original parameter field of interest (which resolves fine-scale structure) are independent. We will use \EditsText{this notion} to design a multiscale Bayesian inference approach. Our approach will decompose the inverse problem into a coarse-scale inference problem, where coarse-scale quantities are inferred from data, and a fine-scale inference problem, where fine-scale parameters are conditioned on realizations of the coarse scale. This strategy will reduce the effective parameter dimension of the inverse problem, increase parameter identifiability, and provide significant opportunities for parallel computation. Our framework will accommodate nonlinear, non-deterministic, and non-Gaussian relationships between the coarse- and fine-scale quantities. The specific coarse-scale quantity that we employ is entirely \textit{flexible} in that it is a consequence of the multiscale solution method chosen for the forward problem; our framework is thus, in principle, applicable to a variety of multiscale modeling methods and a broad array of problems containing multiscale structure.
 

\EditsText{In the Bayesian setting, knowledge about the values and structure of the parameters is represented via the assignment of probabilities \cite{Jaynes,Jeffreys,TarantolaBook,KaipioBook,Stuart2010,Gelman2014}. The prior probability distribution represents an initial state of knowledge, which is updated via the likelihood function to obtain the posterior distribution. 
The likelihood often couples a physics-based forward model (e.g., a PDE) with a probabilistic description of observation and model errors. 
Except in simple cases, the posterior distribution cannot be characterized in closed form and one must instead resort to sampling approaches, e.g., Markov chain Monte Carlo (MCMC) \cite{Robert2004,Brooks2011,Liu2004}, importance sampling \cite{Liu2004}, sequential Monte Carlo \cite{Doucet2001}, or variational approaches including transport maps \cite{Moselhy2011}. High-dimensional parameter spaces and computationally expensive models make all of these methods more challenging to apply. Significant research effort has thus been devoted to dimension reduction approaches that can make sampling more efficient, and reduce the number of forward model evaluations required.}


For example, in the Bayesian approach to inverse problems, Karhunen-Lo\`{e}ve (KL) expansions based on the prior \cite{Li2006,Efendiev2006,Marzouk2009} have proven useful in reducing parameter dimension. But this approach is practically limited to rather smoothing priors, and does not account for the forward model and the data in identifying relevant parameter directions. More recent work has introduced the notion of a likelihood-informed subspace (LIS), which contains parameter directions where the posterior distribution is most different from the prior \cite{CuiMartin2014,Spantini2014,CuiLaw2014}. Construction of the LIS requires finding the dominant eigenmodes of the Hessian of the log-likelihood, preconditioned by the prior covariance, at many points in the parameter space. These approaches rely on a heuristic globalization of Hessian information, but yield only a linear subspace of the original parameter space.

Other approaches to dimension reduction in Bayesian inversion explicitly take \EditsText{advantage} of multiscale structure. Existing approaches generally do so in two ways: (1) by using multiple discretizations of the parameter field itself \EditsText{\cite{Higdon2002,Holloman2006,Wan2011}}, or (2) by employing efficient multiscale numerical solvers for the forward model. 
%
\EditsText{\cite{Efendiev2006} and \cite{Dostert2006} are in this second category and are most related to the present work.} These efforts perform inference on a single discretization of the parameter field, but use a multiscale forward solver to drive a ``delayed-acceptance'' MCMC scheme \cite{Christen2005}, where proposed samples are first accepted or rejected according to an approximation of the forward model solution. These approaches improve MCMC efficiency but still require, per accepted sample, at least one new forward solve that explicitly resolves the fine scales. \cite{Nolen2009, Nolen2012} analyze the impact of the homogenization of elliptic operators on parameter estimates. This analysis requires specific forms for the parameter fields, however, and focuses on \textit{maximum a posteriori} (MAP) estimation rather than the full Bayesian posterior.
%



With the exception of \cite{Nolen2009} and \cite{Nolen2012}, the multiscale sampling approaches described above seek to accelerate asymptotically exact sampling of a posterior distribution on a fine-scale representation of the parameters.
Repeated calls to a forward solver that resolves the fine scales then remain an ``online'' part of each inference algorithm---in the sense that these calls must be performed after observations are obtained. This requirement carries significant computational expense.
We will exploit multiscale structure in a different manner, \textit{approximating} the posterior according to the conditional independence assumptions described earlier and reducing the online time required for posterior exploration.
%
We generate a posterior on the coarse-scale representation and use the conditional distribution of the fine scales to ``prolong'' samples of the coarse-scale posterior to the original high-dimensional parameter space containing fine-scale structure.  In contrast with many previous approaches, we do not require particular forms for the forward model or prior distribution; we avoid globally resolving the fine scales in any forward solves; and we allow for rather general multiscale models whose coarse or fine features may not even lie on a mesh.

To capture the stochastic and in general non-Gaussian relationship among the scales, we will rely on \textit{transport maps}---deterministic functions that push forward a reference distribution to a more complex target distribution. We will construct transport maps in a particular form that enables efficient marginalization and conditioning, both of which are basic operations for our multiscale inference framework. Importantly, these transport maps can be constructed \textit{a priori}, before any data is observed, significantly reducing online computational effort.  We will also take advantage of stationarity and locality, when present in the problem, to accelerate the construction of transport maps that represent the joint distribution of the fine and coarse scales.

The remainder of this paper is organized as follows. Section \ref{sec:multiover} introduces a multiscale decomposition of the inverse problem in a Bayesian setting. Section \ref{sec:maps} introduces transport maps and shows how they can be used to render the decomposition of Section \ref{sec:multiover} into an effective algorithm. Section \ref{sec:small} provides a small illustrative example, Section \ref{sec:application} describes strategies for applying the multiscale framework to elliptic PDE inverse problems, and Section \ref{sec:numerics} demonstrates the efficiency of our approach on two large subsurface flow applications.

 \section{A framework for multiscale Bayesian inference}\label{sec:multiover}


Let $\frv$ be a random variable taking values in
$\real^{\pd_\frv}$. In the discussion below, $\frv$ will represent the
parameter field we wish to infer, which may contain fine scale
structure. For simplicity, we will assume that all probability
distributions have densities with respect to Lebesgue measure. To keep
notation straightforward, we will also specify density functions by
their arguments, except where this might be ambiguous.

In the Bayesian context, the prior probability density $\pi(\frv)$
represents prior knowledge about the random variable
$\frv$. Observations are represented by an $\real^{\pd_\drv}$-valued
random variable $\drv$; conditioning on a particular value of these
observations then yields a posterior probability density according to
Bayes' rule:
\begin{equation} 
\pi(\frv  \vert  \drv) = \frac{\pi(\drv,\frv)}{\pi(\drv)} = 
\frac{\pi(\drv \vert \frv)\pi(\frv)}{\pi(\drv)}\propto \pi(\drv \vert \frv)\pi(\frv).
\label{eq:bayesRule} 
\end{equation} 
Here $\pi(\drv \vert \frv)$, viewed as a function of $\frv$, is the
likelihood function. The normalizing constant $\pi(\drv)$ is called the
evidence. In the Bayesian setting for inverse problems, the likelihood
function typically contains a deterministic \textit{forward} model
$f$, and compares $f(\frv)$ to $\drv$ according to a statistical
model for observation and model errors \cite{KaipioBook,Stuart2010}.

\subsection{A statistical definition for multiscale models}\label{sec:multiover:def}

Many systems contain parameterized features or dynamics that span
multiple scales of spatial or temporal variation. Of course, this is a
rather general observation. To be more precise, our framework employs
a specific definition of ``multiscale'' structure. We will say that a
model mapping $\frv$ to $\drv$ has multiscale structure if there is a
quantity $\crv$ such that $\drv$ and $\frv$ are conditionally
independent given $\crv$, i.e.:
\begin{equation}
\pi(\drv \vert \crv,\frv) = \pi(\drv \vert \crv)\label{eq:multidef}.
\end{equation} 
Very often $\crv$ will be naturally suggested by the forward model,
and will represent a coarse-scale quantity derived from the parameters
$\frv$ and the forward operator. To make the introduction of the
coarse-scale quantity useful, the dimension of $\crv$, denoted by
$\pd_\crv$, should be smaller than the dimension of $\frv$.

Even though this definition might seem abstract, many real systems
exhibit behavior that approximately satisfies \eqref{eq:multidef}.
For example, any deterministic model $f(\frv)$ that can be written as
$f(\frv)=\tilde{f} \left (g(\trv) \right )$, where
$\tilde{f}:\real^{\pd_\crv}\rightarrow\real^{\pd_\drv}$ and
$g:\real^{\pd_\frv}\rightarrow\real^{\pd_\crv}$, will yield a
posterior that satisfies \eqref{eq:multidef}.  In subsurface flow
applications, for instance, $\frv$ could be a conductivity field
discretized on a mesh that resolves the fine scales and
$\crv=g(\frv)$ could be a coarse ``upscaled'' conductivity field,
where $f$ and $\tilde{f}$ provide predictions of hydraulic
head. Further examples exist in any application where observables
depend on some aggregate, integrated, or homogenized behavior
of the fine scale parameter $\frv$.

Of course, in many systems, the equality \eqref{eq:multidef} is only
approximately satisfied. With the resulting likelihood approximation
$\pi(\drv \vert \crv,\frv) \approx \pi(\drv \vert \crv)$, only approximate
posterior samples of $\frv$ can be obtained. In applications where multiscale forward
modeling has been successful, however, this approximation error can be
quite small. Moreover, in practice, the computational advantages of
using the coarse parameter $\crv$ may greatly outweigh the drawbacks
of a posterior approximation.  As we will demonstrate in Section
\ref{chap:maps:twoD}, exploiting multiscale structure can allow us to
tackle very large problems where directly sampling $\pi(\frv  \vert  \drv)$
is otherwise intractable.

\subsection{Algorithmic building blocks: decoupling the coarse and
  fine scales}
In the usual ``single-scale'' setting, we obtain a posterior $\pi(\frv
\vert \drv)$ according to Bayes' rule as shown in
(\ref{eq:bayesRule}). Let us instead consider the joint posterior
distribution of the coarse- and fine-scale parameters $(\frv,\crv)$:
\begin{eqnarray}
\label{eq:bayesexpand} 
 \pi(\frv,\crv \vert \drv) & \propto & \pi(\drv \vert
  \frv,\crv)\pi(\crv,\frv) \\
  & \approx  &\pi(\drv \vert \crv)\pi(\crv,\frv) \nonumber \\
 & = & \pi(\drv \vert \crv)\pi(\crv)\pi(\frv \vert
 \crv) \, . \nonumber
\end{eqnarray}
In moving from the first line to the second, we applied the
conditional independence assumption \eqref{eq:multidef}. Then we
expanded the joint prior $\pi(\crv,\frv)$ into the marginal prior of
the coarse quantity $\crv$ and the conditional prior distribution of
the fine scales given the coarse. This resulting expression is the
foundation of our multiscale inference framework.  The three densities
on the right hand side of \eqref{eq:bayesexpand} can be understood as
a coarse likelihood, a coarse prior, and a downscaling
``prolongation'' density. Notice that only the downscaling density
involves the high-dimensional fine scale parameters $\frv$.

It is trivial to remove $\frv$ from \eqref{eq:bayesexpand} via
marginalization, leaving the posterior density of the coarse
parameters alone: $\pi(\crv \vert \drv)\propto \pi(\drv \vert
\crv)\pi(\crv)$. We can now break sampling the fine-scale posterior
$\pi(\frv \vert \drv)$ into two steps:
(1) coarse-scale inference, which samples the posterior $\pi(\crv
\vert \drv)$ directly and ignores the fine scale parameters; and (2)
fine-scale conditioning, which generates one or more samples of the
fine-scale parameter from $\pi(\frv \vert \crv^{(i)})$, for each
posterior sample $\crv^{(i)}$. Since $\pi(\frv \vert
\drv,\crv)=\pi(\frv \vert \crv)$ under the conditional independence
assumption, the combination of these two steps
will generate samples from the joint posterior $\pi(\frv, \crv \vert
\drv) = \pi( \frv \vert \crv, \drv) \pi( \crv \vert
\drv)$. Marginalizing out the coarse parameter (i.e., simply ignoring
the coarse component of each joint sample) will produce samples of the
fine-scale posterior $\pi(\frv \vert \drv) = \int \pi(\frv, \crv \vert
\drv) d \crv$.

While this two-step process is conceptually simple, two important
issues \EditsText{remain:}
\begin{enumerate}
\item Sampling the coarse-scale posterior in principle requires \EditsText{evaluating
  the prior density} $\pi(\crv)$ of the coarse parameter $\crv$. But the original
  inference problem only specifies a prior on the fine-scale parameter
  $\frv$.
\item Generating fine-scale posterior samples requires that we sample
  from the conditional density $\pi(\frv \vert \crv)$; in general
  cases, this task may be nontrivial.
\end{enumerate}
Both of these issues will be addressed through the construction of
transport maps that represent the joint prior distribution of the
coarse- and fine-scale quantities, with a particular  structure
described in the next section.

\section{Transport maps for multiscale inference}
\label{sec:maps}
Transport maps are deterministic nonlinear variable transformations
between (probability) measures
\cite{Villani2003,Villani2009}. Transport maps have recently been
used to accelerate Bayesian inference, coupled with MCMC
\cite{Parno2014} or in a standalone approach
\cite{Moselhy2011}. Compositions of many simple transport maps have
also been used for density estimation in
\cite{Tabak2010,Tabak2013}.
Here we will use transport maps to transform the joint prior density
$\pi(\crv,\frv)$ into a standard normal distribution that can be
easily sampled. Constructing this transformation with the appropriate
structure will enable easy characterization of the coarse prior density
$\pi(\crv)$ and sampling from downscaling density $\pi(\frv \vert
\crv)$.


\subsection{Transport maps}
\label{sec:maps:exact}
To define a transport map, consider two Borel probability measures on
$\real^{\pd}$, denoted by $\tmeas$ and $\rmeas$. We will call these
the \textit{target} and \textit{reference} measures, respectively, and
associate them with random variables $\trv \sim \tmeas$ and $\rrv \sim
\rmeas$. An \textit{exact} transport map $\emap: \real^{\pd}
\rightarrow \real^{\pd}$ is a deterministic transformation that pushes
forward $\tmeas$ to $\rmeas$, yielding
\begin{equation}
\EditsMath{\rmeas(A) = \tmeas \left ( \emap^{-1}(A) \right )} 
\label{eq:measconst}
\end{equation}
for any Borel measurable set $A \subseteq \real^{\pd}$.
\EditsText{This pushforward relationship is denoted concisely by $\rmeas = \emap_{\sharp} \tmeas$.}
 In terms of the random variables, we may write $\rrv \eqd \emap (\trv)$, where $\eqd$ denotes equality 
in distribution.

Existence of a $\emap$ satisfying \eqref{eq:measconst} is guaranteed
when $\tmeas$ has no atoms \cite{Brenier1991,McCann1995}, but there
can be infinitely many transport maps between two arbitrary
probability measures.  To regularize the problem, and for additional
reasons described below, we restrict our attention to maps with the
following lower triangular structure:
\begin{equation}
\emap(\trv_1,\trv_2,\ldots ,\trv_\pd) = \left[\begin{array}{l}\emap_1(\trv_1)\\ 
\emap_2(\trv_1,\trv_2)\\ \vdots \\ \emap_\pd(\trv_1,\trv_2,\ldots 
,\trv_\pd)\end{array}\right],\label{eq:lowtriform}
\end{equation}
where subscripts denote components of $\trv \in \real^{\pd}$.  This
lower triangular map is known as the Knothe-Rosenblatt (K-R)
rearrangement. For absolutely continuous target and reference
measures, the K-R map exists and is uniquely defined (up to ordering
of the coordinates), and has a lower triangular Jacobian with positive
diagonal entries ($\mu$-a.e.).\footnote{Application-specific orderings
  will be discussed in Section \ref{sec:application}. More general
  comments on useful orderings of the coordinates can be found in
  \cite{Parno2014thesis}.}  The map is thus a bijection between the
ranges of $\trv$ and $\rrv$. This lower triangular structure will be
particularly useful for our multiscale approach, as we explain in the
next section.
 
\subsection{Exact multiscale inference}
\label{sec:maps:exactinference}

Let the reference random variable be composed of independent standard
Gaussians, $\rrv\sim N(0,I)$, and let the target measure be the
\textit{joint prior} on $\trv := (\crv, \frv)$. In this section, we suppose
that we have a lower triangular map
$\emap:\real^{\pd_\crv+\pd_\frv}\rightarrow\real^{\pd_\crv+\pd_\frv}$
that pushes forward the prior to the reference, i.e.,
\begin{equation}
\rrv = \left[\begin{array}{c}\rrv_c \\ \rrv_f\end{array}\right]  \eqd  
\left[\begin{array}{l}\emap_c(\crv)\\ \emap_f(\crv,\frv)\end{array}\right] = 
\emap(\crv,\frv) ,
\label{eq:maptoref}
\end{equation}
where $\rrv_c$ and $\rrv_f$ are standard Gaussian random variables with
dimensions $\pd_\crv$ and $\pd_\frv$, respectively. In
Section~\ref{sec:maps:construction} we will discuss how to construct such a
map, but for now we proceed as if we have an exact transport map in
hand.  It will be convenient to define
$\efmap:\real^{\pd_\crv+\pd_\frv}\rightarrow\real^{\pd_\crv+\pd_\frv}$
as the lower triangular inverse of $\emap(\crv,\frv)$, such that
 \begin{equation}
\efmap(\rrv) = \left[\begin{array}{l}\efmap_c(\rrv_c) \\ 
\efmap_f(\rrv_c,\rrv_f)\end{array}\right]  \eqd  \left[\begin{array}{l}\crv\\ 
\frv\end{array}\right] .
\label{eq:maptotarget}
\end{equation}

We now wish to use the upper block of \eqref{eq:maptotarget} to rewrite the prior and posterior on the coarse-scale quantity
$\crv$.\footnote{\EditsText{Recall that we do not, in general, have a direct way of evaluating the prior density of the coarse-scale quantity $\pi_\crv(\crv)$. As we will show in Section~\ref{sec:maps:construction}, implementing our method only requires the ability to generate joint prior \textit{samples} of the fine-scale and coarse-scale parameters $(\frv, \crv)$.}}  \EditsText{Effectively we will ``transfer'' the coarse-scale inference problem to the reference variable $\rrv_c$ via the bijection $\emap_c = \efmap_c^{-1}$.} The pullback of the prior marginal $\pi_{\crv}(\crv)$ through $\efmap_c$ has the following density: 
$$
\pi_\crv \left ( \efmap_c(\rrv_c) \right ) | \det \nabla \efmap_c(\rrv_c) | = p(\rrv_c),
$$
where $p$ denotes the standard normal density. Thus, knowing the
transport map $\efmap_c$ enables us to evaluate the coarse-scale prior
density $\pi_\crv$. But we can go one step further, applying the same variable transformation to
the coarse posterior $\pi_{\crv \vert \drv}(\crv \vert \drv) \propto
\pi_{\drv \vert \crv}(\drv \vert \crv)\pi_{\crv}(\crv)$ in order to
obtain a posterior on $\rrv_c$:
\begin{eqnarray}
\label{eq:transformedcoarsepost}
\pi(r_c \vert \drv) & = & \pi_{\crv \vert \drv} \left( \efmap_c(\rrv_c)
  \vert \drv \right ) | \det \nabla \efmap_c(\rrv_c) | \\
& \propto &   \pi_{\drv \vert \crv} \left( \drv \vert
  \efmap_c(\rrv_c) \right ) \pi_\crv \left ( \efmap_c(\rrv_c) \right )
| \det \nabla \efmap_c(\rrv_c) | \nonumber \\
& = & \pi_{\drv \vert \crv} \left( \drv \vert \efmap_c(\rrv_c) \right
) p(\rrv_c) \nonumber \\
& = & \pi( \drv \vert \rrv_c) \, p(\rrv_c). \nonumber
\end{eqnarray}
Parameterizing the coarse-scale inference problem in terms of $\rrv_c$
is convenient as now the prior is simply standard normal.

Next, we can use the maps \eqref{eq:maptoref} and
\eqref{eq:maptotarget} to generate samples from the fine-scale
conditional density $\pi(\frv \vert \crv)$. Since $\emap_c$ is a
bijection, sampling from $\pi(\frv \vert \crv^\ast)$ for a fixed value
$\crv^\ast$ is equivalent to sampling $\pi(\frv \vert \rrv_c^\ast)$
when $\emap_c(\crv^\ast)=\rrv_c^\ast$.  
%
%
With the help of $\efmap_f$, we can simulate the random variable $\frv \vert
\rrv_c^\ast$ whose density is $\pi(\frv \vert \rrv_c^\ast)$ using the
fact that
\begin{equation}
\frv  \vert  \rrv_c^\ast \eqd \efmap_f(\rrv_c^\ast,\rrv_f).
\end{equation} 
According to this expression, samples of $\frv \vert \rrv_c^\ast$ can
be generated by first sampling the standard Gaussian $\rrv_f$ and then
evaluating the map $\efmap_f$.
 
We combine these coarse- and fine-scale sampling strategies to define
our complete multiscale framework.  The conditional independence
property (\ref{eq:multidef}) and the maps $\efmap$ and $\emap$ allow
us to sample the fine-scale posterior $\frv \vert \drv$ in two steps:
\begin{enumerate}

\item Use MCMC or any other standard sampling strategy to
  sample the coarse posterior $\pi(\rrv_c \vert \drv)$ defined in
  \eqref{eq:transformedcoarsepost}.

\item For each coarse posterior sample $\rrv_c^\ast$, generate one or
  more samples of $\rrv_f$ from a standard normal distribution and
  evaluate $\efmap_f(\rrv_c^\ast,\rrv_f)$ to obtain fine-scale
  posterior samples.

\end{enumerate}
This procedure is detailed on lines 10--14 of Algorithm
\ref{alg:completealg}.  Clearly, the maps $\emap$ and $\efmap$ are
critical to our approach. Next we will discuss how to construct these
transformations.

\subsection{Constructing transport maps from samples}\label{sec:maps:construction}
Our construction of the ``forward'' (reference to target) map $\efmap$
will depend on the ``inverse'' (target to reference) map $\emap$, so
we will first focus on the construction of $\emap$. Consider the
pullback of the standard normal reference measure through a candidate
map $\emap$; this pullback distribution has density
$\tilde{\pi}(\crv,\frv)$:
\begin{equation}
\tilde{\pi}(\crv,\frv) = \rd\left(\emap(\crv,\frv)\right)\left|\det \demap(\crv,\frv)\right|,
\end{equation}
where $\demap$ is the Jacobian of $\emap$ and $\det \demap $ is the
determinant of the Jacobian.  To simplify notation, we again let
$\jrv=(\crv,\frv)$, so that the target density is $\pi(\crv,\frv)=\pi(\jrv)$.

We evaluate the difference between the target density $\pi$ and
map-induced density $\tilde{\pi}$ using the Kullback-Leibler (KL)
divergence:
\begin{eqnarray}
\label{eq:klexpr} 
D_{\text{KL}}(\pi \Vert \tilde{\pi}) & = & \Ex{\pi} 
\left[\log{\left(\frac{\pi(\jrv)}{\tilde{\pi}(\jrv)}\right)}\right] \\
& = & \Ex{\pi} \left[ 
\log{\left ( \frac{\pi(\jrv)}{ \rd\left (\emap(\jrv)\right) \left|
        \det \demap(\jrv)\right |
}\right)}\right] \nonumber \\[10pt]
& = & \Ex{\pi} \left[\log{\pi(\jrv)} - \log{\rd\left(\emap(\jrv)\right)} - 
\log{\left|\det \demap(\jrv)\right|}\right], \nonumber 
\end{eqnarray}
where $\Ex{\pi}$ denotes the expectation with respect to the joint \textit{prior} 
density $\pi(\jrv)$.  Using (\ref{eq:klexpr}), we will find transport maps by 
solving the minimization problem
\begin{equation}
\min_{\emap \in \espace} \Ex{\pi} \left[- \log{\rd\left(\emap(\jrv)\right)} - 
\log{\left|\det \demap(\jrv)\right|}\right]. \label{eq:exactMin}
\end{equation}
Note that $\log{\pi(\jrv)}$ was removed from the objective because it does not 
depend on $\emap$.  $\espace$ is the space of monotone, continuously 
differentiable, and lower triangular maps from $\real^{\pd_\crv+\pd_\frv}$ to 
$\real^{\pd_\crv+\pd_\frv}$.  For a sufficiently smooth joint density $\pi(\jrv)$,  
$\espace$ will contain the Knothe-Rosenblatt map and hence the solution to 
(\ref{eq:exactMin}) will be an exact measure transformation, for which
$D_{\text{KL}}(\pi \Vert \tilde{\pi})=0$.   

In most situations, the minimization problem (\ref{eq:exactMin}) must
be approximated in two respects: first, because the expectation cannot
be computed exactly, and second, because the map might be represented
in a space $\ispace \subset \espace$ that does not include the exact
Knothe-Rosenblatt map. In these situations, we instead find an
approximate lower triangular map $\imap$ such that $\imap(\jrv)
\approxeqd \rrv$.
%
Suppose we have $K$ samples $x^{(k)}$ from the joint prior
$\pi(\jrv)$. We can use these samples to define a Monte Carlo
approximation of the expectation in (\ref{eq:exactMin}). As detailed
in \cite{Parno2014}, one can define the approximate map $\imap$ as the
solution of the minimization problem
\begin{equation}
\begin{aligned}
& \underset{\imap \in \ispace}{\min}
& & -\frac{1}{K}\sum_{k=1}^K \left[\log{\rd\left(\imap(\jrv^{(k)})\right)} + 
\sum_{i=1}^{\pd_{\crv}+\pd_{\frv}}\log{\frac{\partial 
\imap(\jrv^{(k)})}{\partial\jrv_i}} \right], \\
& \text{s.t.} 
& & \frac{\partial \imap_i(\jrv^{(k)})}{\partial \jrv_i}  \geq 
\lambda_{\text{min}} > 0, \quad k\in\{1,2,\ldots ,K\}, \quad 
i\in\{1,2,\ldots,\pd_{\crv}+\pd_{\frv}\},
\end{aligned}\label{eq:approxMin1}
\end{equation}
where $\lambda_{\text{min}}$ is a small positive scalar introduced to
ensure monotonicity and $\ispace$ is the space of maps spanned by
a finite set of basis functions $\{\basis_1,\basis_2,
\ldots,\basis_N\}$.  Note that the log-determinant term has been
expanded into a sum and that the absolute value has been removed; this
is a result of using a monotonically increasing lower triangular map.
In this work, we use multivariate Hermite polynomials to represent the map.  
%
%
These polynomial basis functions will also be combined with
problem-specific structure (discussed in Section
\ref{sec:application}) to facilitate map construction in high
dimensions.

While \cite{Parno2014} solved (\ref{eq:approxMin}) directly (though in a
different context, with samples generated via MCMC), the
applications in this paper are of much higher dimension. We have found
that additional constraints on $\imap$ can help obtain accurate maps
with fewer samples.  Using the fact that the reference density
$p(\rrv)$ is a standard normal, we constrain the output of $\imap$ to
have unit sample variance and zero sample mean.  With these
constraints, the optimization problem for $\imap_i$ becomes
\begin{equation}
\begin{aligned}
& \underset{\imap \in \ispace_i}{\min}
& & -\frac{1}{K}\sum_{k=1}^K 
\left[\sum_{i=1}^{\pd_{\crv}+\pd_{\frv}}\log{\frac{\partial 
\imap_i(\jrv^{(k)})}{\partial\jrv_i}}\right]\\
& \text{s.t.} 
& & \frac{\partial \imap_i(\jrv^{(k)})}{\partial \jrv_i}  \geq 
\lambda_{\text{min}} > 0, \quad k\in\{1,2,\ldots ,K\}, \quad 
i\in\{1,2,\ldots,\pd_{\crv}+\pd_{\frv}\}\\
& & &  \frac{1}{K}\sum_{k=1}^K \imap_i(\jrv^{(k)}) = 0, \quad 
i\in\{1,2,\ldots,\pd_{\crv}+\pd_{\frv}\}\\
& &  & \frac{1}{K}\sum_{k=1}^K  \imap_i^2(\jrv^{(k)}) = 1, \quad 
i\in\{1,2,\ldots,\pd_{\crv}+\pd_{\frv}\}.
\end{aligned}\label{eq:approxMin}
\end{equation}
Notice that we have removed the $\log{\rd\left(\imap(\jrv^{(k)})\right)} $ term from 
the objective.  Because the reference density $p(\cdot)$ is standard
normal, $\sum_{k=1}^K \log{\rd\left(\imap(\jrv^{(k)})\right)}\propto -
\sum_{k=1}^K \imap^2(\jrv^{(k)})$; hence, satisfying the variance constraint ensures that this term is
constant. The additional constraints in (\ref{eq:approxMin}) make it slightly
more difficult to solve than (\ref{eq:approxMin1}), but in our
experience, imposing the mean and variance information yields more
accurate maps, warranting the extra effort required during
optimization.  We should note that \eqref{eq:approxMin} is non-convex
because of the quadratic constraint. However, we have not found this
property to cause convergence issues in practice; future work might be able to explain this fact by extending the global minimum ideas from \cite{More1993}.

The structure of the constrained optimization problem
(\ref{eq:approxMin}) also enables several efficient solution
approaches. First, when the map parameterization is independent across
each dimension (e.g., each component of the map is represented with
its own expansion), the optimization problem in (\ref{eq:approxMin})
is \textit{separable}.  Thus we can independently solve
$\pd_{\crv}+\pd_{\frv}$ smaller optimization problems instead of one
large optimization problem. Next, we will ensure that each $\imap_i$
is linear in the coefficients of its basis representation (e.g., a
polynomial expansion); as a result, the objective can be evaluated using
efficient linear algebra routines.

Let each component $\imap_i$ of the map be expressed in the form
\begin{equation}
\imap_i(\jrv) = \sum_{\mathbf{j}\in\mathcal{J}_i } \alpha_{i,\mathbf{j}}\basis_\mathbf{j}(\jrv)\label{eq:mapform},
\end{equation}
where $\alpha_{i,\mathbf{j}}$ is the coefficient for the basis
function $\basis_\mathbf{j}(x)$, $\basis_\mathbf{j}(x)$ is a
multivariate Hermite polynomial whose degrees in each coordinate are
specified by the multi-index $\mathbf{j}$, and the multi-index set
$\mathcal{J}_i$ defines the basis functions used for output dimension
$i$ of the map. Note that the triangular structure of the map can be
encoded through the choices of $\mathcal{J}_i$. With this
representation, optimization over $\imap$ is equivalent to optimizing
over the map coefficients.  Using (\ref{eq:mapform}), we now define
two Vandermonde matrices $A_i$ and $G_i$ containing evaluations of the
basis functions and their derivatives, respectively.  
Let $\mathcal{J}_i = \{j_{i,1}, j_{i,2},\ldots, j_{i,|\mathcal{J}_i|}\}$, where $|\mathcal{J}_i|$ is the cardinality
of $\mathcal{J}_i$. Then these matrices take the form
\begin{equation}
\label{eq:vmA}
A_i = \left[\begin{array}{cccc} \basis_{j_{i,1}}(\jrv^{(1)}) & 
\basis_{j_{i,2}}(\jrv^{(1)}) & \ldots & \basis_{j_{i,|\mathcal{J}_i|}}(\jrv^{(1)}) \\ 
 \basis_{j_{i,1}}(\jrv^{(2)}) & \basis_{j_{i,2}}(\jrv^{(2)}) & \ldots & \basis_{j_{i,|\mathcal{J}_i|}}(\jrv^{(2)}) \\ 
 \vdots & \vdots & & \vdots \\
\basis_{j_{i,1}}(\jrv^{(K)}) & \basis_{j_{i,2}}(\jrv^{(K)}) & \ldots & \basis_{j_{i,|\mathcal{J}_i|}}(\jrv^{(K)})\end{array}\right],
\end{equation} 
and 
\begin{equation}
\label{eq:vmG} 
G_i = \left[\begin{array}{cccc} \frac{\partial\basis_{j_{i,1}}}{\partial \jrv_i}(\jrv^{(1)}) & \frac{\partial\basis_{j_{i,2}}}{\partial \jrv_i}(\crv^{(1)}) & \ldots &\frac{\partial\basis_{j_{i,|\mathcal{J}_i|}}}{\partial \jrv_i}(\jrv^{(1)}) \\ 
 \frac{\partial\basis_{j_{i,1}}}{\partial \jrv_i}(\jrv^{(2)}) & \frac{\partial\basis_{j_{i,2}}}{\partial \jrv_i}(\jrv^{(2)}) & \ldots & \frac{\partial\basis_{j_{i,|\mathcal{J}_i|}}}{\partial \jrv_i}(\jrv^{(2)}) \\ 
 \vdots & \vdots & & \vdots \\
 \frac{\partial\basis_{j_{i,1}}}{\partial \jrv_i}(\jrv^{(K)}) & \frac{\partial\basis_{j_{i,2}}}{\partial \jrv_i}(\jrv^{(K)}) & \ldots & \frac{\partial\basis_{j_{i,|\mathcal{J}_i|}}}{\partial \jrv_i}(\jrv^{(K)})\end{array}\right].
\end{equation} 
With these matrices in hand, the optimization problem from (\ref{eq:approxMin}) 
can be written simply as
\begin{equation}
\begin{aligned}
& \underset{\alpha_i}{\min}
& - c^\top \log{\left(G_i\alpha_i\right)}&\\
& \text{s.t.} 
& G_i\alpha_i&  \geq \lambda_{\text{min}},\\
& &  c^\top A_i\alpha_i &= 0, \\
& &  \alpha_i^\top A_i^\top A_i\alpha_i& = K,
\end{aligned}\label{eq:approxMinMatrix}
\end{equation}
for each $i \in \{ 1, \ldots, \pd_{\crv}+\pd_{\frv} \}$, where the
$\log$ is taken componentwise, $c$ is a length-$K$ vector of ones, and
$\alpha_i$ is a vector of the expansion coefficients.  This problem
can now be solved easily using any technique for constrained
optimization.  In particular, we use an augmented Lagrangian method
\cite{Andreani2007,Lewis2010} to handle the constraints and a full
Newton optimizer with backtracking line search on the unconstrained
subproblems. More advanced methods or
implementations like IPOPT \cite{Wachter2006} or ADMM \cite{Boyd2011}
might reduce the computational effort needed to solve the optimization
problem in (\ref{eq:approxMinMatrix}). In our experience, however, the
main computational cost of map construction does not lie in solving
(\ref{eq:approxMinMatrix}), but rather in generating the $K$ prior
samples of $x=(\crv,\frv)$ needed to define \eqref{eq:vmA} and \eqref{eq:vmG}.

\subsection{From inverse map to forward map}\label{sec:maps:inverse}
Recall that the posterior sampling procedure in
Section~\ref{sec:maps:exactinference} requires evaluations of the maps
$\efmap_c$ and $\efmap_f$ in \eqref{eq:maptotarget}, which push
forward the standard normal reference distribution to the joint prior
distribution $\pi(\jrv)$. As proposed in \cite{Parno2014}, taking
advantage of the lower triangular structure, it is possible to
evaluate the inverse of $\emap$ (that is, $\efmap$) using a sequence
of one-dimensional polynomial solves.  However, when many evaluations
of $\ifmap$ are required, it is more efficient to approximate $\efmap$
directly and to evaluate this approximation without explicitly
inverting $\emap$.  We now define a regression procedure that
constructs an approximation $\ifmap$ of $\efmap$, using the map
$\imap$ and the samples $\{ \jrv^{(k)} \}$.

Using each joint sample $\jrv^{(k)}$, we can compute $\rrv^{(k)} =
\imap\left(\jrv^{(k)}\right)$ to obtain sample pairs corresponding to
the input and output of $\efmap$.  With these pairs, we can construct
$\efmap$ with standard least squares regression.  The least squares
objective is
\begin{equation}
  \min_{\ifmap} \sum_{k=1}^{K} \left(\ifmap(\rrv^{(k)}) - 
    \jrv^{(k)}\right)^2\label{eq:lsqmap}.
\end{equation}  
As with $\imap$, we represent the triangular map $\ifmap$ using an
expansion of multivariate Hermite polynomials.  This allows us to find
the coefficients of the map that minimizes (\ref{eq:lsqmap}) using
standard linear least squares techniques (i.e., the QR decomposition
of a Vandermonde matrix).  The convergence properties of similar
regression-based maps were studied by \cite{Stavropoulou2011} in a
discrete optimal transport setting.

\subsection{A complete algorithm for multiscale inference}
The entire process of generating fine-scale posterior samples using
our multiscale inference framework is described in Algorithm \ref{alg:completealg}.

\begin{algorithm}
\footnotesize
\DontPrintSemicolon
\KwIn{\EditsText{A way to sample the fine-scale prior distribution $\pi(\frv)$; a way to sample the upscaling distribution 
$\pi(\gamma \vert \frv)$;} the number of coarse posterior samples $N$; and the number of fine samples per coarse sample $M$.}
\KwOut{Samples of the fine scale posterior $\pi(\frv \vert \drv)$}
\BlankLine
\BlankLine
\tcc{Step 1: Generate prior samples}
\For{$k \gets 1 \textbf{ to } K$}{
Sample $\frv^{(k)}$ from $\pi(\frv)$\;
\EditsText{Sample $\crv^{(k)}$ given  $\frv^{(k)}$}\;
}
\BlankLine
\BlankLine
\tcc{Step 2: Compute the forward map $\imap$}
\For{$i\gets 1 \textbf{ to } \pd_\frv+\pd_\crv$}{
Solve (\ref{eq:approxMinMatrix}) to get $\imap_i$\;
}
\BlankLine
\BlankLine
\tcc{Step 3: Generate sample pairs and solve for the inverse map $\ifmap$}
\For{$k \gets 1 \textbf{ to } K$}{
$\left(\rrv_c^{(k)},\rrv_f^{(k)}\right)=\imap\left(\crv^{(k)},\frv^{(k)}\right)$\;
}
Solve (\ref{eq:lsqmap}) to get $\ifmap_c$ and $\ifmap_f$\;
\BlankLine
\BlankLine
\tcc{Step 4: Sample the coarse posterior}
Generate $\left\{\rrv_c^{(1)},\rrv_c^{(2)},\ldots ,\rrv_c^{(N)}\right\}$ from $\pi(\rrv_c \vert \drv)$ \eqref{eq:transformedcoarsepost}, 
using MCMC or another sampling method\;
\BlankLine
\BlankLine
\tcc{Step 5: Generate fine scale posterior samples}
\For{$i \gets 1 \textbf{ to } N$}{
\For{$j \gets 1 \textbf{ to } M$}{
Sample $\rrv_f^{(i,j)}$ from a standard Gaussian\;
$\frv^{(iM+j)} \gets \ifmap_f\left(\rrv_c^{(i)},\rrv_f^{(i,j)}\right)$\;
}
}
\Return{Posterior samples 
$\left\{\frv^{(1)},\frv^{(2)},\ldots ,\frv^{(NM)}\right\}$}\;
\caption{Overview of the entire multiscale inference framework.}
\label{alg:completealg}
\end{algorithm}

\subsection{Choosing the number of fine scale samples}\label{sec:trimap:var}
After sampling the coarse-scale posterior (line 9 of Algorithm
\ref{alg:completealg}), we have a set of (possibly correlated) samples
$\{\rrv_c^{(1)},\rrv_c^{(2)},\ldots,\rrv_c^{(N)}\}$ from
$\pi(\rrv_c \vert \drv)$.  The next step is to ``prolong'' these samples
back to the fine scale by sampling $\pi(\frv \vert \rrv_c^{(i)})$ for each
$i$. If our goal is to minimize the computational effort required to
estimate the posterior expectation of a $\frv$-dependent quantity with
a certain accuracy, it is useful to consider how the variance of the
estimator depends on the number of fine scale samples $M$ produced for
each coarse scale sample. There is a tradeoff between reducing the
coarse-scale contribution to the variance (by increasing $N$) and
reducing the conditional fine-scale contribution to the variance (by
increasing $M$). The optimal choice depends on the computational cost
of generating each kind of sample, on the degree of correlation
among the coarse-scale samples, and on the magnitudes of the variances
on the coarse and fine scales.

For simplicity, suppose that we are interested in the fine-scale posterior mean
$\Ex{}[\frv \vert \drv]$.\footnote{The analysis in this section can be extended to
  the estimation of posterior expectations of more general functions
  $h(\frv)$ of the fine scale parameters.}  Using $M$ fine scale
samples for each of the $N$ coarse samples, a Monte Carlo estimator
of $\Ex{}[\frv \vert \drv]$ is
\begin{equation}
\widehat{{\frv}}(\mathbf{\rrv_c},\mathbf{\rrv_f}) = 
\frac{1}{NM}\sum_{i=1}^{N}\sum_{j=1}^{M} 
\ifmap_f\left(\rrv_c^{(i)},\rrv_f^{(i,j)}\right),
\label{eq:postmeanest}
\end{equation}
where $\mathbf{\rrv_c}$ is a set of $N$ correlated samples of
$\pi(\rrv_c \vert \drv)$ and $\mathbf{\rrv_f}$ is a set containing the $NM$
fine scale samples.

We wish to choose $M$ in order to minimize the variance of
the estimator $\widehat{\frv}$ for a given computational effort. This 
variance can be expanded (using the law of total variance) as
\begin{eqnarray}
\label{eq:finevar2}
\Var{\mathbf{\rrv_c},\mathbf{\rrv_f}} 
\left[\widehat{{\frv}}(\mathbf{\rrv_c},\mathbf{\rrv_f})\right] &=& 
\Var{\mathbf{\rrv_c}}\left[\Ex{\mathbf{\rrv_f}}\left\{\widehat{\frv}(\mathbf{
\rrv_c},\mathbf{\rrv_f}) \vert \mathbf{\rrv_c}\right\}\right] + 
\Ex{{\mathbf{\rrv_c}}}\left[\Var{\mathbf{\rrv_f}}\left\{\widehat{\frv}
(\mathbf{\rrv_c},\mathbf{\rrv_f}) \vert \mathbf{\rrv_c}\right\}\right]\\
&=&\frac{C_1}{N} + \frac{C_2}{NM}, \nonumber
\end{eqnarray}
where $C_1$ and $C_2$ are constants depending on the form of $\pi(\frv
\vert \rrv_c)$ and $\pi(\rrv_c \vert \drv)$, and on the degree of
autocorrelation in $\mathbf{\rrv_c}$ (e.g., how well the coarse MCMC
chain mixes).  More details on derivation of $C_1$ and $C_2$ can be
found in \cite{Parno2014thesis}. In Section \ref{sec:numerics}, we
will discuss the estimation of $C_1$ and $C_2$.

The expression in (\ref{eq:finevar2}) is quite intuitive: part of the
estimator variance stems from limited coarse posterior sampling (the
$C_1$ term) and part of the variance is a result of limited
coarse-to-fine sampling (the $C_2$ term).  However, the coarse samples
have a different computational cost than the fine samples.  With this
in mind, we now try to find the values of $N$ and $M$ that minimize
$\Var{\mathbf{\rrv_c},\mathbf{\rrv_f}}
[\widehat{{\frv}}(\mathbf{\rrv_c},\mathbf{\rrv_f})]$ for a fixed
computational cost.

Let $t_{\text{tot}}$ be the total sampling time, which for this discussion is fixed \textit{a priori}.  Let $t_{\text{c}}$ be the average time it takes to generate one coarse sample and let $t_{\text{f}}$ 
be the average time required to generate a fine sample from $\pi(\frv \vert \rrv_c)$ using 
$\ifmap_f (\rrv_c,\rrv_f)$.  Because $t_{\text{tot}}$ is fixed, $t_{\text{c}}$ and $t_{\text{f}}$ must satisfy the constraint
\begin{equation}
 t_{\text{tot}}=t_{\text{c}}N + t_{\text{f}}NM\label{eq:tottime}.
 \end{equation}
Solving \eqref{eq:tottime} for $M$, plugging the result into
\eqref{eq:finevar2}, and minimizing over $N$, we find that minimum
variance under the constraint \eqref{eq:tottime} occurs at the
following optimal value of $N$:
\begin{equation}
N^\ast =  \frac{t_{\text{tot}}\left(C_1t_{\text{c}} - \sqrt{C_1C_2t_{\text{c}}t_{\text{f}}}\right)}{C_1t_{\text{c}}^2-C_2t_{\text{c}}t_{\text{f}}}\label{eq:optN},
\end{equation}
which corresponds to an optimal value of $M$:
 \begin{equation}
M^\ast = \frac{t_{\text{c}}}{t_{\text{f}}}\left[\frac{C_1t_{\text{c}}-C_2t_{\text{f}}}{\left(C_1t_{\text{c}} - 
\sqrt{C_1C_2t_{\text{c}}t_{\text{f}}}\right)}-1\right]\label{eq:optK2}.
\end{equation}
Notice that the optimal number of fine samples $M^\ast$ does not depend on the number of coarse samples $N$ or the total time $t_{\text{tot}}$.  As we will show in Section \ref{sec:numerics}, the values of $C_1$ and $C_2$ can be estimated and \eqref{eq:optK2} can be used as a guideline for choosing $M$.  The qualitative behavior of \eqref{eq:optK2} is also informative.  Figure \ref{fig:optimalMs} illustrates $M^\ast$ for varying $t_{\text{c}}$ and $t_{\text{f}}$, with fixed $C_1$ and $C_2$.  As one would expect, when fine samples are less expensive \EditsText{than} coarse samples ($t_{\text{f}}<t_{\text{c}}$), it is usually advantageous to produce more than one fine sample per coarse sample. This advantage diminishes as $t_{\text{f}}\rightarrow t_{\text{c}}$.

\begin{figure}
\centering
%
%
%
%
\begin{tikzpicture}

\definecolor{mycolor1}{rgb}{0,0.75,0.75}
\definecolor{mycolor2}{rgb}{0.75,0,0.75}

\begin{axis}[%
view={0}{90},
width=0.8\textwidth,
height=0.3\textwidth,
scale only axis,
xmin=0.2, xmax=4,
xlabel={$t_{\text{f}}$},
ymin=0, ymax=7,
ylabel={$M^\ast$},
legend style={nodes=right}]
\addplot [
color=blue,
solid, very thick
]
coordinates{
 (0.2,1.87092223248322)(0.277551020408163,1.58817852567391)(0.355102040816327,1.40408742400638)(0.43265306122449,1.27204021978659)(0.510204081632653,1.17138260188548)(0.587755102040816,1.09137130228188)(0.66530612244898,1.0257935727951)(0.742857142857143,0.970773878447974)(0.820408163265306,0.923752756006095)(0.897959183673469,0.882962858273726)(0.975510204081633,0.847139291427035)(1.0530612244898,0.815349713337019)(1.13061224489796,0.786889755534167)(1.20816326530612,0.761216021298213)(1.28571428571429,0.737901679689705)(1.36326530612245,0.716606168824694)(1.44081632653061,0.697053999070647)(1.51836734693878,0.67901959288881)(1.59591836734694,0.66231622939785)(1.6734693877551,0.646787841718692)(1.75102040816327,0.632302836034187)(1.82857142857143,0.618749368686546)(1.90612244897959,0.606031691526857)(1.98367346938775,0.594067291223862)(2.06122448979592,0.582784626420497)(2.13877551020408,0.57212132047627)(2.21632653061224,0.562022705216291)(2.29387755102041,0.552440637863462)(2.37142857142857,0.543332532583735)(2.44897959183673,0.534660562101476)(2.5265306122449,0.526390995179571)(2.60408163265306,0.518493643458827)(2.68163265306122,0.510941396944024)(2.75918367346939,0.503709831822405)(2.83673469387755,0.496776877669493)(2.91428571428571,0.490122533698682)(2.99183673469388,0.483728625735619)(3.06938775510204,0.477578597185177)(3.1469387755102,0.471657328511136)(3.22448979591837,0.465950980743357)(3.30204081632653,0.460446859322093)(3.37959183673469,0.455133295228025)(3.45714285714286,0.449999540862943)(3.53469387755102,0.445035678565556)(3.61224489795918,0.440232539989421)(3.68979591836735,0.435581634850999)(3.76734693877551,0.431075087787433)(3.84489795918367,0.426705582255275)(3.92244897959184,0.422466310560727)(4,0.418350929244815) 
};

\addlegendentry{$t_{\text{c}}=1.0001$};

\addplot [
color=green!50!black,
solid, very thick
]
coordinates{
(0.2,3.37272330617263)(0.277551020408163,2.86301944298015)(0.355102040816327,2.53115725316107)(0.43265306122449,2.29311492544982)(0.510204081632653,2.11165880293195)(0.587755102040816,1.9674219286007)(0.66530612244898,1.84920454213435)(0.742857142857143,1.7500201921912)(0.820408163265306,1.66525491825908)(0.897959183673469,1.59172271240371)(0.975510204081633,1.52714334255267)(1.0530612244898,1.46983606967084)(1.13061224489796,1.41853112427663)(1.20816326530612,1.37224892167585)(1.28571428571429,1.33022001104245)(1.36326530612245,1.2918304593207)(1.44081632653061,1.25658363961289)(1.51836734693878,1.22407290186738)(1.59591836734694,1.19396164317404)(1.6734693877551,1.16596852078579)(1.75102040816327,1.13985630969869)(1.82857142857143,1.11542338864554)(1.90612244897959,1.09249715183455)(1.98367346938775,1.07092885196314)(2.06122448979592,1.0505895209758)(2.13877551020408,1.03136671210939)(2.21632653061224,1.0131618747003)(2.29387755102041,0.995888221460685)(2.37142857142857,0.979468982638964)(2.44897959183673,0.963835966766822)(2.5265306122449,0.948928366330406)(2.60408163265306,0.934691760584245)(2.68163265306122,0.921077278168997)(2.75918367346939,0.908040890123399)(2.83673469387755,0.89554280995407)(2.91428571428571,0.883546982116879)(2.99183673469388,0.872020643913113)(3.06938775510204,0.860933948664336)(3.1469387755102,0.850259640287264)(3.22448979591837,0.839972771183158)(3.30204081632653,0.830050456789118)(3.37959183673469,0.820471661290429)(3.45714285714286,0.811217009924029)(3.53469387755102,0.802268624059371)(3.61224489795918,0.793609975860505)(3.68979591836735,0.78522575983976)(3.76734693877551,0.77710177903085)(3.84489795918367,0.769224843854771)(3.92244897959184,0.761582682038979)(4,0.754163858189983) 
};

\addlegendentry{$t_{\text{c}}=3.25$};

\addplot [
color=red,
solid, very thick
]
coordinates{
 (0.2,4.38750213675162)(0.277551020408163,3.72443950579855)(0.355102040816327,3.29272722620731)(0.43265306122449,2.98306315754239)(0.510204081632653,2.74701084817661)(0.587755102040816,2.55937624643844)(0.66530612244898,2.40558982856864)(0.742857142857143,2.27656307250138)(0.820408163265306,2.16629377770959)(0.897959183673469,2.07063733600858)(0.975510204081633,1.98662744326317)(1.0530612244898,1.91207766274597)(1.13061224489796,1.84533617905205)(1.20816326530612,1.78512867183293)(1.28571428571429,1.73045417930284)(1.36326530612245,1.68051404934915)(1.44081632653061,1.63466223087987)(1.51836734693878,1.59236972171828)(1.59591836734694,1.55319864248521)(1.6734693877551,1.51678300054151)(1.75102040816327,1.48281419505713)(1.82857142857143,1.45102994132616)(1.90612244897959,1.42120570024127)(1.98367346938775,1.39314796968308)(2.06122448979592,1.36668897792301)(2.13877551020408,1.3416824454211)(2.21632653061224,1.31800016977002)(2.29387755102041,1.29552925128122)(2.37142857142857,1.2741698218604)(2.44897959183673,1.25383317271478)(2.5265306122449,1.23444020067675)(2.60408163265306,1.2159201109863)(2.68163265306122,1.19820932795873)(2.75918367346939,1.18125057527928)(2.83673469387755,1.1649920955671)(2.91428571428571,1.14938698495175)(2.99183673469388,1.13439262315356)(3.06938775510204,1.11997018327997)(3.1469387755102,1.10608420848713)(3.22448979591837,1.09270224498833)(3.30204081632653,1.07979452275517)(3.37959183673469,1.06733367675542)(3.45714285714286,1.05529450278266)(3.53469387755102,1.04365374291665)(3.61224489795918,1.03238989645632)(3.68979591836735,1.02148305282681)(3.76734693877551,1.01091474350456)(3.84489795918367,1.00066781045409)(3.92244897959184,0.990726288943914)(4,0.981075302920219) 
 };

\addlegendentry{$t_{\text{c}}=5.5$};

\addplot [
color=mycolor1,
solid, very thick
]
coordinates{
 (0.2,5.20817506426195)(0.277551020408163,4.42108798078296)(0.355102040816327,3.90862483901742)(0.43265306122449,3.54103876601949)(0.510204081632653,3.26083337507454)(0.587755102040816,3.03810212081947)(0.66530612244898,2.85555028111491)(0.742857142857143,2.70238934520489)(0.820408163265306,2.57149441374087)(0.897959183673469,2.45794563840693)(0.975510204081633,2.35822187419882)(1.0530612244898,2.26972771605732)(1.13061224489796,2.19050238002503)(1.20816326530612,2.11903318684712)(1.28571428571429,2.05413194696825)(1.36326530612245,1.99485062209956)(1.44081632653061,1.94042231866868)(1.51836734693878,1.89021908576871)(1.59591836734694,1.84372113961547)(1.6734693877551,1.80049402942637)(1.75102040816327,1.76017143124355)(1.82857142857143,1.72244200056562)(1.90612244897959,1.6870391987236)(1.98367346938775,1.65373332943892)(2.06122448979592,1.62232523963856)(2.13877551020408,1.59264128850629)(2.21632653061224,1.56452929364758)(2.29387755102041,1.53785523772766)(2.37142857142857,1.51250057253798)(2.44897959183673,1.48835999649502)(2.5265306122449,1.46533961035222)(2.60408163265306,1.44335537734052)(2.68163265306122,1.42233183007883)(2.75918367346939,1.40220097883976)(2.83673469387755,1.3829013851346)(2.91428571428571,1.3643773718239)(2.99183673469388,1.34657834659547)(3.06938775510204,1.3294582200692)(3.1469387755102,1.31297490327415)(3.22448979591837,1.29708987201194)(3.30204081632653,1.28176778783378)(3.37959183673469,1.26697616713646)(3.45714285714286,1.25268509132045)(3.53469387755102,1.23886695212106)(3.61224489795918,1.22549622717692)(3.68979591836735,1.21254928168254)(3.76734693877551,1.20000419261637)(3.84489795918367,1.18784059256895)(3.92244897959184,1.1760395306398)(4,1.16458334824091) 
 };

\addlegendentry{$t_{\text{c}}=7.75$};

\addplot [
color=mycolor2,
solid, very thick
]
coordinates{
 (0.2,5.91607978309962)(0.277551020408163,5.02201037785608)(0.355102040816327,4.43989230479309)(0.43265306122449,4.02234325773161)(0.510204081632653,3.70405183549043)(0.587755102040816,3.45104654014144)(0.66530612244898,3.24368191915268)(0.742857142857143,3.06970306757461)(0.820408163265306,2.9210166566553)(0.897959183673469,2.79203412326113)(0.975510204081633,2.67875572189276)(1.0530612244898,2.57823327526922)(1.13061224489796,2.48823948607689)(1.20816326530612,2.40705606891882)(1.28571428571429,2.33333333333333)(1.36326530612245,2.26599438192643)(1.44081632653061,2.20416808354307)(1.51836734693878,2.14714113503594)(1.59591836734694,2.09432310265442)(1.6734693877551,2.04522048426772)(1.75102040816327,1.99941716449294)(1.82857142857143,1.95655948031232)(1.90612244897959,1.91634466463151)(1.98367346938775,1.87851180043197)(2.06122448979592,1.84283466538927)(2.13877551020408,1.80911601710857)(2.21632653061224,1.77718298828483)(2.29387755102041,1.74688334570101)(2.37142857142857,1.71808242785836)(2.44897959183673,1.69066062038877)(2.5265306122449,1.66451126108764)(2.60408163265306,1.63953889075394)(2.68163265306122,1.61565778434157)(2.75918367346939,1.59279071083451)(2.83673469387755,1.57086788091188)(2.91428571428571,1.54982604969517)(2.99183673469388,1.5296077482722)(3.06938775510204,1.51016062270971)(3.1469387755102,1.49143686322651)(3.22448979591837,1.47339270934445)(3.30204081632653,1.45598801934788)(3.37959183673469,1.43918589440258)(3.45714285714286,1.42295234931805)(3.53469387755102,1.40725602326355)(3.61224489795918,1.39206792483148)(3.68979591836735,1.37736120673021)(3.76734693877551,1.36311096612082)(3.84489795918367,1.34929406721817)(3.92244897959184,1.33588898328056)(4,1.3228756555323) 
};

\addlegendentry{$t_{\text{c}}=10.0$};

\addplot[color=black, dotted, very thick] coordinates{ (0.2,1)(4.0,1)};

\end{axis}
\end{tikzpicture}
\caption{Optimal $M$ from \eqref{eq:optK2} for varying sample costs $t_{\text{c}}$ and $t_{\text{f}}$. 
$C_1=1$ and $C_2=0.7$ are fixed in this illustration.}
\label{fig:optimalMs}
\end{figure}
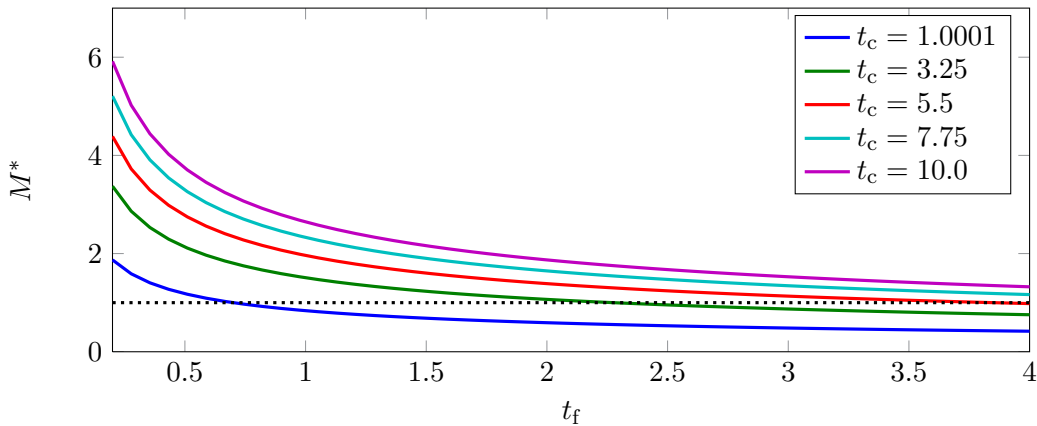

\newcommand{\PlotSize}{0.25\textwidth}

\section{A small proof-of-concept example}\label{sec:small}
For illustration, we now consider a small multiscale inference problem
with only two fine scale parameters, $\frv_1$ and $\frv_2$.  The fine
scale prior is a bivariate Gaussian with zero mean and \EditsText{identity covariance}.
The fine-to-coarse model defines the coarse-scale quantity $\crv$ as
a modified harmonic mean of two \textit{a priori} log-normal random
variables associated with the fine scale:
\begin{equation}
\crv = \frac{1}{1 + \exp(-\frv_1) + \exp(-\frv_2)} + \eta_f,
\end{equation}
where $\eta_f\sim N(-0.3, 1.5\times 10^{-3})$. The data $\drv \in \real$ are related nonlinearly to the coarse quantity through
\begin{equation}
\drv = \text{atan}(\gamma)  + \eta_c,
\end{equation}
where $\eta_c\sim N(0, 10^{-2})$. We have thus defined $\pi(\frv_1,
\frv_2)$, $\pi(\crv \vert \frv_1, \frv_2)$, and $\pi(\drv \vert
\crv)$. 

We solve this problem using Algorithm \ref{alg:completealg}, producing
samples from the posterior $\pi(\frv_1, \frv_2 \vert \drv)$. The maps
$\imap$ and $\ifmap$ are represented with total-degree truncated
multivariate Hermite polynomials. Figure \ref{fig:smallresults} \EditsText{and Table \ref{tab:SmallConvTable}}
compare the true posterior density with the posterior density obtained using third,
fifth, and seventh degree maps.  \EditsText{Complete implementation details and
   code for this example can be found at \url{http://muq.mit.edu/examples}.}
 As the map degree is increased, the
posterior density converges to the truth. \EditsText{This convergence is also seen in Table \ref{tab:SmallConvTable}. Note that some of the small wiggles in the plots are artifacts of the kernel density estimator used to visualize the multiscale posterior densities.}
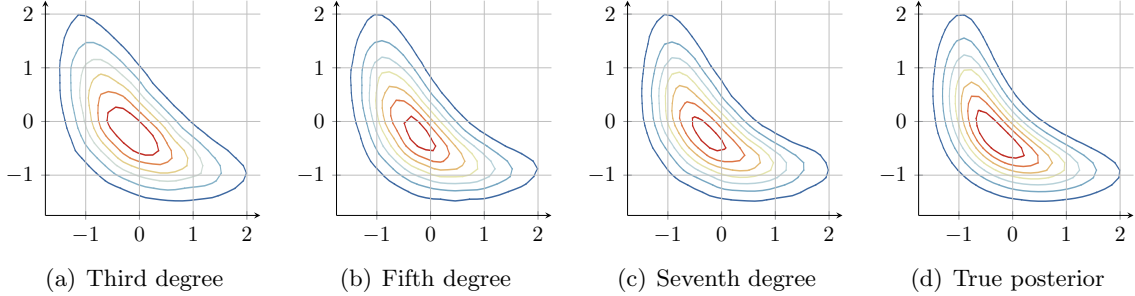
\begin{figure}
\centering

\subfigure[Third degree]{
\begin{tikzpicture}[scale=0.75]

\begin{axis}[%
view={0}{90},
width=\PlotSize,
height=\PlotSize,
scale only axis,
axis on top=true,
xmin=-1.75, xmax=2.25,
xmajorgrids,
ymin=-1.75, ymax=2.25,
ymajorgrids,
axis lines=left]

\addplot[thick,blue] graphics[xmin=-1.5,ymin=-1.5,xmax=2,ymax=2] {pgf_figures/ThirdSmallPosterior.eps};

\end{axis}
\end{tikzpicture}
}\subfigure[Fifth degree]{
\begin{tikzpicture}[scale=0.75]

\begin{axis}[%
view={0}{90},
width=\PlotSize,
height=\PlotSize,
scale only axis,
axis on top=true,
xmin=-1.75, xmax=2.25,
xmajorgrids,
ymin=-1.75, ymax=2.25,
ymajorgrids,
axis lines=left]

\addplot[thick,blue] graphics[xmin=-1.5,ymin=-1.5,xmax=2,ymax=2] {pgf_figures/FifthSmallPosterior.eps};

\end{axis}
\end{tikzpicture}
}\subfigure[Seventh degree]{
\begin{tikzpicture}[scale=0.75]

\begin{axis}[%
view={0}{90},
width=\PlotSize,
height=\PlotSize,
scale only axis,
axis on top=true,
xmin=-1.75, xmax=2.25,
xmajorgrids,
ymin=-1.75, ymax=2.25,
ymajorgrids,
axis lines=left]

\addplot graphics[xmin=-1.5,ymin=-1.5,xmax=2,ymax=2] {pgf_figures/SeventhSmallPosterior.eps};

\end{axis}
\end{tikzpicture}
}\subfigure[True posterior]{
\begin{tikzpicture}[scale=0.75]

\begin{axis}[%
view={0}{90},
width=\PlotSize,
height=\PlotSize,
scale only axis,
axis on top=true,
xmin=-1.75, xmax=2.25,
xmajorgrids,
ymin=-1.75, ymax=2.25,
ymajorgrids,
axis lines=left]

\addplot[thick,blue] graphics[xmin=-1.5,ymin=-1.5,xmax=2,ymax=2] {pgf_figures/TrueSmallPosterior.eps};

\end{axis}
\end{tikzpicture}
}

\caption{\EditsText{Convergence of the multiscale posterior to the true posterior as the total polynomial degree is increased.  In all cases, $150\,000$ samples of $\pi(\gamma,\theta)$ were used to build the maps.}}
\label{fig:smallresults}
\end{figure}

\begin{table}
\centering
\caption{\EditsText{Convergence of posterior from multiscale method on small test problem.  The KL divergence between the true posterior $\pi$ and the multiscale posterior $\tilde{\pi}$ was computed using quadrature with the exact posterior density and a kernel density estimate of the multiscale posterior.  The values reported here are the average KL divergences obtained from $30$ runs of the multiscale method.}}
\label{tab:SmallConvTable}
\begin{tabular}{|l|cccc|}\hline
Map Order & 1 & 3 & 5 & 7\\\hline
Error, $D_{KL}\left[\pi \| \tilde{\pi}\right]$ &  2.366e-01 & 5.549e-2 & 3.507e-2 & 3.407e-2\\\hline
\end{tabular}
\end{table}

\section{Application in simple groundwater flow}\label{sec:application}

To illustrate the accuracy and performance of our multiscale approach, we now 
consider an inverse problem from subsurface hydrology. Our goal is to 
characterize subsurface structure by inferring a spatially distributed 
conductivity field using limited observations of hydraulic head.  An elliptic 
equation, commonly called the pressure equation, will serve as a
simple steady-state model of groundwater flow in a confined aquifer: 
\begin{equation}
-\nabla\cdot\left(\kappa(x)\nabla h(x)\right) = f(x)\label{eq:press1},
\end{equation}
where $x \in \Omega \subset \real^d$ is a spatial coordinate in $d=1$
or $d=2$ dimensions, $\kappa(x)$ is the hydraulic 
conductivity field we wish to characterize, $f(x)$ contains well or recharge 
terms, and $h(x)$ is the hydraulic head that we can measure at several 
locations throughout the domain.  See \cite{BearPM} for a derivation 
of this model and a comprehensive discussion of flow in porous media.

The elliptic model in (\ref{eq:press1}) acts as a nonlinear lowpass
filter, removing high frequency features of $\kappa(x)$ from $h(x)$.
This means that some features of $\kappa(x)$ cannot be estimated even
when $h(x)$ is observed with high precision. Variational
discretization methods such as multiscale finite element methods
(MsFEM)~\cite{msfem,Aarnes08}, multiscale finite volume methods
~\cite{Jenny06}, variational multiscale
methods~\cite{Hughes98,Juanes08}, heterogeneous multiscale
methods~\cite{Weinan07}, and subgrid upscaling~\cite{Arbogast00} take
advantage of this smoothing to create a smaller, easy-to-solve, coarse
scale problem. The common idea behind all of these strategies is to
(implicitly or explicitly) coarsen the elliptic operator in a way that
allows for more efficient, yet accurate, solutions. In the examples
below, MsFEM will be used to solve (\ref{eq:press1}) and to define the
coarse parameter $\crv$.

\subsection{Multiscale finite element method}
Here, we present a very brief description of MsFEM, with only enough
detail to understand its use in the multiscale inference setting. See
\cite{msfem} for a comprehensive discussion of MsFEM theory and
implementation.

Let $\Omega$ be the spatial domain on which we want to solve
(\ref{eq:press1}), and let $\mathcal{P}$ be a mesh discretization of
$\Omega$. 
Then, for a suitable function space $X$, the weak formulation of
(\ref{eq:press1}) seeks $u\in X$ such that
\[
\int_\Omega \kappa(x) \nabla u \cdot \nabla v dx = \int_\Omega f v dx, \quad 
\forall v\in X.
\]
Projecting $u$ and $v$ onto a finite number of \textit{nodal} basis functions 
$\{\phi_1,\phi_2,\ldots,\phi_B\}$ associated with $\mathcal{P}$ yields a finite-dimensional linear system $A\tilde{u} = 
b$, where $\tilde{u}$ is a vector of coefficients of the basis functions and 
\begin{equation}
 A_{ij} = \int_\Omega\kappa(x) \nabla \phi_i \cdot \nabla \phi_j dx = \sum_{C\in\mathcal{P}} 
\int_C \kappa(x) \nabla \phi_i \cdot \nabla \phi_j dx \,.
\end{equation}
Here $C$ is an element in the discretization $\mathcal{P}$ of
$\Omega$. We refer to the quantity
\[
e_{ij, C}=\int_C \kappa(x) \nabla \phi_i \cdot \nabla \phi_j \, dx
\] 
as an \textit{elemental integral}.  Note that the elemental integrals
are assembled to construct $A$, and thus are sufficient to
characterize the coefficients $\tilde{u}$---and hence the values of the
pressure head $h(x)$ at the coarse element nodes. This property will
be important for the use of MsFEM in the context of inference.

For the pressure equation in (\ref{eq:press1}), the key difference
between an MsFEM formulation and a standard continuous Galerkin
formulation lies in the choice of basis functions $\{ \phi_i \}$. In
the standard Galerkin setting, one might use hat functions or another
standard basis. In the MsFEM context, however, the basis functions are
chosen to satisfy a homogeneous version of (\ref{eq:press1}) over each
element $C$. These basis functions depend on---and thus encode---local
fine-scale spatial variation in the coefficient $\kappa(x)$.
The advantage of this choice is that a good approximation of
$h(x)$ can be achieved with a coarse discretization $\mathcal{P}$ and
subsequently a smaller linear system.  See \cite{msfem},
\cite{Jenny03}, or \cite{Jenny06} for important implementation
details, such as the choice of boundary conditions when solving for
the MsFEM basis functions.

\subsection{Multiscale framework applied to MsFEM}
Recall that our goal is to accelerate inference of the conductivity
field $\kappa(x)$, which may contain fine-scale spatial features that
are represented with a high-dimensional discretization. The MsFEM
approach implies that elemental integrals are sufficient to describe
the head $h(x)$,\footnote{Given the elemental integrals $\{e_{ij,C}
  \}$, we can solve directly for $h(x)$ at the nodes of the coarse
  discretization $\mathcal{P}$. We will thus construct the coarse mesh
  so that observations occur only at the coarse nodes. Representing
  $h(x)$ at other points in space requires explicit access to the
  MsFEM basis functions $\{ \phi_i(x) \}$.}  which is equivalent to the
conditional independence assumption in (\ref{eq:multidef}).  The
elemental integrals can therefore be used to define the coarse
parameters $\crv$ in the multiscale inference framework. The exact
relationship between the elemental integrals $e_{ij,C}$ and the coarse
parameters $\crv$ will depend on the spatial dimension (one or two in
our case) and will be discussed below.  In all cases, however, the
fine scale parameter will be defined as $\frv=\log{\kappa}$. Here and
below, when we omit the argument $x$ and write $\kappa$ rather than
$\kappa(x)$, we refer to a discretized version of the conductivity,
$\kappa \in \real^{\pd_\frv}$. 

The large dimension of $\frv$ makes this problem interesting, but at
the same time makes construction of the transport map difficult.  For
example, a map of total degree three in $110$ dimensions will have
$234136$ polynomial coefficients!  Such a general form for the map is
infeasible, and a more judicious choice of map parameterization is
required.

\subsubsection{Strategies for building maps in one spatial dimension}
\label{sec:onedim}
In one spatial dimension, MsFEM produces one independent elemental integral per coarse 
element. Thus the coarse parameter dimension $\pd_\crv$ is equal to the number of coarse elements. The log-conductivity $\frv$ has finer scale features, so its dimension $\pd_\frv$ is typically much higher. Let $\finc$ be the number of fine elements in each coarse element, so that $\pd_\frv=\finc\pd_\crv$.  In the numerical example of Section~\ref{chap:multiscale:results1d}, we will use $\pd_\crv = 10$ and $\pd_\frv=100$, so $\finc=10$.

Now consider the impact of these dimensions on the map representation. A polynomial map of total degree three in $O(10)$ dimensions is straightforward to handle, for example, and thus we do not need to employ any special truncations or structure in defining the coarse-scale maps $\imap_c$ and $\ifmap_c$; polynomial maps of moderate degree will suffice in practice. The coarse-to-fine map $\imap_f$, however, would be a function from $\real^{110}$ to $\real^{100}$ in the example mentioned above. A generic total-order polynomial representation for such a  function is not tractable. Instead, we will take advantage of spatial \textit{locality} to construct a much more efficient parameterization of $\imap_f$.

First, let us endow $\frv$ with a Gaussian prior. (This does not sacrifice generality, as other prior distributions can be written as deterministic transformations of this Gaussian; indeed, we are actually using a log-normal prior on the conductivity $\kappa(x)$, since $\frv = \log \kappa$.)
Now each of $\frv$, $\rrv_c$, and $\rrv_f$ are multivariate Gaussians.  While this does not imply that  $\frv$, $\rrv_c$, and $\rrv_f$ are \textit{jointly} Gaussian, it does suggest that a linear map may characterize much of the joint structure between these random variables, allowing localized nonlinearities in the map to characterize non-Gaussian features of the joint distribution.  Consider the vector of reference random variables
\begin{equation*}
r = \left[\rrv_c, \rrv_f\right]^\top = \left[ \rrv_{c,1}, \rrv_{c,2},\dots, 
\rrv_{c,\pd_\crv}, \rrv_{f,1}, \rrv_{f,2},\dots, 
\rrv_{f,\pd_\frv}\right]^\top\label{eq:combinedref1d}.
\end{equation*}
We begin with a linear representation of the map $\imap_f$ and then enrich the collection of linear terms with selected nonlinear terms.  The initial set of linear multi-indices for output $i$ of $\imap_f$ is
\begin{equation}
\setf{J}_i^1 = \left\{\mathbf{j} : \mathbf{j} \in \mathbb{N}_0^{\pd_\crv+i},\,\, \|\mathbf{j}\|_1 \leq 1 \right\} \quad \text{for   } i=1\ldots\pd_\frv.
\end{equation}
Now, to introduce some nonlinear structure, we will take advantage of spatial locality in the definition of the coarse quantities $\crv$ according to MsFEM.  Every component of $\rrv$ is spatially related to one element in the coarse mesh and thus one component of $\crv$. Specifically, for our structured mesh, component $i$ of the fine scale field $\frv$ is related to component $\rho(i)$ of the coarse parameter, where
\begin{equation}
\rho(i) = \lfloor{i/\finc}\rfloor+1.
\end{equation}
Thus, to introduce local nonlinear terms for the $i^{\text{th}}$ output of $\ifmap_f$, we will include nonlinear terms in the inputs $\rrv_{c,\rho(i)}$ and $\rrv_{f,i}$.  Similarly, the $i^{\text{th}}$ output of $\imap_f$ will be nonlinear in $\crv_{\rho(i)}$ and $\frv_{i}$.
Combining these terms with the linear multi-indices yields a multi-index set tuned to the coarse/fine quantities in our one-dimensional application of MsFEM:
\begin{equation}
\setf{J}_i = \setf{J}_i^1 \cup \left\{\mathbf{j} : \mathbf{j}\in \mathbb{N}_0^{\pd_\crv+i},\,\, \|\mathbf{j}\|_1 \leq P,\,\, j_k=0 \text{  for  } k\notin\{\rho(i),i+\pd_\crv\} \right\}\label{eq:1dtuned}.
\end{equation}
where $P$ is the maximum polynomial degree of the nonlinear terms.  \EditsText{To help assure monotonicity of the map, $P$ should be odd. When constructing maps from a finite number of samples $K$, $P$ should also be chosen no larger than needed to adequately Gaussianize the target samples; this value is problem-dependent, of course, but in practice is often as low as $P=3$.}\footnote{\EditsText{For more information on assessing the quality of transports, e.g., via tests of Gaussianity, see \cite{Marzouk2016}.}} This localized multi-index for $\imap_i$ will result in nonlinear terms and interactions between $\crv_{\rho(i)}$ and $\frv_i$, and linear terms for all components of $\crv$ and components $\frv_k$ with $k\leq i$.  The localized set $\setf{J}_i$ will contain $i+\pd_\crv+(P+1)(P+2)/2$ terms, whereas a standard total degree multi-index set would require $\frac{(i+\pd_\crv+P)!}{(i+\pd_\crv)!P!}$ terms.

A further simplification occurs if we \EditsText{temporarily} restrict our attention to $P=1$ in the multi-index set above. Now the fine-scale maps $\ifmap_f$ and $\imap_f$ are completely linear. (The smaller coarse-scale maps $\ifmap_c$ and $\imap_c$ can remain nonlinear.) While the optimization and regression approach from Section \ref{sec:maps} can still be used, directly appealing to cross-covariance matrices can enable more efficient construction of $\ifmap_f$ when the prior $\pi(\frv)$ is Gaussian, $\frv \sim N(\mu_\frv, \Sigma_{\frv \frv})$.  Recall that when constructing $\ifmap_c$ using regression, samples of $\crv$ are pushed through $\imap_c$ to obtain corresponding samples of $\rrv_c$:  $\rrv_c^{(i)} = \imap_c(\crv^{(i)})$. Furthermore, because prior-distributed samples $\crv^{(i)}$ are generated by sampling the conditional $\crv^{(i)} \sim \pi(\crv|\frv^{(i)})$, each reference sample $\rrv_c^{(i)}$ is matched with a sample $\frv^{(i)}$ from the prior $\pi(\frv)$. 
We know that $\rrv_c$ and $\frv$ are individually Gaussian. \EditsText{For this special linear--map case, we can make the further assumption that they are \textit{jointly} Gaussian, or at least can be well-approximated as such. Under this temporary assumption,} an empirical estimate of the cross-covariance of $\rrv_c$ and $\frv$, denoted by $\Sigma_{\rrv_c\frv}$, can be used to define $\ifmap_f$. Then the conditional distribution of $\frv$ given $\rrv_c$ is simply
\begin{equation}
\pi(\frv | \rrv_c) = N\left(\mu_\frv + \Sigma_{\rrv_c\frv}^\top\rrv_c, 
\Sigma_{\frv\frv} - \Sigma_{\rrv_c\frv}^\top\Sigma_{\rrv_c\frv}\right),
\end{equation}
which implies that $\ifmap_f$ is given by
\begin{equation}
\frv = \ifmap_f(\rrv_c,\rrv_f) = \mu_\frv + \Sigma_{\rrv_c\frv}^\top\rrv_c + 
\left(\Sigma_{\frv\frv} - \Sigma_{\rrv_c\frv}^\top\Sigma_{\rrv_c\frv}\right)^{1/2}\rrv_f.\label{eq:xcovMap}
\end{equation}
The map in \eqref{eq:xcovMap} is similar to an ensemble Kalman filter (EnKF) update \cite{Enkf}.  In fact, the Kalman gain would be given by $\Sigma_{\rrv_c\frv}$ and the observation matrix by $H=\Sigma_{\rrv_c\frv}\Sigma_{\frv\frv}^{-1}$.  However, our approach estimates $H$ from samples, while the EnKF estimates $\Sigma_{\frv\frv}$.

In our numerical experiments, we have observed that this method of constructing a linear $\ifmap_f$ is much more efficient than the more general optimization and regression approach from Section \ref{sec:maps}.  Just as importantly, in our applications, this linear $\ifmap_f$ map seems to give the same posterior accuracy as a linear $\ifmap_f$ produced with optimization and regression.  The accuracy and efficiency tables in Section~\ref{sec:numerics} will illustrate the performance of this cross-covariance
approach for constructing the map, \EditsText{and compare it to the nonlinear $P>1$ case.}

\subsubsection{Strategies for building maps in two spatial dimensions}
\label{sec:twodim}
In two spatial dimensions, we use quadrilateral coarse elements, resulting in ten elemental integrals per coarse element. Among these ten quantities, however, there are only six degrees of freedom. The six degrees of freedom are coefficients of an orthogonal basis for the elemental 
integrals.\footnote{We find this basis by taking an SVD of the matrix containing prior samples of the elemental integrals. This matrix is rank-deficient. The reason that there are only six degrees of freedom can be understood by considering various symmetries of the problem.}  As the number of coarse elements increases, the number of coarse quantities quickly becomes too large to tackle with a simple total-degree map.  We will again use the problem structure to find a more tractable representation for $\imap_c$ and $\ifmap_c$.  In particular, we will restrict our attention to problems with \textit{stationary} prior distributions and use this stationarity in combination with the locality of MsFEM.  For convenience of notation, let $V=\pd_\crv/6$ denote the number of coarse elements in our 2D discretization. 

As in the one-dimensional formulation above, we will again combine $\rrv_c$ and $\rrv_f$ into a 
single vector, but now the expression
\begin{equation}
\rrv = \left[\rrv_c, \rrv_f\right]^\top = \left[ \rrv_{c,1}, \, \rrv_{c,2},\,\ldots, \, \rrv_{c,V}, \, \rrv_{f,1}, \, 
\rrv_{f,2}, \, \ldots,\, \rrv_{f,\pd_\frv}\right]^\top, \nonumber
\end{equation}
represents the coarse reference random variables in blocks. That is, $\rrv_{c,i}\in \real^6$ contains the six components of $\rrv_c$ corresponding to coarse element $i$.  On the other hand, each $\rrv_{f,k}$ is a scalar.  Similar to $\rrv_c$, $\crv = \left[\crv_1,\crv_2,\ldots,\crv_V\right]$, with each $\crv_i \in \real^6$, represents a block definition of $\crv$. Each of the six scalar components of $\crv_i$ (i.e., $\crv_{i,k} \in \real$, $k=1\ldots 6$) represents a particular coefficient of the orthogonal basis for the elemental integrals in coarse element $i$. With a stationary prior on $\frv$, we will have a stationary prior on $\crv$. Stationarity implies that the marginal distribution of any $\crv_{i,k}$ will be the same for all $i$ and, moreover, that the six-dimensional distribution of the coefficients is the same across elements:
\begin{equation}
\crv_i \eqd \crv_j,
\end{equation}
for $i,j\in\{1,2,\ldots,V\}$.  We will exploit this structure to build $\ifmap_c$.  

First, consider a map $\ifmap_c^m$ that pushes a 6-dimensional standard Gaussian to 
the prior marginal distribution $\pi(\crv_i)$.  This map is $6$-dimensional 
regardless of how many coarse elements are used, and it captures the nonlinear 
dependence among the six coarse degrees of freedom.  Now, assume that we have 
constructed $\ifmap_c^m$ and its inverse $\imap_c^m$ using the optimization and 
regression approach from Section \ref{sec:maps}.  Using $\imap_c^m$, we can define 
the random variable $\rrv_c^m \in \real^{\pd_\crv}$ as
\begin{equation}
\rrv_c^m = \left[\imap_c^m(\crv_1),\imap_c^m(\crv_2),\ldots,\imap_c^m(\crv_V)\right]^\top .
\end{equation}
Notice that each block of $\rrv_c^m$ is marginally a standard Gaussian with iid components, but the entire variable $\rrv_c^m$ is not; correlations between coarse elements remain in $\rrv_c^m$ (due to correlations in the fine scale field).  To remove these correlations, we use a lower triangular Cholesky 
decomposition of the covariance of $\rrv_c^m$ given by
\begin{equation}
\cov\left[\rrv_c^m\right] = LL^{\top}.
\end{equation}
The lower triangular Cholesky factor $L$ can itself be divided into blocks 
corresponding to each of the coarse elements
\begin{equation}
L = \left[\begin{array}{ccccc}L_{1,1}& 0 & 0 & \cdots & 0\\ L_{2,1} & L_{2,2} & 0 & 
\cdots & 0\\ \vdots & & \ddots & & \vdots \\ L_{(V-1),1} & L_{(V-1),2} & \cdots & 
L_{(V-1),(V-1)} & 0\\ L_{V,1} & L_{V,2} & \cdots & L_{V,(V-1)} 
&L_{V,V}\end{array}\right],
\end{equation}
where each diagonal entry is a $6\times 6$ lower triangular matrix.  Notice that 
applying $L^{-1}$ to $r_c^m$ will remove linear correlations from $r_c^m$,
leading to
\begin{equation}
\rrv_c = L^{-1}r_c^m, \ \  \Rightarrow L\rrv_c = r_c^m.
\end{equation}
Combining $L$ with the local nonlinear map $\ifmap_c^m$, the entire coarse map $\ifmap_c$ is 
defined as
\begin{equation}
\crv \eqd \ifmap_c(\rrv_c) = \left[\begin{array}{l}\ifmap_c^m\left(L_{1,1}\rrv_{c,1}\right)\\ 
\ifmap_c^m\left(L_{2,1}\rrv_{c,1} + L_{2,2}\rrv_{c,2}\right)\\ \vdots \\ \ifmap_c^m\left(L_{V,1}\rrv_{c,1} + 
L_{V,2}\rrv_{c,2} + \ldots + L_{V,V}\rrv_{c,V}\right)\end{array}\right]\label{eq:coarsemap2d}.
\end{equation}
Importantly, constructing this map only requires building a nonlinear map in six dimensions, which can be accomplished easily with total degree polynomial expansions.  

In the two-dimensional example below, $\ifmap_f$ is constructed using the cross covariance approach described in Section~\ref{sec:onedim} for the one-dimensional problem.  The samples of $\rrv_c$ used in the sample covariance $\Sigma_{\rrv_c\frv}$ are computed using the nonlinear inverse map $\imap_c^m$ composed with $L^{-1}$.

\section{Numerical results}\label{sec:numerics}

\subsection{One spatial dimension}\label{chap:multiscale:results1d}
Here we apply our multiscale framework and the previous section's problem-specific map structure to our first ``large-scale'' inference problem.  The goal of this section is to analyze the efficiency and accuracy of our multiscale inference strategy by comparing it with a standard MCMC approach.  The inverse problem involves inferring the spatially distributed conductivity field in \eqref{eq:press1} from noisy observations of the hydraulic head, using MsFEM as the forward model. The spatial domain is one-dimensional, $\Omega = [0,1]$. While our multiscale approach can handle much larger problems (see Section~\ref{chap:maps:twoD}), the problem size is restricted in this example in order to enable comparison with full-dimensional MCMC.

We wish to sample the posterior distribution $\pi(\theta | \drv)$, where $\theta=\log \kappa$ is the discretized log conductivity field and the data $\drv$ are a set of pointwise head observations.  We use a Gaussian prior on $\frv$ with zero mean and exponential covariance kernel
\begin{equation}
\cov\left(\theta(x_1),\theta(x_2)\right) = \sigma_\theta^2 \exp{\left[ -\frac{\left|x_1-x_2\right|}{L}\right] }.
\end{equation}
We set the correlation length to $L=0.1$ and the prior variance to $\sigma_\theta^2 = 1.0$.  An exponential kernel was chosen for two reasons: (i) this class of covariance kernel yields rough fields that are often found in practice, but difficult to handle with dimension reduction techniques such as the Karhunen-Lo\`{e}ve decomposition; and (ii) MsFEM is more accurate for problems with stronger scale separation, which is the case for rougher coefficient fields.  We use $10$ coarse elements and $\finc = 10$ fine elements per coarse element.  Thus $\theta$ is a $100$-dimensional random variable.  The data $\drv$ are $9$-dimensional, coming from observations at the interior nodes of the coarse mesh. Zero Dirichlet boundary conditions are imposed: $h(0)=h(1)=0$. To generate the data, a realization of the prior log-conductivity (shown in Figure \ref{fig:PosteriorRealizations1d}) is combined with a very fine-mesh FEM forward solver to produce a representative head field.  The head field is then down-sampled and combined with additive iid Gaussian noise to obtain the data.  The noise has zero mean and variance $10^{-4}$.

Benchmark results are obtained by running MCMC on the full 100 dimensional representation of $\frv$, with MsFEM as the forward model.  We use two variants of MCMC in our tests: the delayed rejection adaptive Metropolis (DRAM) MCMC algorithm \cite{Haario2006} and a preconditioned Metropolis-adjusted Langevin algorithm (preMALA) \cite{Roberts2003}.  The DRAM algorithm is tuned to have an acceptance rate of 35\%.  Two stages are used for the DR part of DRAM, but the second stage was turned off after $7 \times 10^4$ MCMC steps.  We set the covariance of the preMALA proposal to the inverse of the Gauss-Newton Hessian at the posterior MAP point.  The preMALA algorithm also uses gradient information to shift the proposal towards higher density regions of the posterior. For the single-scale posterior here, finite differences were used to compute the Hessian, which may have hindered preMALA performance in Table \ref{tab:SingEfficiencyComp}.  For both preMALA and DRAM, we run long chains of $5 \times 10^6$ samples, discarding the first $10^5$ steps as burn-in after starting the chain from the MAP point.  Because the DRAM chain seems to mix better than the preMALA chain, we use the former for the accuracy comparison in Table \ref{tab:SingAccuracyComp}. 

In the multiscale approach of Algorithm~\ref{alg:completealg}, we must sample the coarse posterior $\pi(\rrv_c | \drv)$. We do so using preMALA with the Hessian at MAP as a preconditioner; for this target distribution, preMALA was found to be more efficient than DRAM. preMALA was tuned to have an acceptance rate of around 55\%.

When the multiscale definition in (\ref{eq:multidef}) is completely satisfied (as it is in this case) and \textit{exact} transport maps are used, posterior samples produced by our multiscale framework will be samples from $\pi(\frv | \drv)$. As described in the preceding sections, however, various approximations are required to efficiently compute the transport maps.  Hence in {this} application, our multiscale method produces only approximate samples of $\pi(\frv | \drv)$.   Table \ref{tab:SingAccuracyComp} and Figure \ref{fig:AccuracyComp} show that this approximation is quite good. Table \ref{tab:SingAccuracyComp} reports quantiles of the marginal posterior $\pi(\frv_i \vert y)$ for $\theta_i$ at particular spatial locations. The ``exact'' quantiles are taken to be those produced by the long full-dimensional DRAM run. These are compared with quantiles computed using $\frv$-samples from our multiscale framework.  The multiscale 95\% intervals in Figure \ref{fig:AccuracyComp}  are computed from the mean and variance of 50 independent runs of the multiscale method.
 Each run used $5\times10^4$ prior samples to construct the maps, and then generated $10^5$ posterior samples of $\frv$, which were used to estimate the quantiles.  Importantly, these posterior quantiles provide a diagnostic that is sensitive to non-Gaussian posterior structure.

As shown in Figure \ref{fig:AccuracyComp}, there is a negative bias in the results near $x=0.3$.  This is likely caused by the approximation of $\ifmap_f$ for parameters near that point.  A coarse element boundary exists at $x=0.3$ and there is large dip in the true value of $\theta$ over this boundary.  Such large dips do not occur in high-density regions of the prior, and restricting $\ifmap_f$ to have only local nonlinearities might prevent the map from adequately capturing the tail behavior necessary to exactly characterize this posterior.  With a more expressive coarse-to-fine map $\ifmap_f$, this bias would decrease.  In all other locations, however, the true MCMC posterior and the multiscale posterior are in excellent agreement.

\begin{table}[h!]
\centering
\caption[Error in posterior reconstruction for the one-dimensional example.]{Estimated bias in posterior quantile estimates for the multiscale inference framework, at different spatial locations $x$. $E_\alpha$ is the average error (i.e., the bias) between the full-dimensional MCMC estimate and multiscale estimate of the $\alpha/100$ quantile.  \EditsText{Focusing on the median $(\alpha = 50)$, biases do not seem to change significantly as the degree of $\ifmap_f$ changes. Values of $E_{50}$ are more sensitive to the degree of $\ifmap_c$. 
}}
\small
\footnotesize
\begin{tabular}{|c|c|c|ccccc|}\hline
\EditsText{$\ifmap_c$} degree & \EditsText{$\ifmap_f$} degree & $x$ & $E_{05}$ & $E_{25}$ & $E_{50}$ & $E_{75}$ & $E_{95}$\\\hline
\multirow{8}{*}{1} & \multirow{4}{*}{1} & 0.1 & 1.11e-02 & 4.72e-02 & 5.92e-02 & 5.67e-02 & 3.35e-02\\
 & & 0.3 & -3.58e-01 & -3.05e-01 & -2.83e-01 & -2.75e-01 & -2.83e-01\\
 & & 0.5 & -1.28e-01 & -1.40e-01 & -1.62e-01 & -1.95e-01 & -2.60e-01\\
 & & 0.9 & 1.86e-02 & 6.46e-02 & 7.85e-02 & 7.92e-02 & 5.02e-02\\
\cline{2-8}
& \multirow{4}{*}{3} & 0.1 & 2.16e-01 & 1.26e-01 & 5.42e-02 & -2.58e-02 & -1.53e-01\\
 & & 0.3 & -1.49e-01 & -2.21e-01 & -2.80e-01 & -3.44e-01 & -4.42e-01\\
 & & 0.5 & 7.50e-02 & -5.98e-02 & -1.61e-01 & -2.67e-01 & -4.23e-01\\
 & & 0.9 & 2.45e-01 & 1.54e-01 & 7.91e-02 & -4.53e-03 & -1.35e-01\\
\hline
 \multirow{8}{*}{3} & \multirow{4}{*}{1} & 0.1 & -9.24e-04 & 3.60e-02 & 4.77e-02 & 4.76e-02 & 2.28e-02\\
 & & 0.3 & -3.60e-01 & -3.05e-01 & -2.84e-01 & -2.73e-01 & -2.79e-01\\
 & & 0.5 & -1.00e-01 & -1.12e-01 & -1.34e-01 & -1.69e-01 & -2.34e-01\\
 & & 0.9 & 2.17e-03 & 4.84e-02 & 6.26e-02 & 6.31e-02 & 3.53e-02\\
\cline{2-8}
 & \multirow{4}{*}{3} & 0.1 & 2.15e-01 & 1.22e-01 & 4.17e-02 & -4.46e-02 & -1.90e-01\\
 & & 0.3 & -4.11e-02 & -9.15e-02 & -1.32e-01 & -1.78e-01 & -2.54e-01\\
 & & 0.5 & 1.06e-01 & -3.52e-02 & -1.39e-01 & -2.49e-01 & -4.15e-01\\
 & & 0.9 & 2.42e-01 & 1.30e-01 & 4.74e-02 & -3.97e-02 & -1.79e-01\\
\hline
\end{tabular}

\label{tab:SingAccuracyComp}
\end{table}

\begin{figure}[h!]
\centering
\subfigure[95\% region of multiscale quantile estimator (shaded blue) using cross-covariance map, compared to ``benchmark'' MCMC quantile.]{
\input{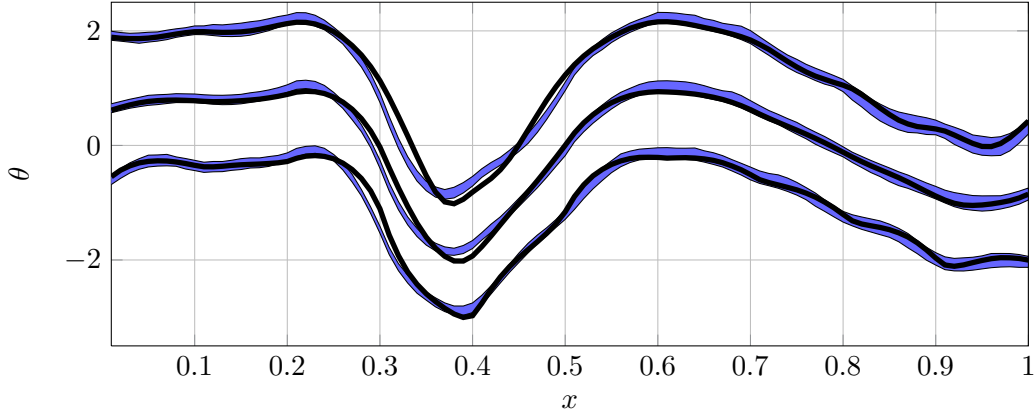}
}

\subfigure[95\% region of multiscale quantile estimator (shaded red) using local polynomial map, compared to ``benchmark'' MCMC quantile.]{
\input{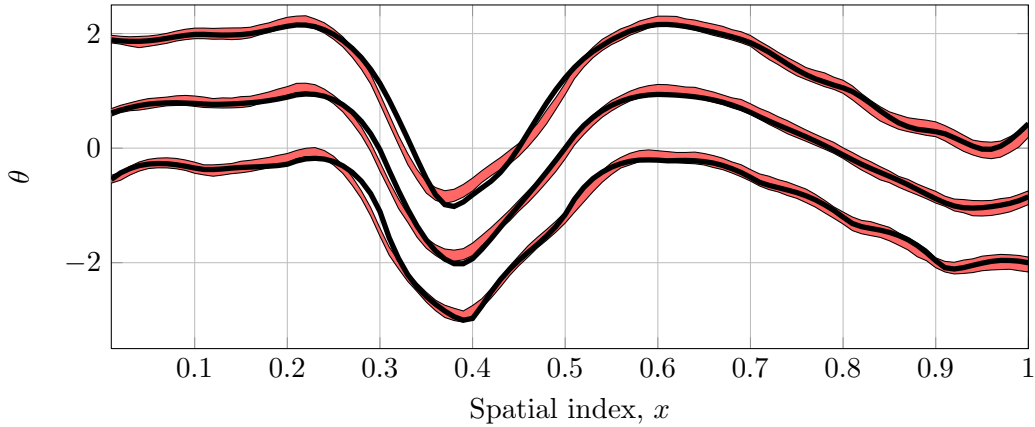}
}
\caption[Comparison of multiscale posterior and MCMC gold-standard.]{Comparison of multiscale estimates of posterior \EditsText{5\%, 50\%, and 95\%} quantiles with a fine-scale MCMC approach.  The fine-scale MCMC chain was run for 4.9 million steps and the resulting quantiles are taken as the ``true'' quantiles in our analysis.  Note that the vertical grid lines correspond to coarse element boundaries.  A quantitative summary of these plots is given in Table \ref{tab:SingAccuracyComp}.}
\label{fig:AccuracyComp}
\end{figure}

Marginal posterior diagnostics such as quantiles only tell part of the story.  Another important feature is the correlation structure of the posterior realizations. As shown in Figure \ref{fig:PosteriorRealizations1d}, our multiscale approach correctly produces posterior samples with the same fine-scale roughness as the prior.

\begin{figure}[h!]
\centering
%
%
%
%
\begin{tikzpicture}

\definecolor{mycolor1}{rgb}{1,0.4,0.4}

\begin{axis}[%
view={0}{90},
width=0.8\textwidth,
height=0.3\textwidth,
xlabel={$x$},
ylabel={$\theta$},
scale only axis,
scaled x ticks={real:100},
xtick scale label code/.code={},
xmin=1, xmax=100,
ymin=-3.25, ymax=3.25,
xtick={0,10,20,30,40,50,60,70,80,90,100},
grid=major]

\addplot [
color=black,
solid,
line width=2.0pt
]
coordinates{
 (1,0.269669619319458)(2,0.579747498624557)(3,0.528837940307898)(4,0.49530533785475)(5,0.895682668994097)(6,1.10997991136204)(7,1.51211373461646)(8,1.22402823248357)(9,2.09766842296398)(10,1.93188238483048)(11,1.72043629205902)(12,2.03217791311232)(13,2.26251669182534)(14,1.57650855999966)(15,1.72495458529599)(16,0.961867728723202)(17,0.679515815849807)(18,0.338753137785097)(19,0.00742016888646635)(20,0.0899882944432449)(21,0.680317659127182)(22,1.62630872288624)(23,2.34197606368608)(24,2.29159525409367)(25,1.91461707683589)(26,1.72473238833072)(27,1.55165721360827)(28,1.44383948673445)(29,1.05782003684245)(30,1.02856199144301)(31,1.29517830107831)(32,1.2971777577895)(33,1.1632793121539)(34,-0.206762887471099)(35,-0.770273962153661)(36,-1.04766309731082)(37,-1.39374135623121)(38,-1.77126249727882)(39,-2.16615023246074)(40,-1.46859018139525)(41,-1.39298718229892)(42,-1.35408664900749)(43,-0.963365270189555)(44,-0.832808225869272)(45,0.074217575210926)(46,0.511633821544153)(47,0.544519221317226)(48,0.921384276819488)(49,1.12493786099789)(50,0.755286350084721)(51,0.44743102919157)(52,0.777291137618191)(53,0.943038207348669)(54,1.73539009875318)(55,1.51224091500752)(56,1.6395321857792)(57,2.61051184461203)(58,2.86742124440512)(59,3.00508277576955)(60,2.957799140293)(61,2.66665655770402)(62,2.24134781055051)(63,2.04227928427519)(64,1.57021367938511)(65,1.68424534589623)(66,1.24814702960197)(67,1.54271371963207)(68,1.70532745391753)(69,1.33827677778465)(70,0.705797422472921)(71,1.39688779282288)(72,1.02073265876119)(73,0.710244116551437)(74,0.654581020752791)(75,1.06491423116093)(76,0.53661105241859)(77,0.967800610094932)(78,1.38151908038421)(79,1.3221990948175)(80,1.02865277624149)(81,0.604446051651992)(82,0.519780960945106)(83,0.392140876262304)(84,0.375773111700194)(85,0.112038243464245)(86,-0.365755742132705)(87,0.377368090608562)(88,0.494525331535962)(89,0.34282775747827)(90,1.15619603002858)(91,0.508329033573297)(92,-0.184111176700673)(93,-0.131710501552563)(94,0.0367536366768948)(95,-0.578164997243381)(96,0.0414460364339606)(97,0.280507786140259)(98,0.632754519017751)(99,-0.14614138697476)(100,-0.27411936305081) 
};

\addlegendentry{True $\log K$};

\addplot [
color=mycolor1,
solid,
line width=1.0pt
]
coordinates{
 (1,0.920030988430325)(2,1.43632229995545)(3,1.93909888767352)(4,2.60343559905874)(5,2.6519228690643)(6,2.92361232065699)(7,2.75620899321713)(8,2.06387833420503)(9,1.60917489907051)(10,1.04941223562392)(11,1.93118526806301)(12,2.43431807013383)(13,2.23224214124306)(14,1.6817123843139)(15,1.18061051466229)(16,1.71826885284313)(17,1.21432454693172)(18,1.53595243014282)(19,1.90266019826024)(20,1.5408424944472)(21,1.51942979990309)(22,1.29424318736365)(23,0.767286464621753)(24,1.02970765544929)(25,1.0257327659256)(26,1.01267220249322)(27,1.28169021955641)(28,1.70218574412303)(29,0.873237763895295)(30,0.666527123293824)(31,0.24479794847269)(32,-0.823521614904098)(33,-0.75744972926639)(34,-1.33471322214361)(35,-0.690515319957954)(36,-1.00965860260149)(37,-0.361635571931564)(38,-0.574112025067273)(39,-0.330872957737015)(40,-0.249021219059408)(41,0.0463726183999015)(42,-0.248384716615526)(43,0.494848849423806)(44,1.11587142896103)(45,0.82665549143467)(46,0.240311340682081)(47,0.0691611129709722)(48,-0.455424655050423)(49,-0.500276177215641)(50,-0.295834033617083)(51,0.579043913760005)(52,1.11115835041556)(53,0.870972246804006)(54,0.887582607926648)(55,1.15788418730026)(56,1.49807195086268)(57,1.74622488454168)(58,1.95292754370382)(59,1.47254026482886)(60,1.41865263079199)(61,1.29272558134264)(62,1.05092334420492)(63,1.10721870942627)(64,1.12591871875519)(65,2.43522781872155)(66,2.5135588297229)(67,2.66450460921189)(68,2.1085143857748)(69,1.91906187131831)(70,2.10385843333558)(71,2.10370969158232)(72,1.60343529011386)(73,0.899638474487766)(74,0.602308827247009)(75,0.891419225964499)(76,2.09143621723518)(77,1.90152468313915)(78,1.44156610427126)(79,1.29017552411095)(80,0.841173572055729)(81,1.49442124081292)(82,0.0153570069651581)(83,0.303658778426859)(84,0.595350654104812)(85,0.369976583312272)(86,0.724668462404787)(87,0.584137699318818)(88,-0.275244321423341)(89,-0.101219189859073)(90,-0.189423236357332)(91,-0.273916749863209)(92,-1.13191671630152)(93,-1.43357758179296)(94,-1.36920596342884)(95,-0.917265813799321)(96,0.0541759853316938)(97,0.435403560296594)(98,0.847499304484431)(99,1.23186540184204)(100,1.11352335217194) 
};

\addlegendentry{Realization};

\addplot [
color=green!50!black,
solid,
line width=1.0pt
]
coordinates{
 (1,0.146771612833969)(2,0.650378559971491)(3,0.847613807641488)(4,0.90730408307375)(5,0.479391116060783)(6,0.446087844928605)(7,0.996752183380157)(8,1.0011952365564)(9,1.03414304529055)(10,0.821349979854116)(11,1.01415451795378)(12,1.39436316557608)(13,0.749752037269777)(14,0.826516564872388)(15,1.01278565160113)(16,1.26623903728149)(17,0.769220427121793)(18,0.776006230086726)(19,0.95542796342954)(20,1.34319912685496)(21,1.38643645767547)(22,1.57708328263324)(23,0.699871978701641)(24,-0.0907353820579365)(25,-0.361510323175591)(26,-0.358172752758981)(27,0.157834376821698)(28,0.210711439551868)(29,0.0786317372002847)(30,-0.369441658687108)(31,-0.528368798608097)(32,-0.436563979215968)(33,-0.793822985976674)(34,-0.878084232625105)(35,-1.69421487451432)(36,-2.05555800058408)(37,-2.58502550571045)(38,-2.50104918927489)(39,-2.68519256914962)(40,-2.03177021355895)(41,-1.82486964372762)(42,-1.52553348285242)(43,-0.719438070863789)(44,-1.61947442417179)(45,-0.968996237758567)(46,-0.527736309115099)(47,-0.546527157156738)(48,-0.882439853951684)(49,-0.432908911422511)(50,-0.634194622594535)(51,-0.0675069374943229)(52,-0.154295673962698)(53,0.588639897523179)(54,1.01371248510283)(55,0.375177384148103)(56,0.0330894302054315)(57,0.364533226156554)(58,0.375532992511754)(59,1.17612092127442)(60,1.85153714002786)(61,2.27738749641493)(62,1.72152753210113)(63,1.28592440776334)(64,0.624470150864377)(65,1.5222718586391)(66,2.2256777264159)(67,1.78283572815408)(68,1.73543008985919)(69,1.92108740136099)(70,1.92196178604763)(71,1.10198760675855)(72,0.319574467492276)(73,0.124488290928926)(74,0.237511215176369)(75,-0.590680900636277)(76,-0.184163224132008)(77,-0.522131836895547)(78,-0.310307411927136)(79,-0.547572065430619)(80,-0.277139523896986)(81,-0.330371168148948)(82,-0.476405486871892)(83,-0.651217034468876)(84,-0.245887456114226)(85,-0.373591038725634)(86,-1.15785543979343)(87,-0.89975453003211)(88,-0.652842774217815)(89,-0.737472154016976)(90,-0.95472646512282)(91,-0.716882033961702)(92,-0.874197892123779)(93,-1.50235756786966)(94,-1.54817766094489)(95,-1.14440210773178)(96,-0.888339234057916)(97,-0.479763231393481)(98,-1.34798634717712)(99,-1.80674881581693)(100,-2.19370057342556) 
};

\addlegendentry{Realization};

\end{axis}
\end{tikzpicture}
\caption[Posterior summary in one dimensional example.]{Comparison of posterior realizations with true log conductivity.  The posterior samples maintain the same fine-scale correlation structure as the true log conductivity.  As in Figure \ref{fig:AccuracyComp}, the vertical grid lines correspond to coarse element boundaries.  }
\label{fig:PosteriorRealizations1d}
\end{figure}
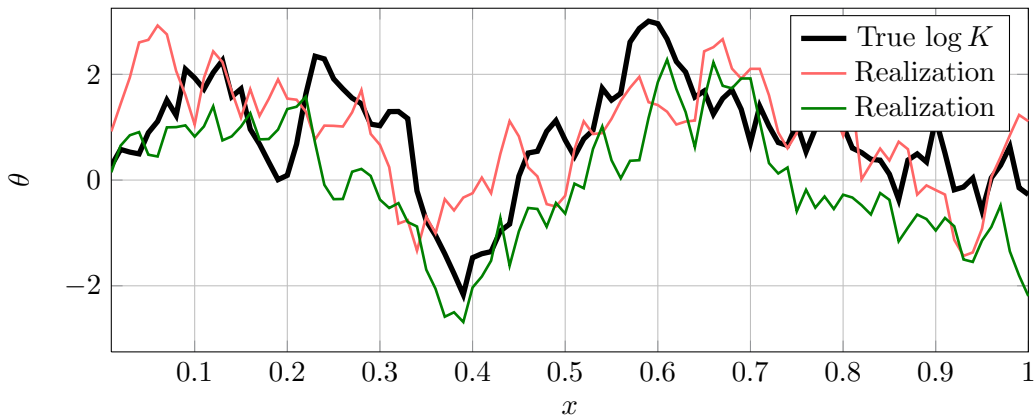

Now consider the computational efficiency of the multiscale method.  The effective sample size (ESS) is one measure of the information contained in a set of posterior samples.  The ESS represents the number of effectively independent samples contained in the set.  In an MCMC context, we can easily compute this quantity for a chain at equilibrium \cite{Wolff2004}.  Here, however, we will use a more direct definition of ESS using the variance of a Monte Carlo estimator.  Suppose that we have a Monte Carlo estimator $\widehat{{\frv}_i}$  of the mean of $\frv_i$.  The ESS for such an estimator is given by the ratio of the variances of the target random variable and the estimator:
\begin{equation}
\text{ESS}_i = \frac{\Var{}(\frv_i)}{\Var{}(\widehat{{\frv}_i})}
\end{equation}
Here $\text{ESS}_i$ denotes the effective sample size for dimension $i$ of $\frv$; ESS can differ for each dimension, and we will typically report the minimum and maximum $\text{ESS}_i$ for $i = 1\ldots \pd_{\frv}$. This expression for ESS is more costly to compute than methods based on MCMC autocorrelation \cite{Wolff2004} because evaluating $\Var{}(\widehat{{\frv}_i})$ requires \textit{many} independent realizations of the Monte Carlo estimator (i.e., running the inference procedure many times). But this approach is less susceptible to errors stemming from autocorrelation integration and does not require us to use ordered samples like those from an MCMC scheme. 

Table \ref{tab:SingEfficiencyComp} shows the efficiency of our multiscale approach.  Comparing full-scale DRAM MCMC with the multiscale results, we see that even when a nonlinear coarse-to-fine map $\ifmap_f$ is used, proper tuning of the method can speed up the number of effectively independent samples generated per second by a factor of 2 to 5, with one fine sample generated per coarse sample ($M=1$). When a linear $\ifmap_f$ is employed, we can see speedups of 4.5 to 9 times, using $M=5$.  These results indicate that as long as minor approximations to the posterior are acceptable, there is a clear advantage to using our multiscale approach.   


\begin{table}[h!]
\centering
\caption[Posterior sampling efficiency comparison on one dimensional example.]{Comparison of posterior sampling efficiency between full-dimensional MCMC and variants of our multiscale framework.  The key column is ESS/$t_{on}$ (effectively independent samples generated per second), where higher numbers indicate better efficiency. The time $t_{on}$ measures the computation that must be performed after a particular value of $d$ is observed; it does not include map-construction time.}
\small
\footnotesize
\begin{tabular}{|l|rc|r|rr|rr|}\hline
& & & & \multicolumn{2}{c|}{ESS} & \multicolumn{2}{c|}{ESS/$t_{on}$} \\
Method & $N$ & $M$ & $t_{on}$ (sec) & Min & Max & Min & Max \\\hline
MCMC-DRAM & 4900000 & NA & 2252.63 & 6340 & 11379 & \mbox{\boldmath$2.8$} & \mbox{\boldmath$5.1$} \\
MCMC-PreMALA & 4900000 & NA & 2773.47 & 274 & 729 & \mbox{\boldmath$0.1$} & \mbox{\boldmath$0.3$}  \\\hline
Cross Covariance & 500000 & 1 & 287.31 & 2987 & 20244 & \mbox{\boldmath$10.4$} & \mbox{\boldmath$70.5$}  \\
Cross Covariance & 500000 & 5 & 314.17 & 3971 & 14597 & \mbox{\boldmath$12.6$} & \mbox{\boldmath$46.5$}\\\hline
Local Cubic & 450000 & 1 &  937.41 & 5408 & 23679 & \mbox{\boldmath$5.8$} & \mbox{\boldmath$25.3$}  \\
Local Cubic & 450000 & 5 & 3555.05 & 5294 & 18759 & \mbox{\boldmath$1.5$} & \mbox{\boldmath$5.3$}   \\\hline
\end{tabular}
\label{tab:SingEfficiencyComp}
\end{table}

Using the timing and ESS data from Table \ref{tab:SingEfficiencyComp} for $M=1$ and $M=5$, we can also compute the optimal number of fine samples to generate for each coarse sample.  To deploy the optimal expression in (\ref{eq:optK2}), we first use a least squares approach to compute the unknown coefficients $C_1$ and $C_2$.  For the linear case, we obtain $C_1=22.7867$ and $C_2=10.2019$, which yield an optimal value of $M=4$.  For the local cubic case, we obtain $C_1=11.6076$ and $C_2=3.3135$, which yield an optimal value of $M=1$.  It is worth generating additional fine-scale samples for the inexpensive linear map, but for the slightly more expensive cubic map, the time to generate more fine-scale samples is not worthwhile. This time would be better spent generating coarse samples. 
These values for $M$ are dependent on the cost of each coarse model evaluation.   In this one-dimensional problem, the coarse model is very cheap to evaluate---on par with the cost of evaluating the cubic map.  However, for problems with more expensive model evaluations or with poorer coarse MCMC mixing, this will not be the case and larger values of $M$ will be optimal.


\subsection{Two spatial dimensions}\label{chap:maps:twoD}
The relatively small dimension of $\frv$ in the one-dimensional problem above allowed us to compare our multiscale approach with very long ``benchmark'' MCMC runs.  However, we expect our multiscale inference approach to yield even larger performance increases on large-scale problems where direct use of MCMC may not be feasible at all.  Here we will again infer a log conductivity field in the elliptic equation \eqref{eq:press1}; however, this example will have two spatial dimensions.  The 2D grid is defined by an $8\times 8$ mesh of coarse elements over $[0,1]\times [0,1]$, with $13\times 13$ fine elements in each coarse element.  The log-conductivity is piecewise constant on each fine element, resulting in a $10816$ dimensional inference problem.  The zero mean Gaussian prior is again defined by an exponential kernel with correlation length $0.1$.  In two dimensions, this kernel takes the form
\begin{equation}
\cov\left(\theta(x_1),\theta(x_2)\right) = \sigma_\theta^2\exp{\left[-\frac{\| x_1 - x_2\|_2}{L}\right]},
\end{equation}
where $\|\cdot\|_2$ is the usual Euclidean norm, $\sigma_\theta^2 = 1.0$, and $L=0.1$.  Notice that this kernel is isotropic but not separable. 

Synthetic data are generated by a full fine-scale simulation using a standard Galerkin FEM with additive iid Gaussian noise added to observations at each of the coarse nodes.  The noise variance is $10^{-6}$.  Homogeneous (i.e., no flow) boundaries are used for $x=0$ and $x=1$.  Dirichlet conditions are used on the top and bottom of the domain.  These conditions are given by
\begin{eqnarray*}
h(x,y=0) & =&  x\\
h(x,y=1) & = & 1-x.
\end{eqnarray*}

Since we cannot reasonably apply any other sampling method to this large problem directly, our confidence in the accuracy of the posterior can be judged from the accuracy of the transport maps $\ifmap_c$ and $\ifmap_f$.  From the one-dimensional results, we know that linear $\ifmap_f$ derived from the cross covariance of $\rrv_c$ and $\frv$ performs quite well; it is reasonable to assume that the same is true in the 2D case. A qualitative validation of the coarse map $\ifmap_c$ is given in Figure \ref{fig:coarsmap2d}.   The figure shows the true prior density of the coarse parameters over one coarse element, as well as the density induced by the coarse map, $\ifmap_c$ in (\ref{eq:coarsemap2d}). We see that the coarse map represents the coarse prior quite well.  In terms of computational effort, the map was constructed using 85000 prior samples and a multivariate Hermite polynomial representation of $\ifmap_c^m$ of total degree seven. Using the MIT Uncertainty Quantification (MUQ) library \cite{MUQ}, and taking advantage of 16 compute nodes, each employing four threads on a cluster with 3.6 GHz Intel Xeon E5-1620 processors, prior sampling, $\ifmap_c$ construction, and $\ifmap_f$ construction took less than one hour for this problem.

\newcommand{\groupPlotWidth}{0.2000\textwidth}
\begin{figure}[h!]
\centering
\subfigure[True coarse density.]{
\begin{tikzpicture}[scale=0.8]
\begin{groupplot}[group style={group size=6 by 6,xlabels at=edge bottom,ylabels at=edge left, xticklabels at=edge bottom,yticklabels at=edge left,vertical sep=1pt,horizontal sep=1pt},height=\groupPlotWidth,width=\groupPlotWidth,ticks=none,enlargelimits=false]

\nextgroupplot[title=$\gamma_1$]
\addplot graphics[xmin=0,ymin=0,xmax=1,ymax=1] {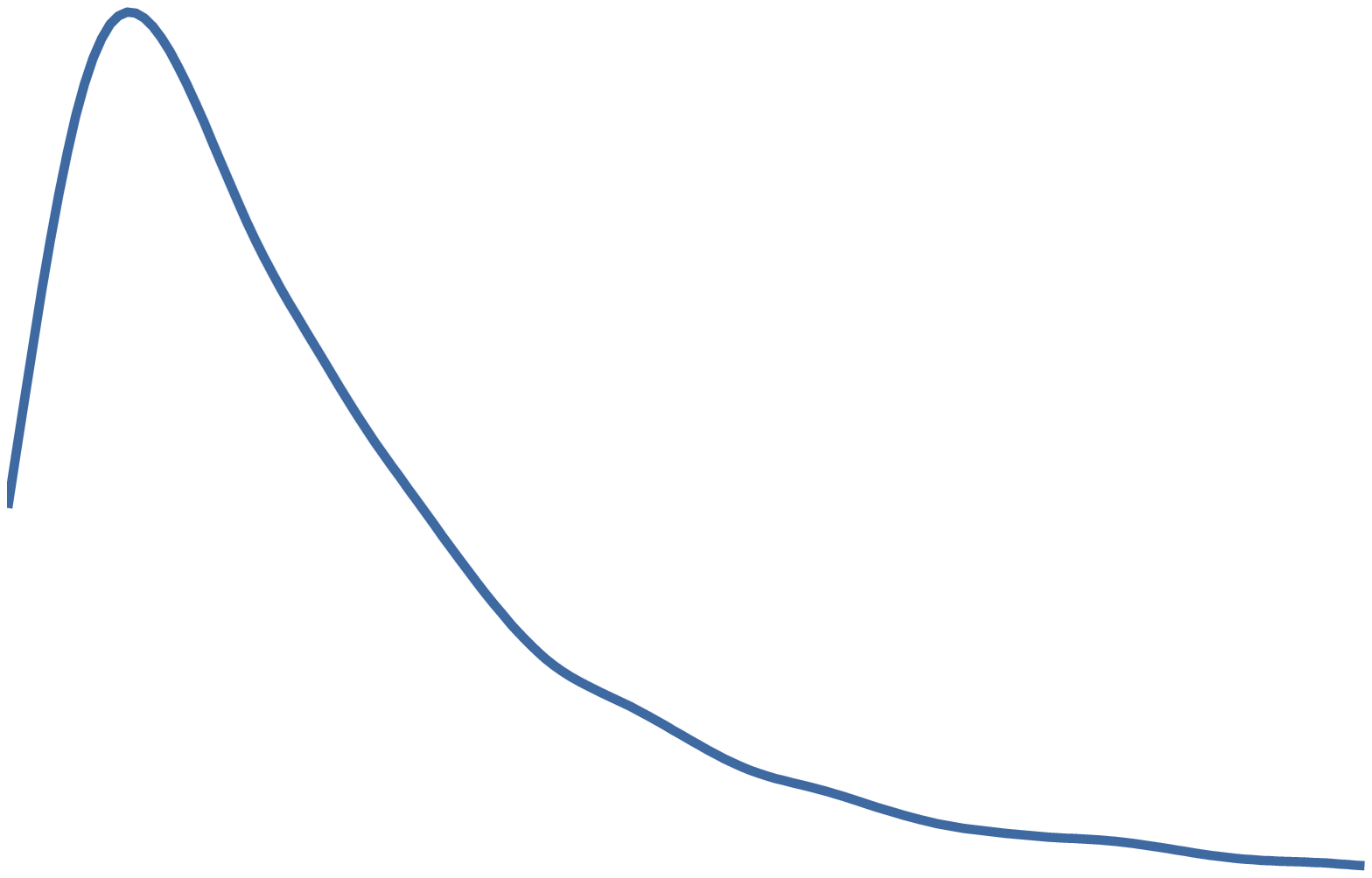};

\nextgroupplot[group/empty plot]
\nextgroupplot[group/empty plot]
\nextgroupplot[group/empty plot]
\nextgroupplot[group/empty plot]
\nextgroupplot[group/empty plot]

\nextgroupplot
\addplot graphics[xmin=0,ymin=0,xmax=1,ymax=1] {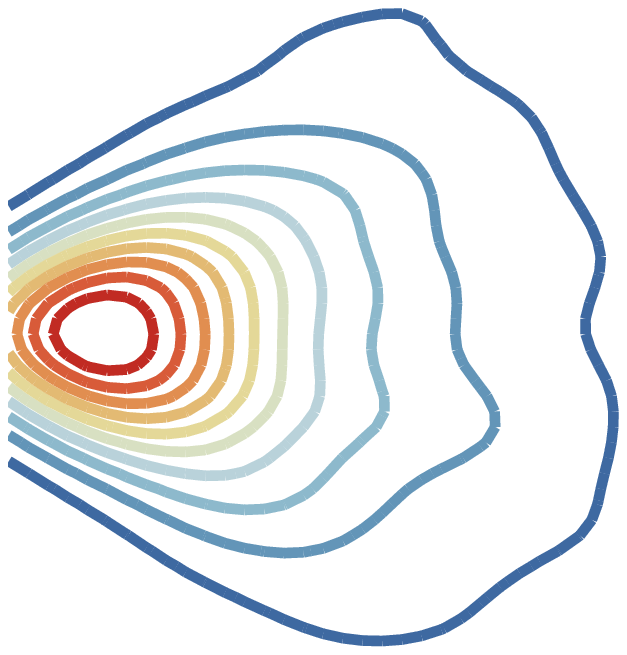};
\nextgroupplot[title=$\gamma_2$]
\addplot graphics[xmin=0,ymin=0,xmax=1,ymax=1] {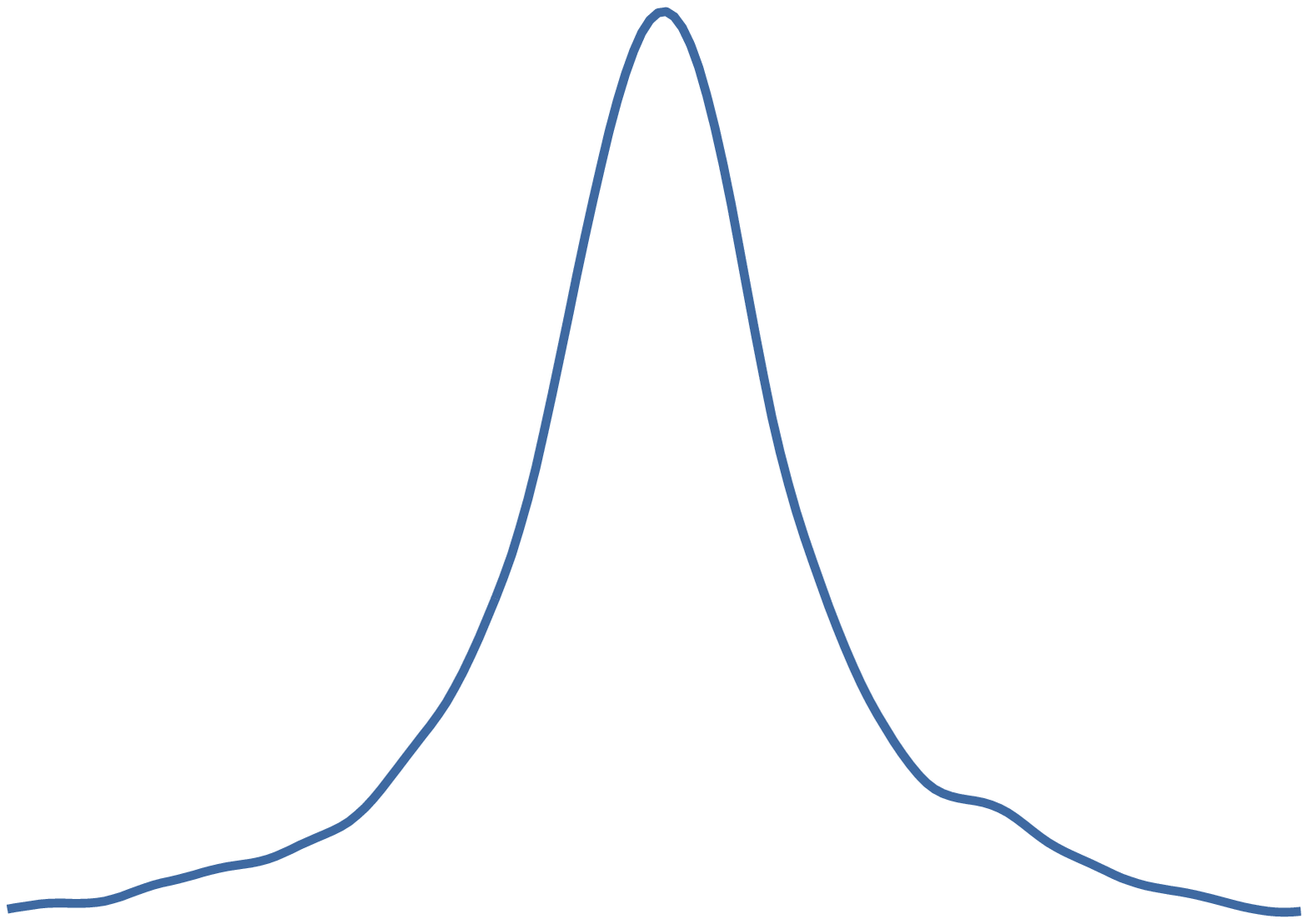};

\nextgroupplot[group/empty plot]
\nextgroupplot[group/empty plot]
\nextgroupplot[group/empty plot]
\nextgroupplot[group/empty plot]

\nextgroupplot
\addplot graphics[xmin=0,ymin=0,xmax=1,ymax=1] {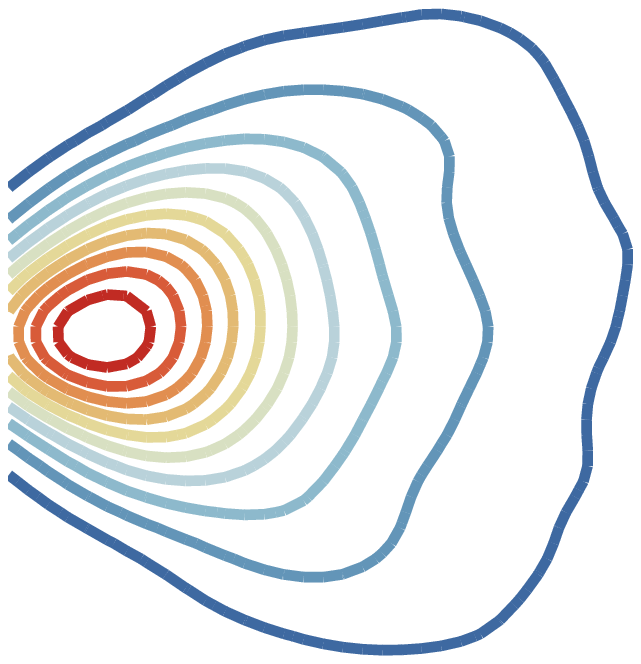};
\nextgroupplot
\addplot graphics[xmin=0,ymin=0,xmax=1,ymax=1] {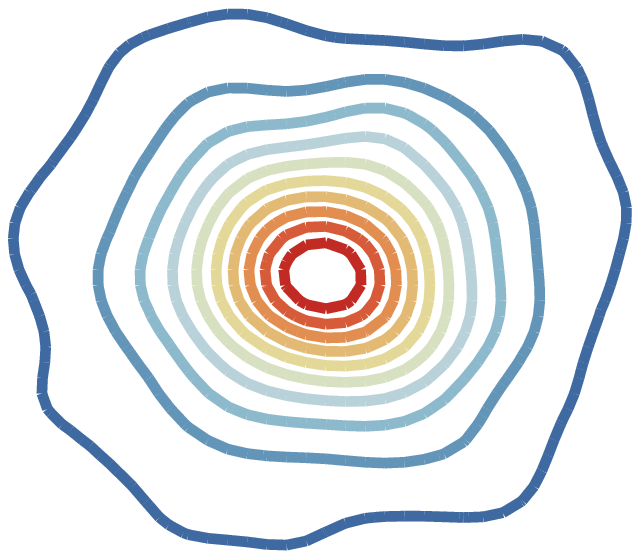};
\nextgroupplot[title=$\gamma_3$]
\addplot graphics[xmin=0,ymin=0,xmax=1,ymax=1] {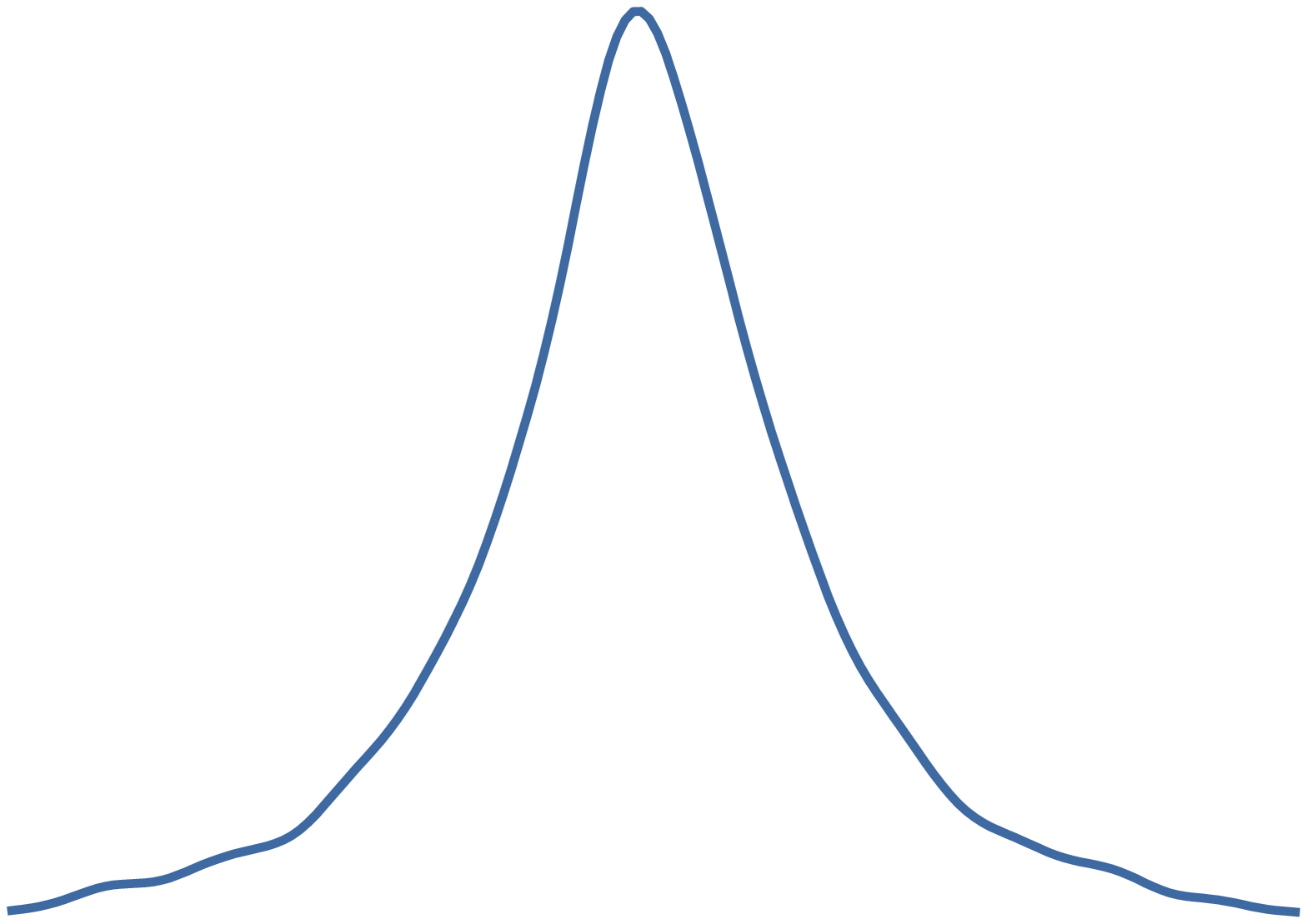};

\nextgroupplot[group/empty plot]
\nextgroupplot[group/empty plot]
\nextgroupplot[group/empty plot]

\nextgroupplot
\addplot graphics[xmin=0,ymin=0,xmax=1,ymax=1] {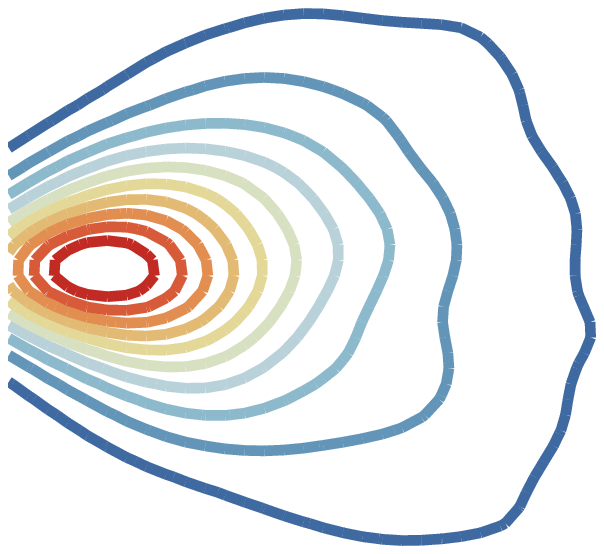};
\nextgroupplot
\addplot graphics[xmin=0,ymin=0,xmax=1,ymax=1] {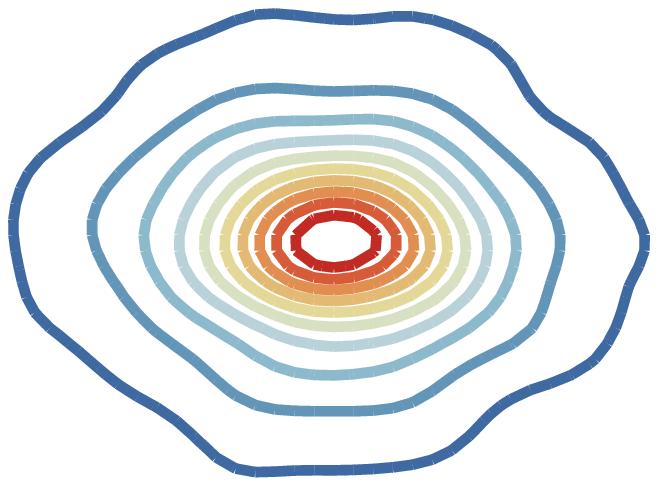};
\nextgroupplot
\addplot graphics[xmin=0,ymin=0,xmax=1,ymax=1] {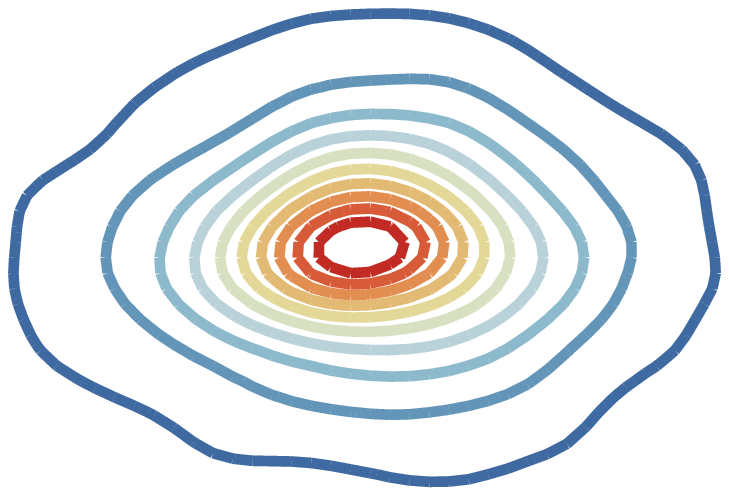};
\nextgroupplot[title=$\gamma_4$]
\addplot graphics[xmin=0,ymin=0,xmax=1,ymax=1] {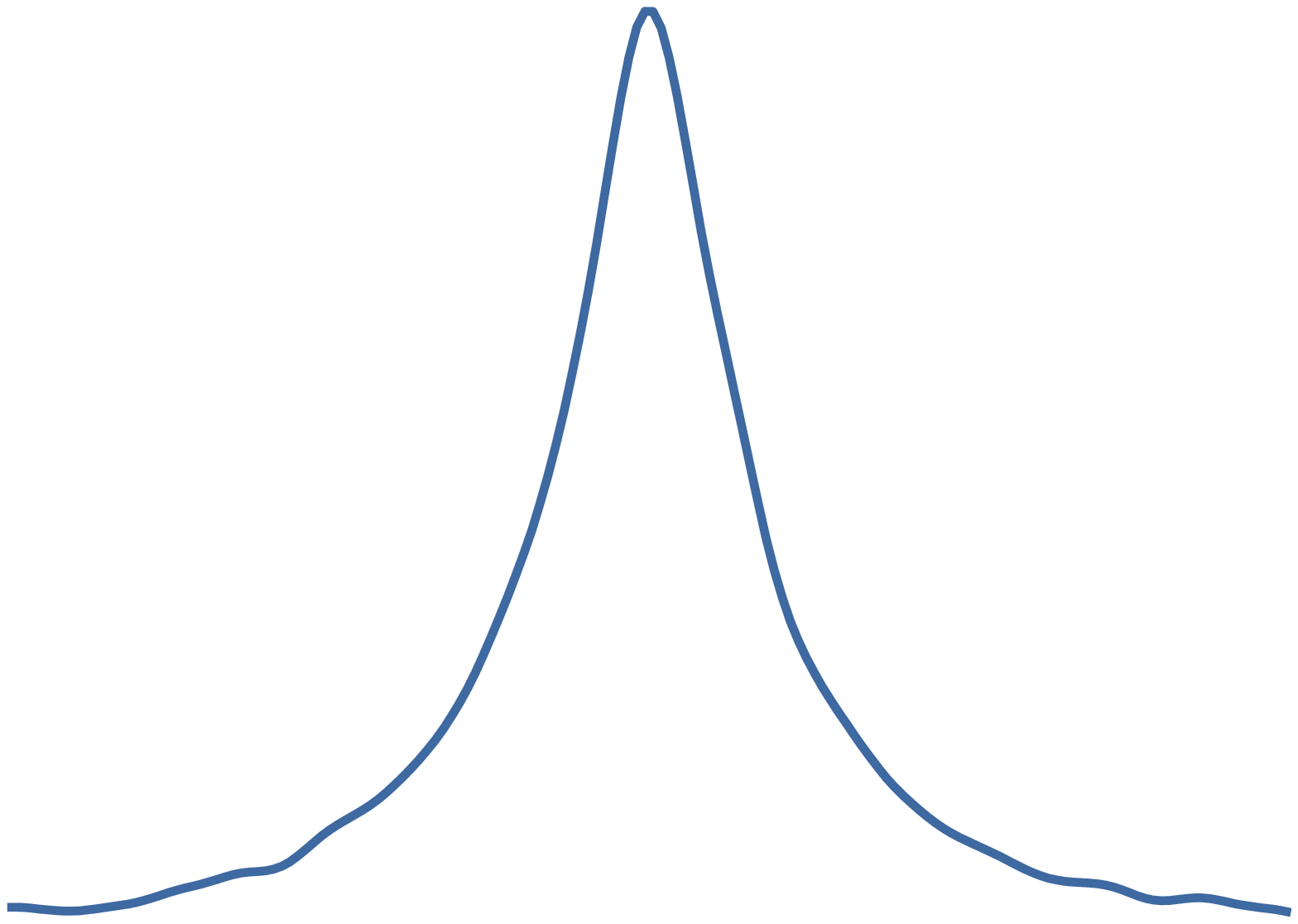};

\nextgroupplot[group/empty plot]
\nextgroupplot[group/empty plot]

\nextgroupplot
\addplot graphics[xmin=0,ymin=0,xmax=1,ymax=1] {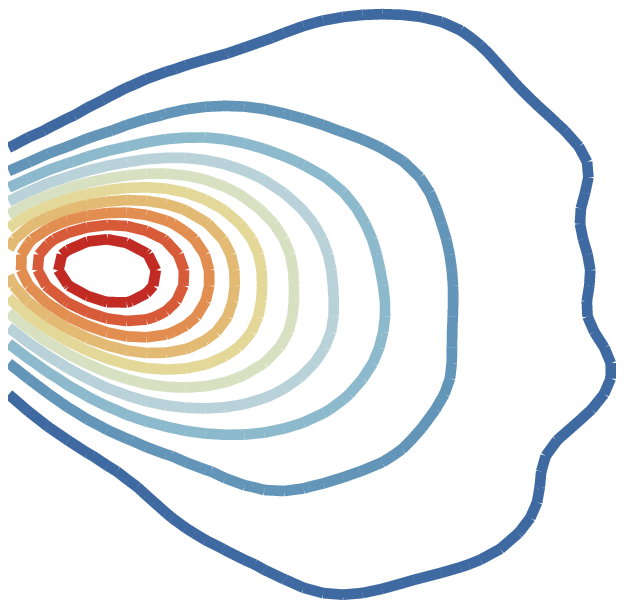};
\nextgroupplot
\addplot graphics[xmin=0,ymin=0,xmax=1,ymax=1] {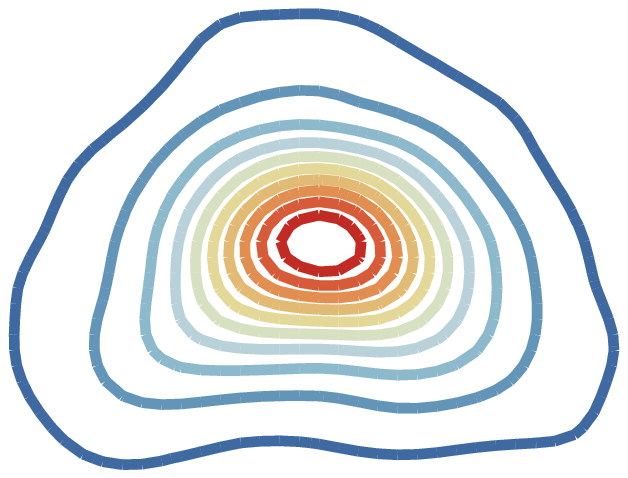};
\nextgroupplot
\addplot graphics[xmin=0,ymin=0,xmax=1,ymax=1] {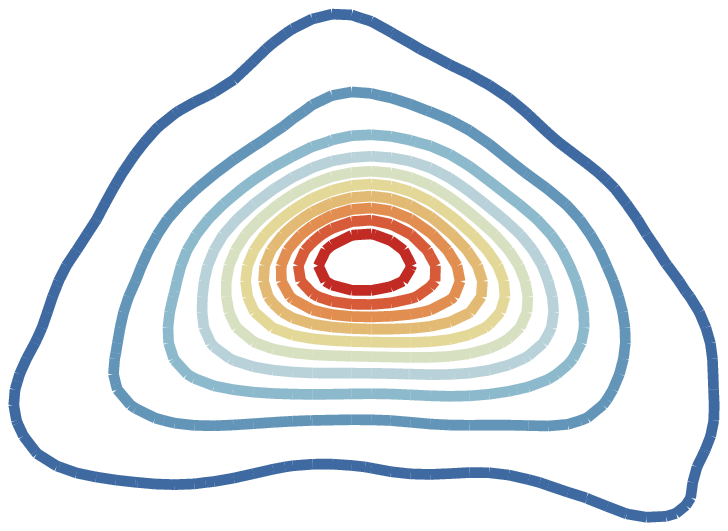};
\nextgroupplot
\addplot graphics[xmin=0,ymin=0,xmax=1,ymax=1] {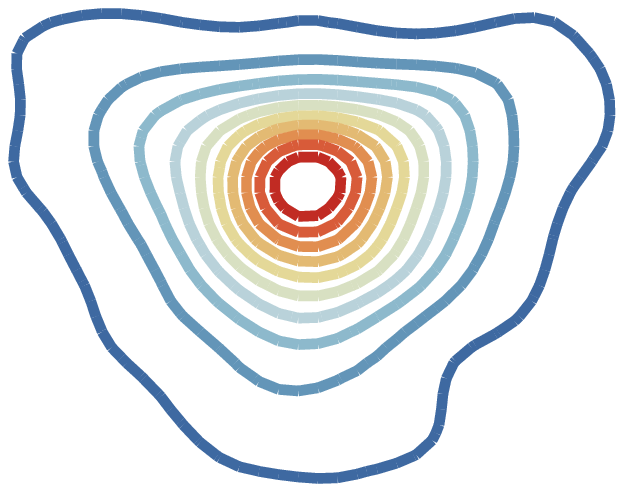};
\nextgroupplot[title=$\gamma_5$]
\addplot graphics[xmin=0,ymin=0,xmax=1,ymax=1] {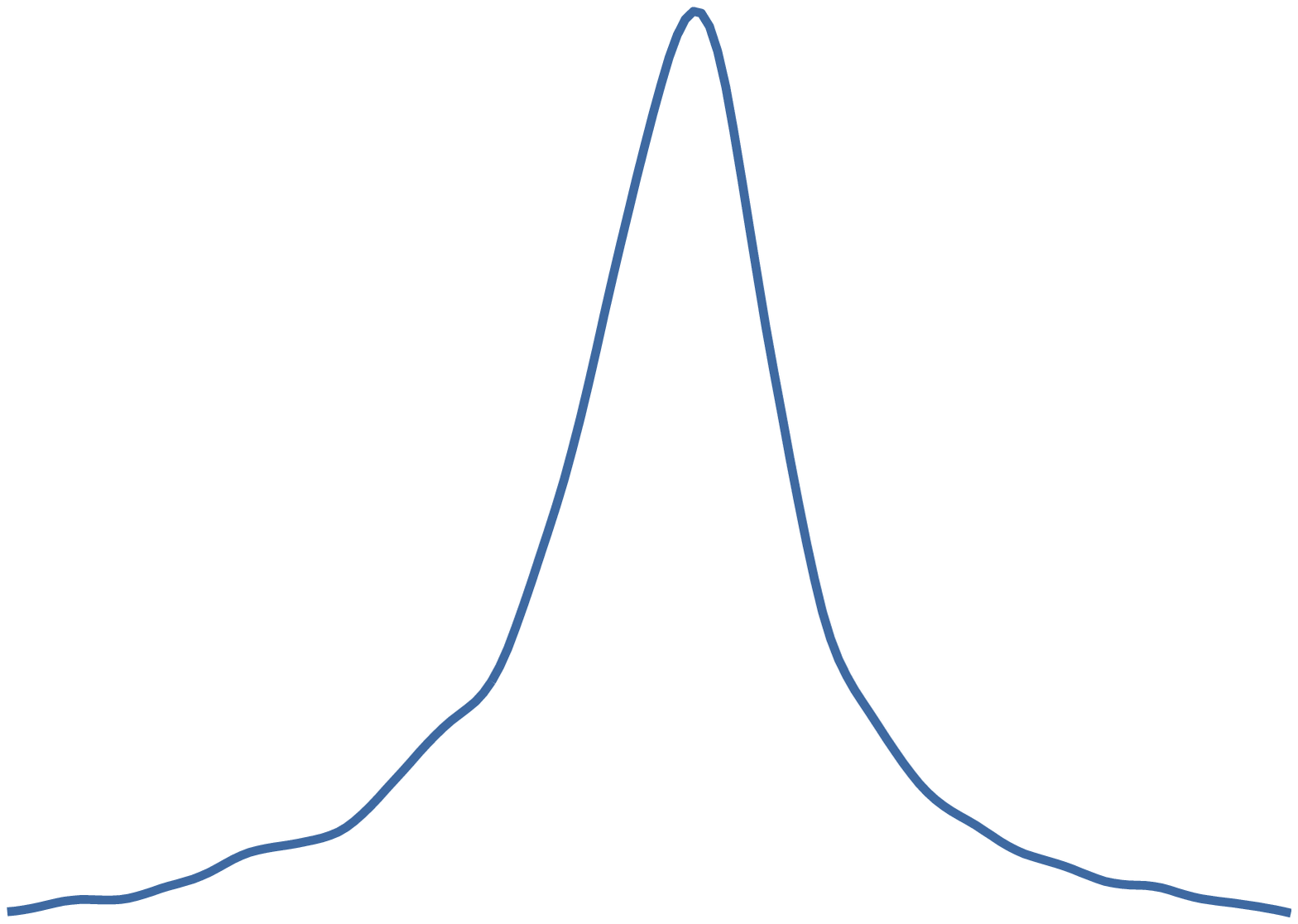};

\nextgroupplot[group/empty plot]

\nextgroupplot
\addplot graphics[xmin=0,ymin=0,xmax=1,ymax=1] {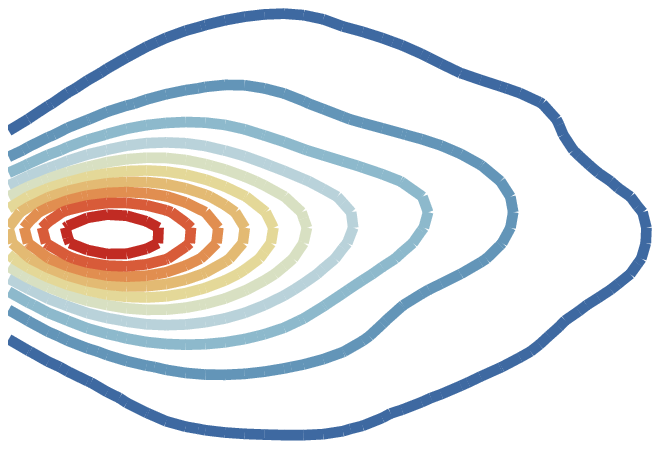};
\nextgroupplot
\addplot graphics[xmin=0,ymin=0,xmax=1,ymax=1] {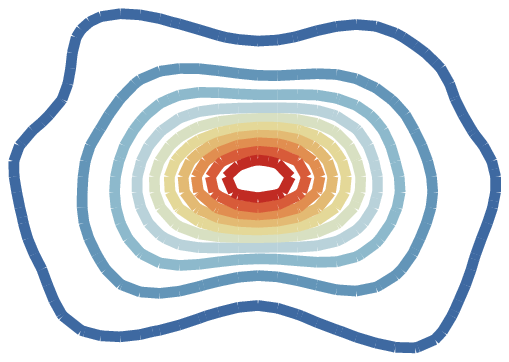};
\nextgroupplot
\addplot graphics[xmin=0,ymin=0,xmax=1,ymax=1] {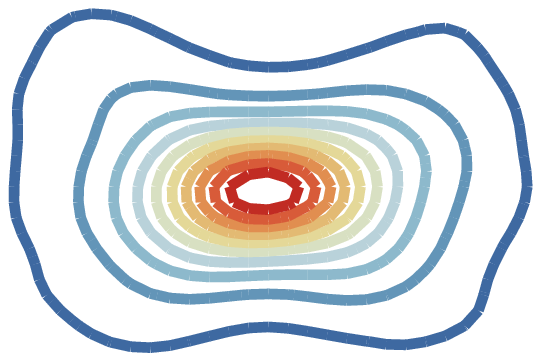};
\nextgroupplot
\addplot graphics[xmin=0,ymin=0,xmax=1,ymax=1] {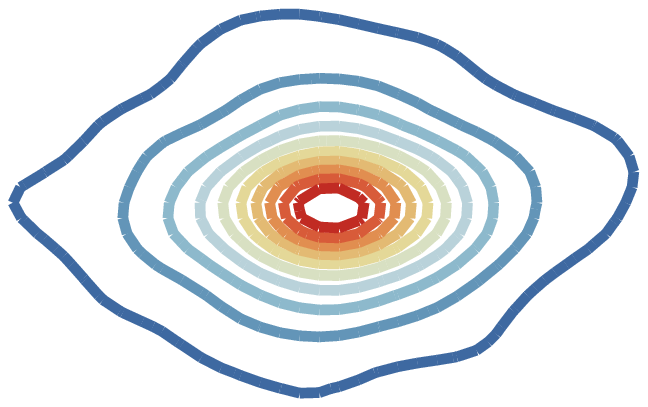};
\nextgroupplot
\addplot graphics[xmin=0,ymin=0,xmax=1,ymax=1] {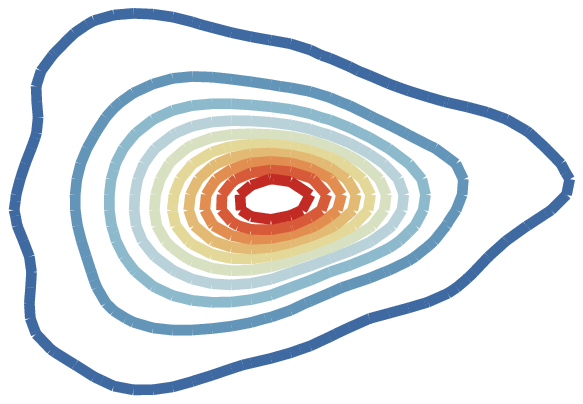};
\nextgroupplot[title=$\gamma_6$]
\addplot graphics[xmin=0,ymin=0,xmax=1,ymax=1] {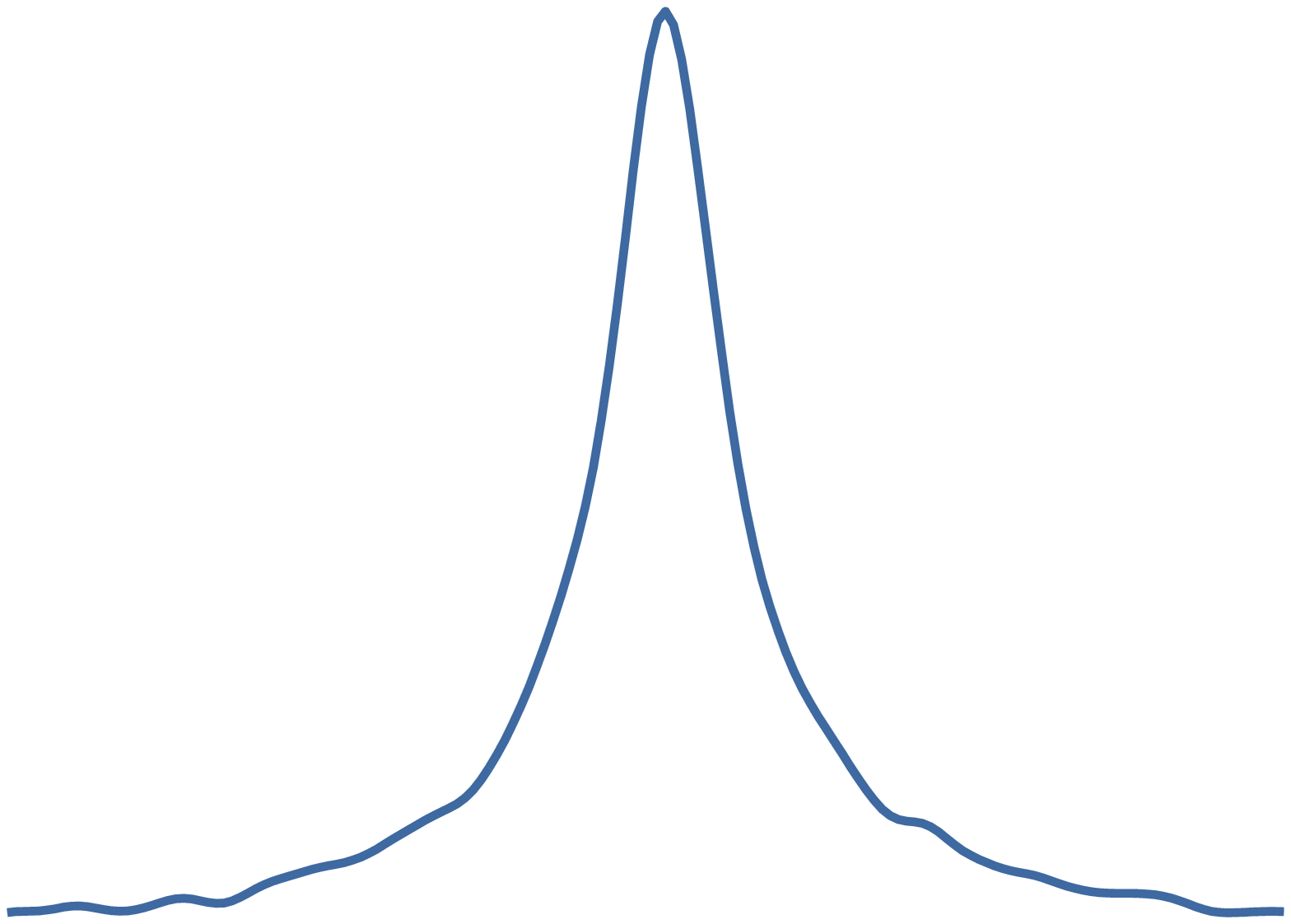};

\end{groupplot}
\end{tikzpicture}
}\subfigure[Map-induced coarse density.]{
\begin{tikzpicture}[scale=0.8]
\begin{groupplot}[group style={group size=6 by 6,xlabels at=edge bottom,ylabels at=edge left, xticklabels at=edge bottom,yticklabels at=edge left,vertical sep=1pt,horizontal sep=1pt},height=\groupPlotWidth,width=\groupPlotWidth,ticks=none,enlargelimits=false]

\nextgroupplot[title=$\gamma_1$]
\addplot graphics[xmin=0,ymin=0,xmax=1,ymax=1] {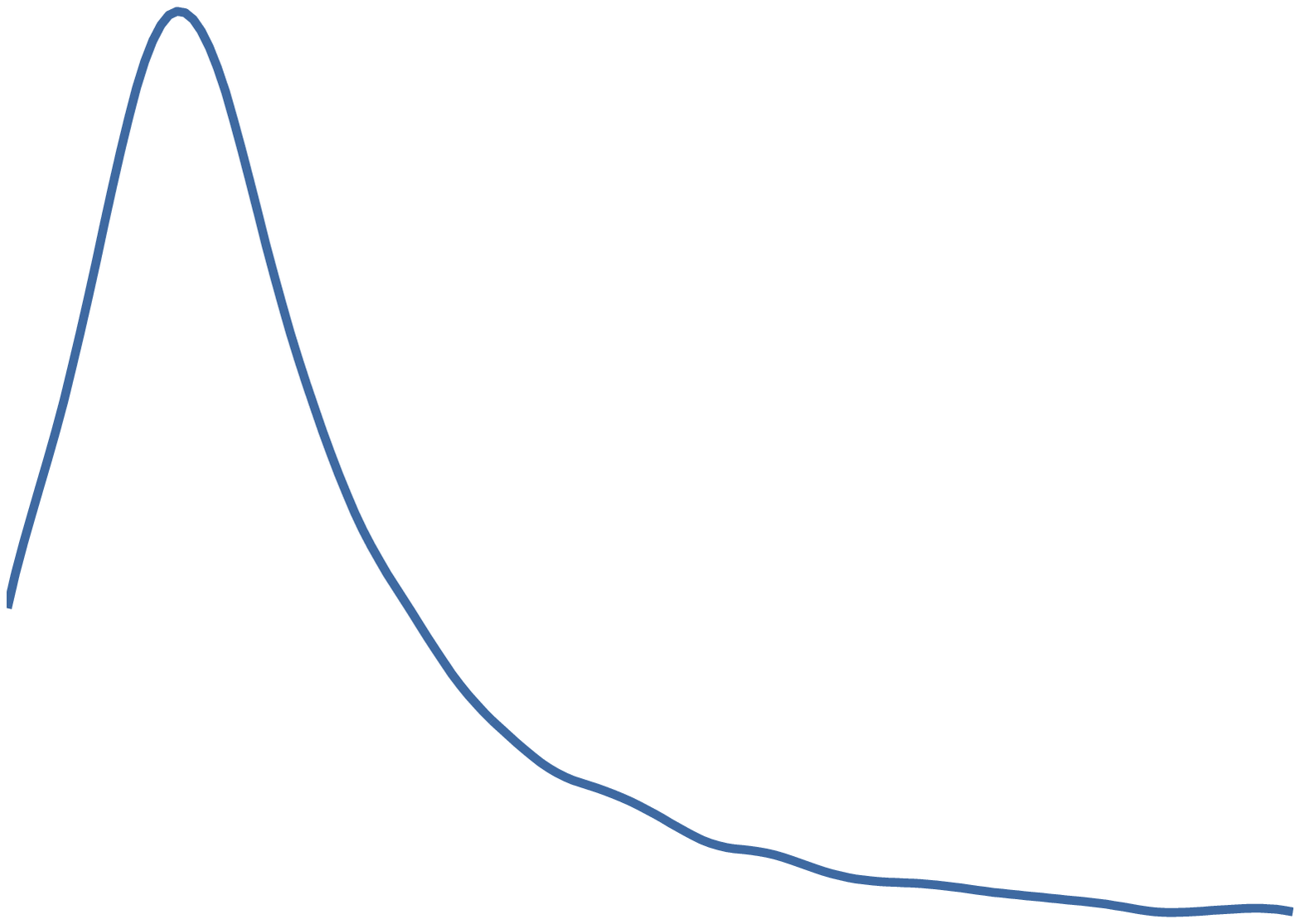};

\nextgroupplot[group/empty plot]
\nextgroupplot[group/empty plot]
\nextgroupplot[group/empty plot]
\nextgroupplot[group/empty plot]
\nextgroupplot[group/empty plot]

\nextgroupplot
\addplot graphics[xmin=0,ymin=0,xmax=1,ymax=1] {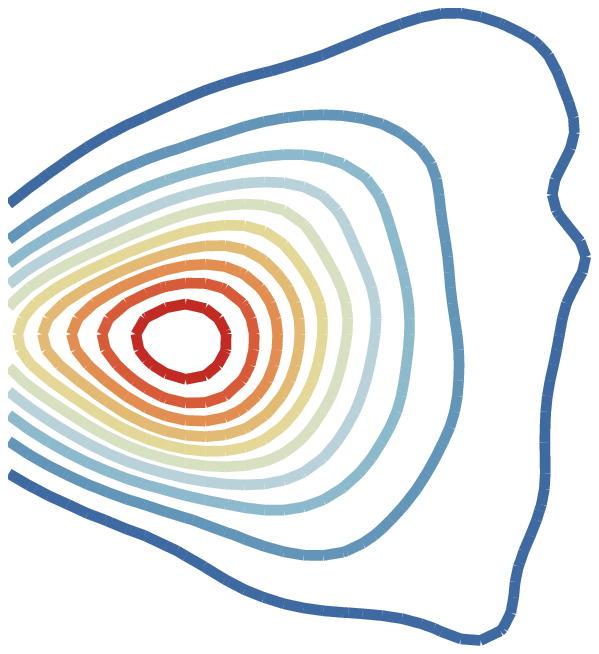};
\nextgroupplot[title=$\gamma_2$]
\addplot graphics[xmin=0,ymin=0,xmax=1,ymax=1] {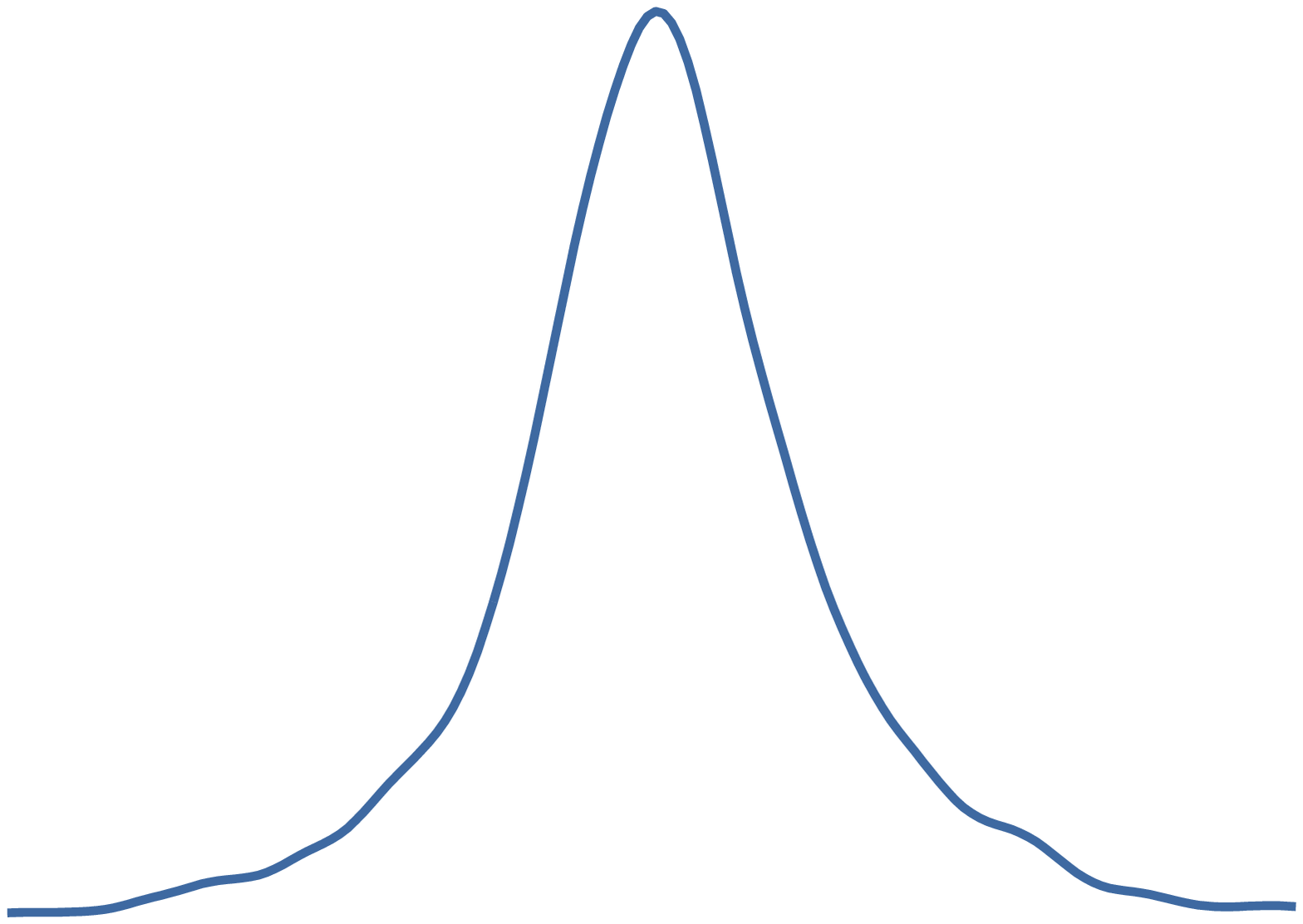};

\nextgroupplot[group/empty plot]
\nextgroupplot[group/empty plot]
\nextgroupplot[group/empty plot]
\nextgroupplot[group/empty plot]

\nextgroupplot
\addplot graphics[xmin=0,ymin=0,xmax=1,ymax=1] {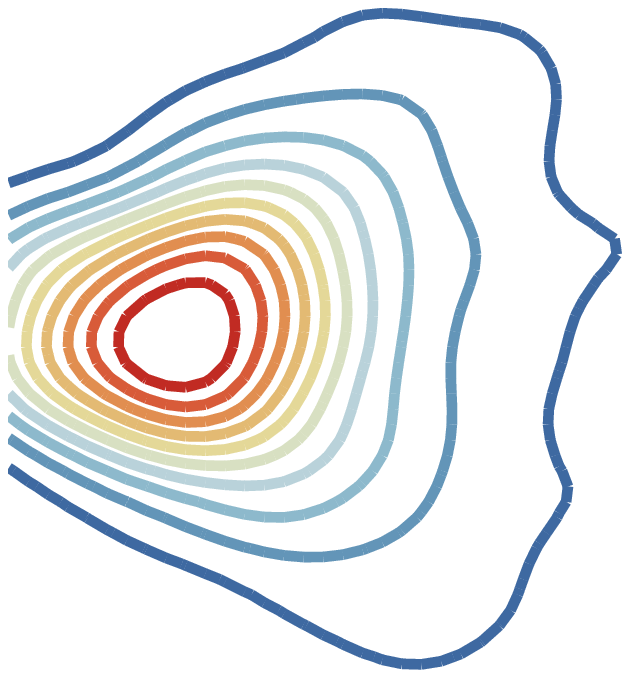};
\nextgroupplot
\addplot graphics[xmin=0,ymin=0,xmax=1,ymax=1] {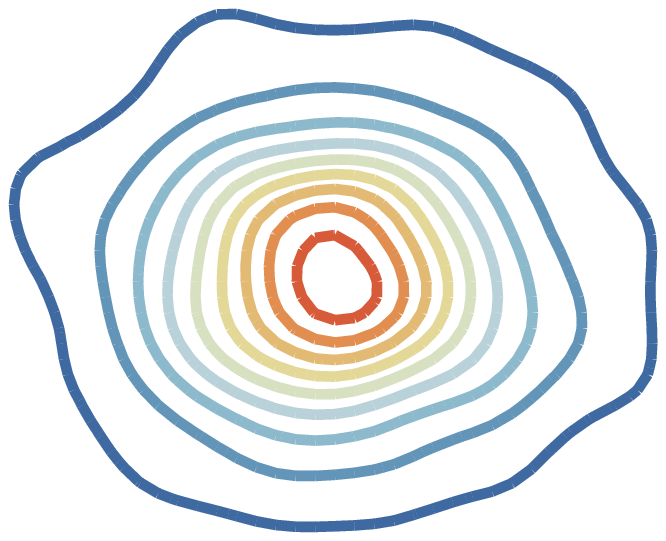};
\nextgroupplot[title=$\gamma_3$]
\addplot graphics[xmin=0,ymin=0,xmax=1,ymax=1] {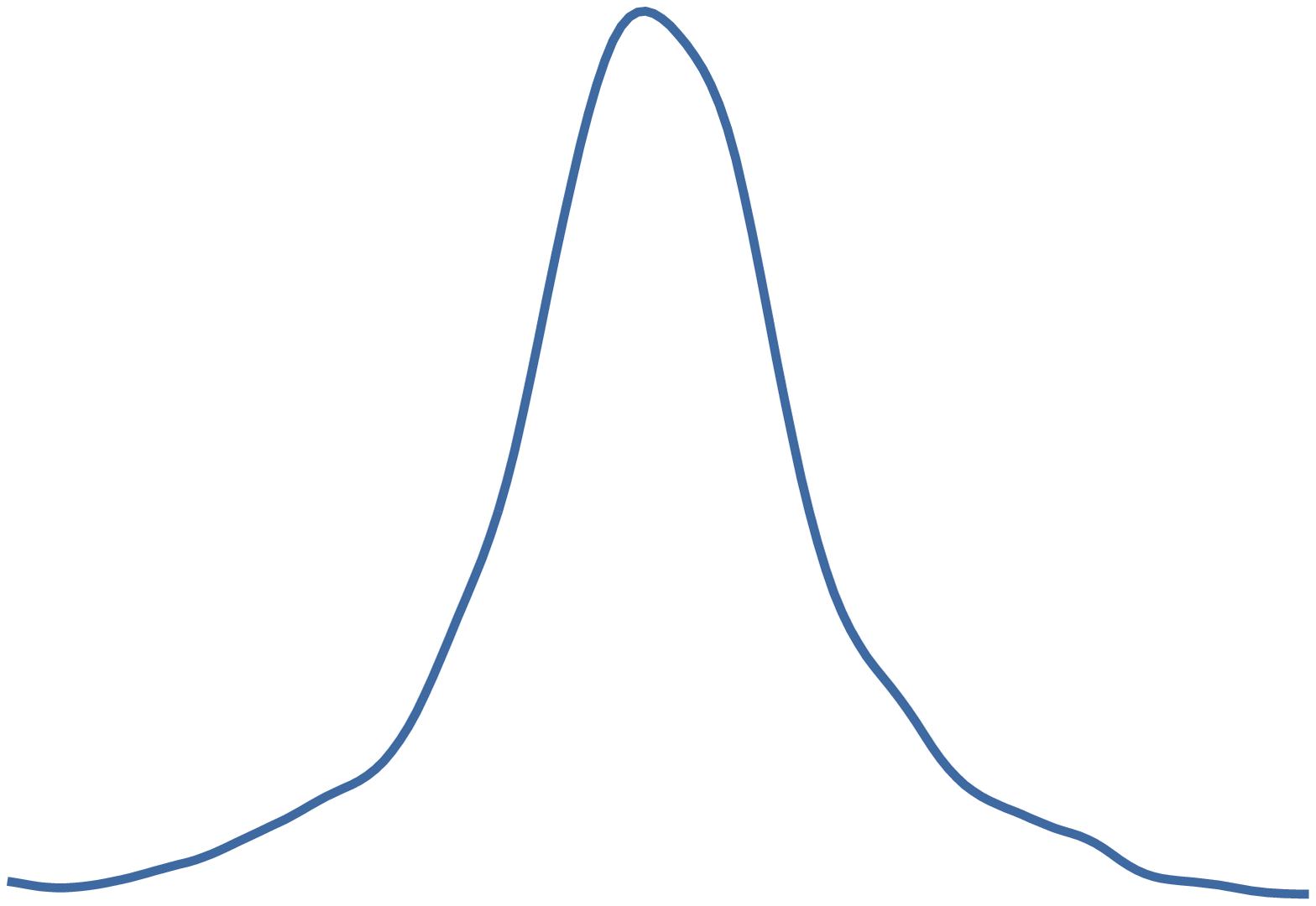};

\nextgroupplot[group/empty plot]
\nextgroupplot[group/empty plot]
\nextgroupplot[group/empty plot]

\nextgroupplot
\addplot graphics[xmin=0,ymin=0,xmax=1,ymax=1] {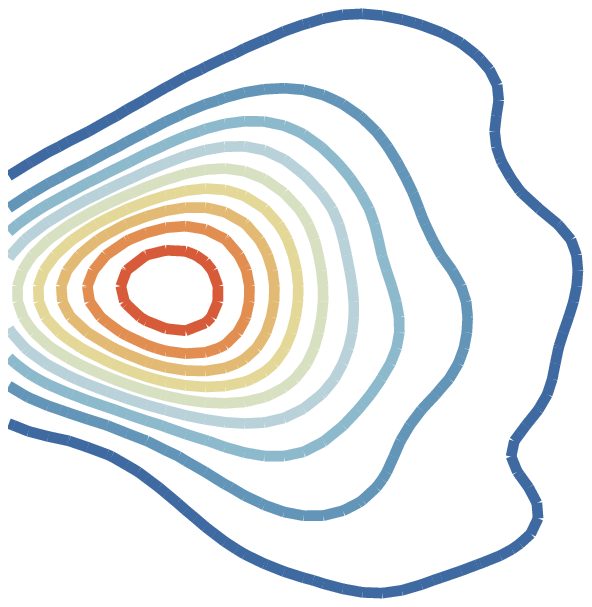};
\nextgroupplot
\addplot graphics[xmin=0,ymin=0,xmax=1,ymax=1] {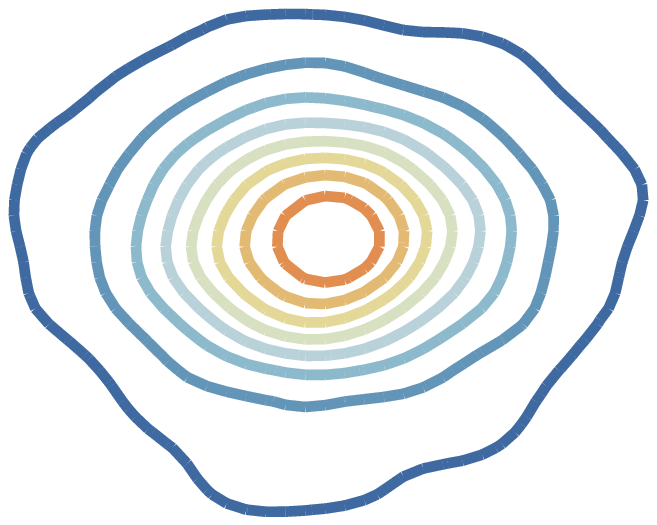};
\nextgroupplot
\addplot graphics[xmin=0,ymin=0,xmax=1,ymax=1]{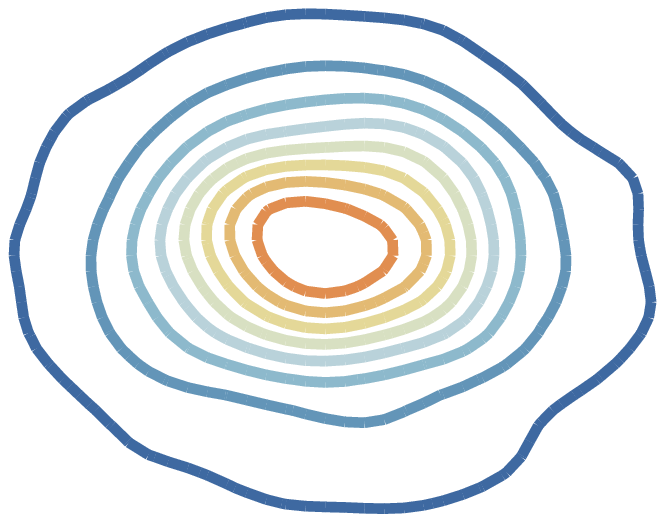};
\nextgroupplot[title=$\gamma_4$]
\addplot graphics[xmin=0,ymin=0,xmax=1,ymax=1] {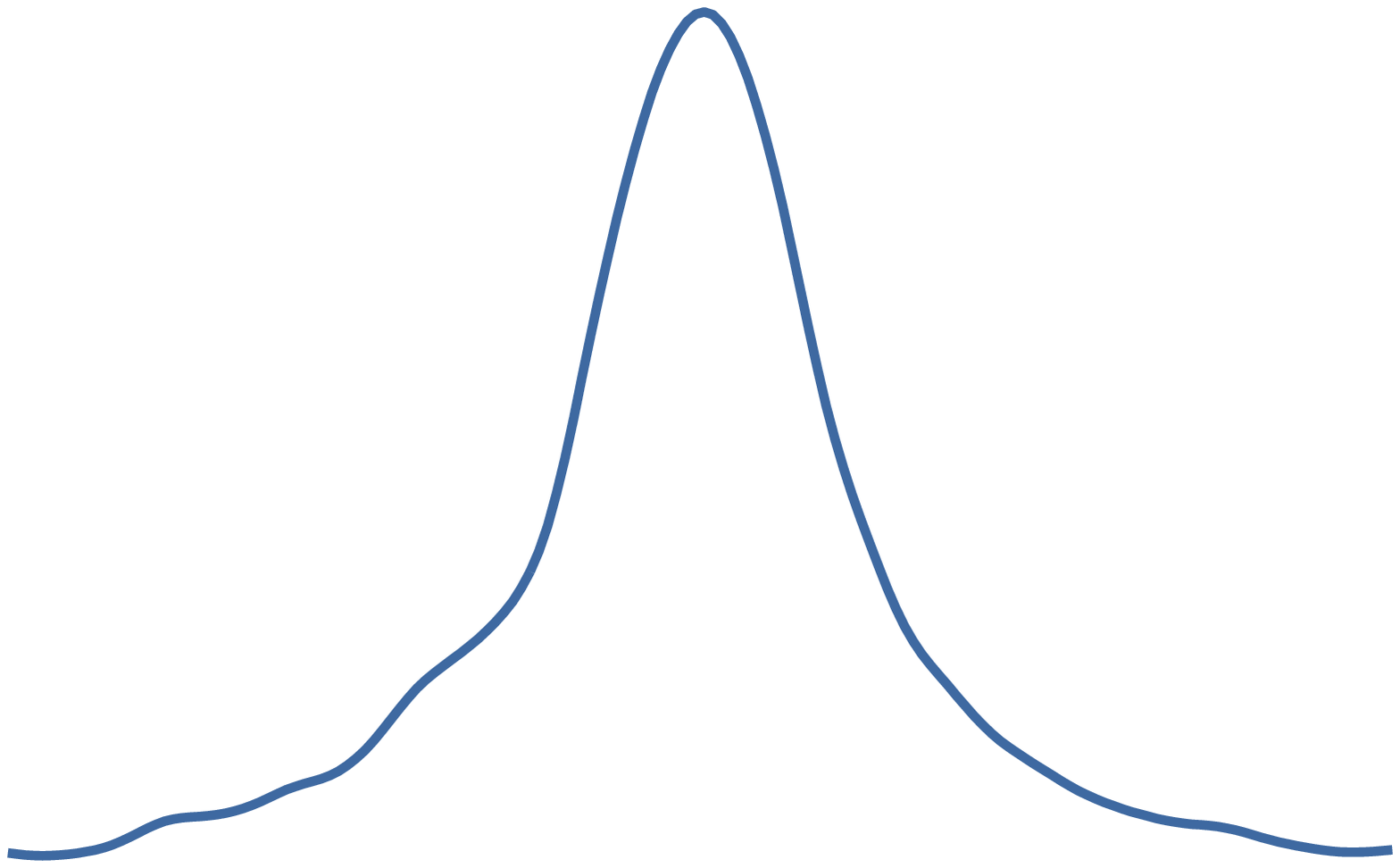};

\nextgroupplot[group/empty plot]
\nextgroupplot[group/empty plot]

\nextgroupplot
\addplot graphics[xmin=0,ymin=0,xmax=1,ymax=1] {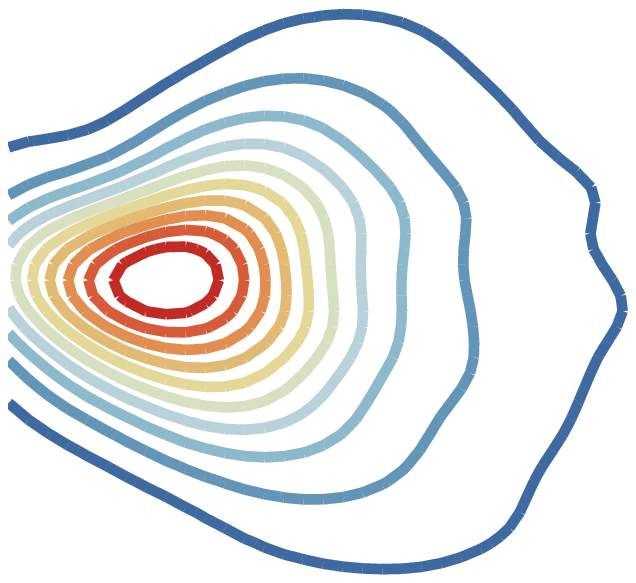};
\nextgroupplot
\addplot graphics[xmin=0,ymin=0,xmax=1,ymax=1] {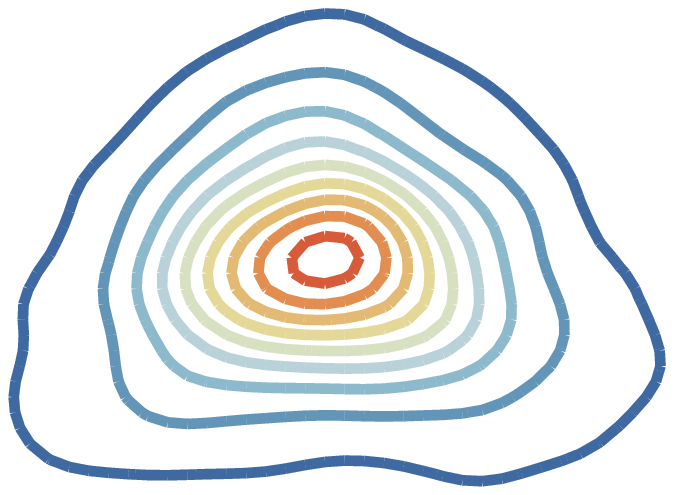};
\nextgroupplot
\addplot graphics[xmin=0,ymin=0,xmax=1,ymax=1] {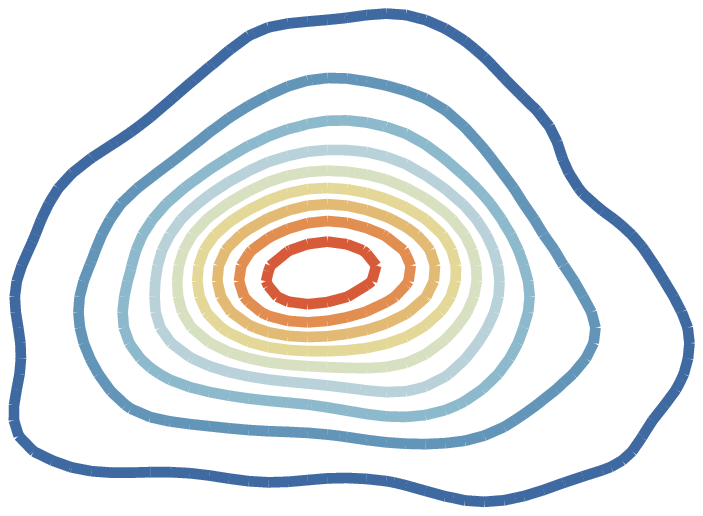};
\nextgroupplot
\addplot graphics[xmin=0,ymin=0,xmax=1,ymax=1]{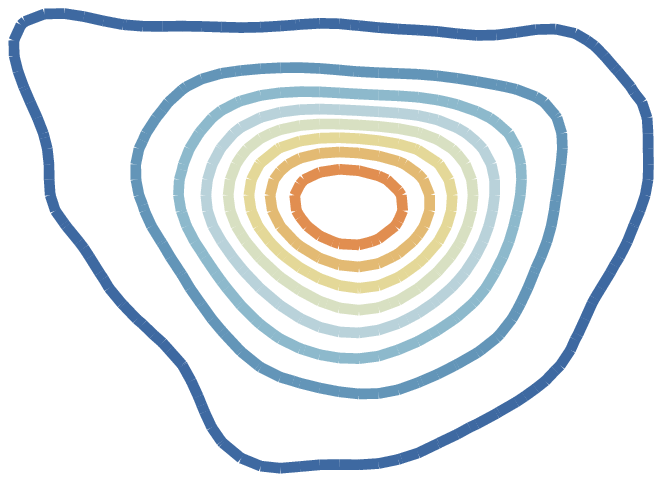};
\nextgroupplot[title=$\gamma_5$]
\addplot graphics[xmin=0,ymin=0,xmax=1,ymax=1] {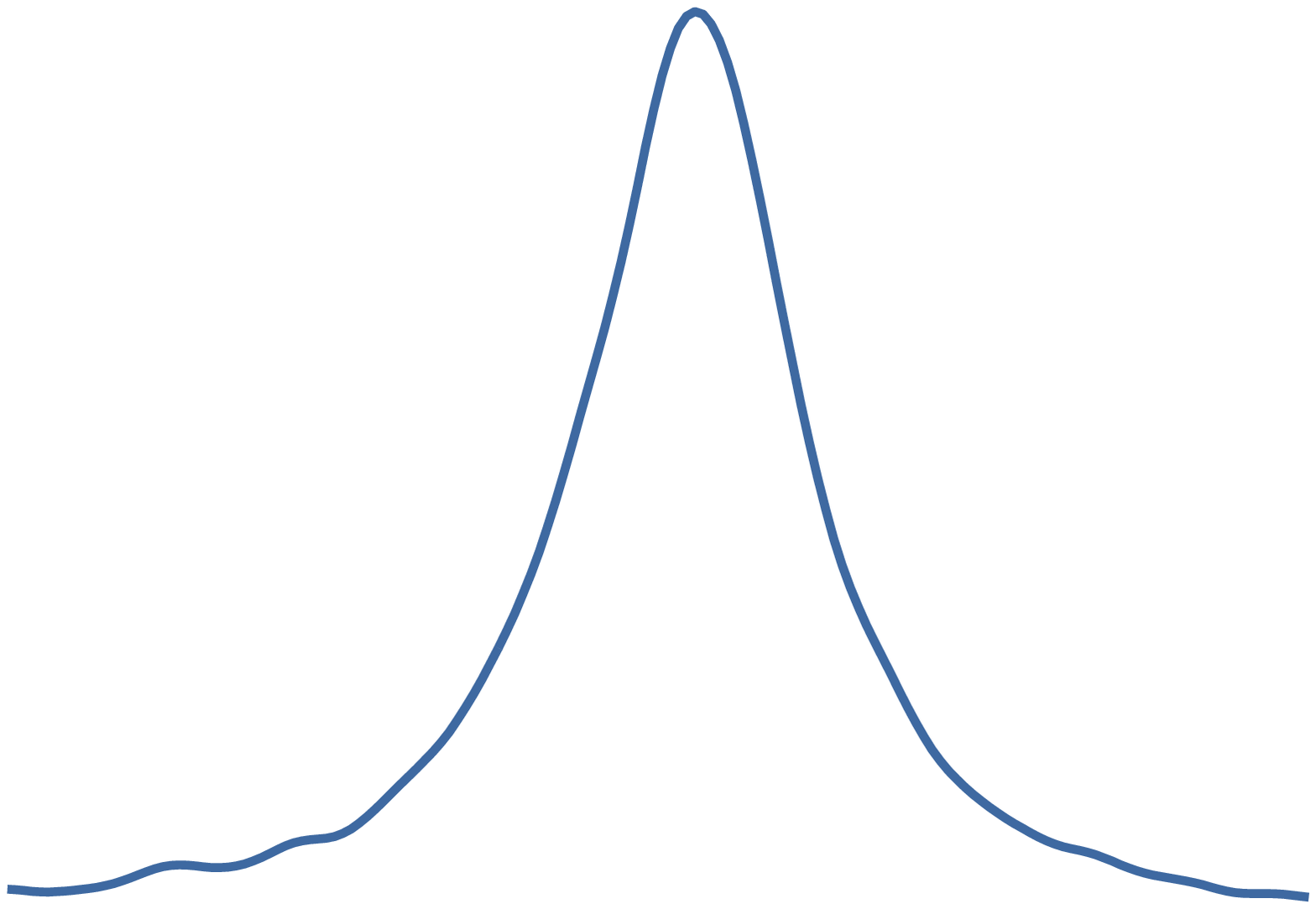};

\nextgroupplot[group/empty plot]

\nextgroupplot
\addplot graphics[xmin=0,ymin=0,xmax=1,ymax=1] {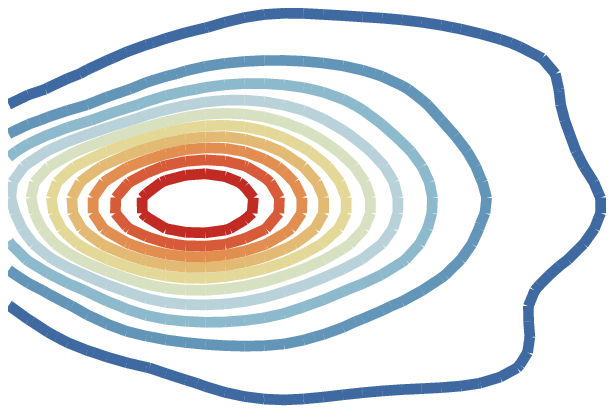};
\nextgroupplot
\addplot graphics[xmin=0,ymin=0,xmax=1,ymax=1] {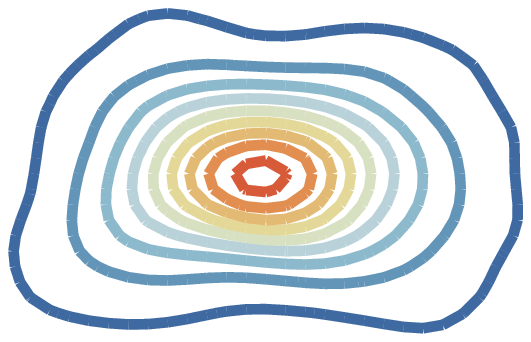};
\nextgroupplot
\addplot graphics[xmin=0,ymin=0,xmax=1,ymax=1]{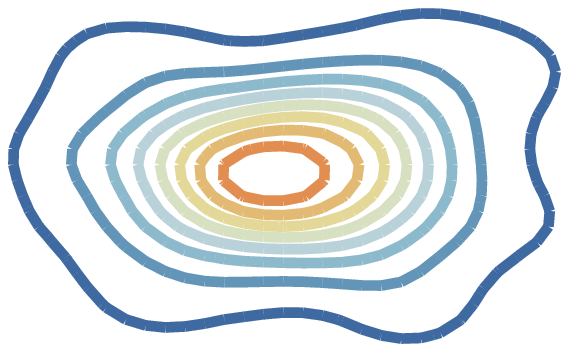};
\nextgroupplot
\addplot graphics[xmin=0,ymin=0,xmax=1,ymax=1]{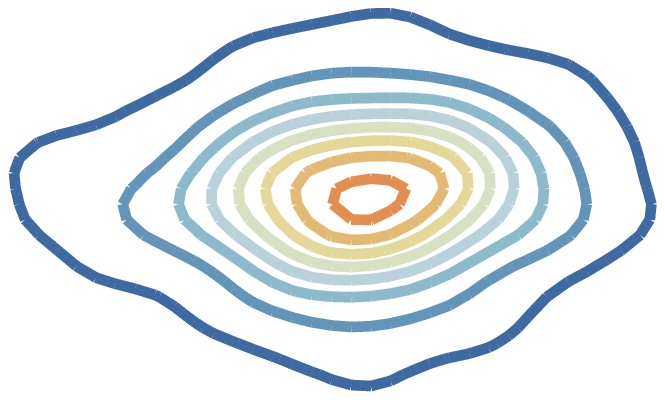};
\nextgroupplot
\addplot graphics[xmin=0,ymin=0,xmax=1,ymax=1] {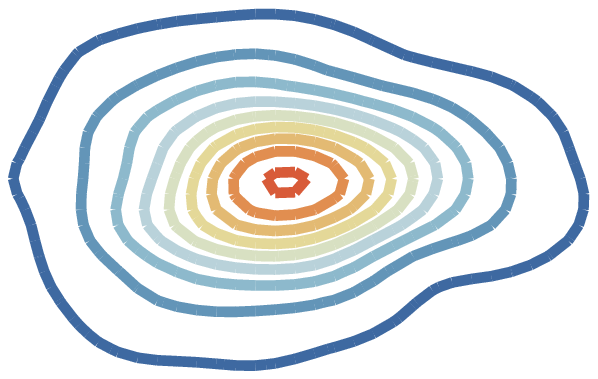};
\nextgroupplot[title=$\gamma_6$]
\addplot graphics[xmin=0,ymin=0,xmax=1,ymax=1] {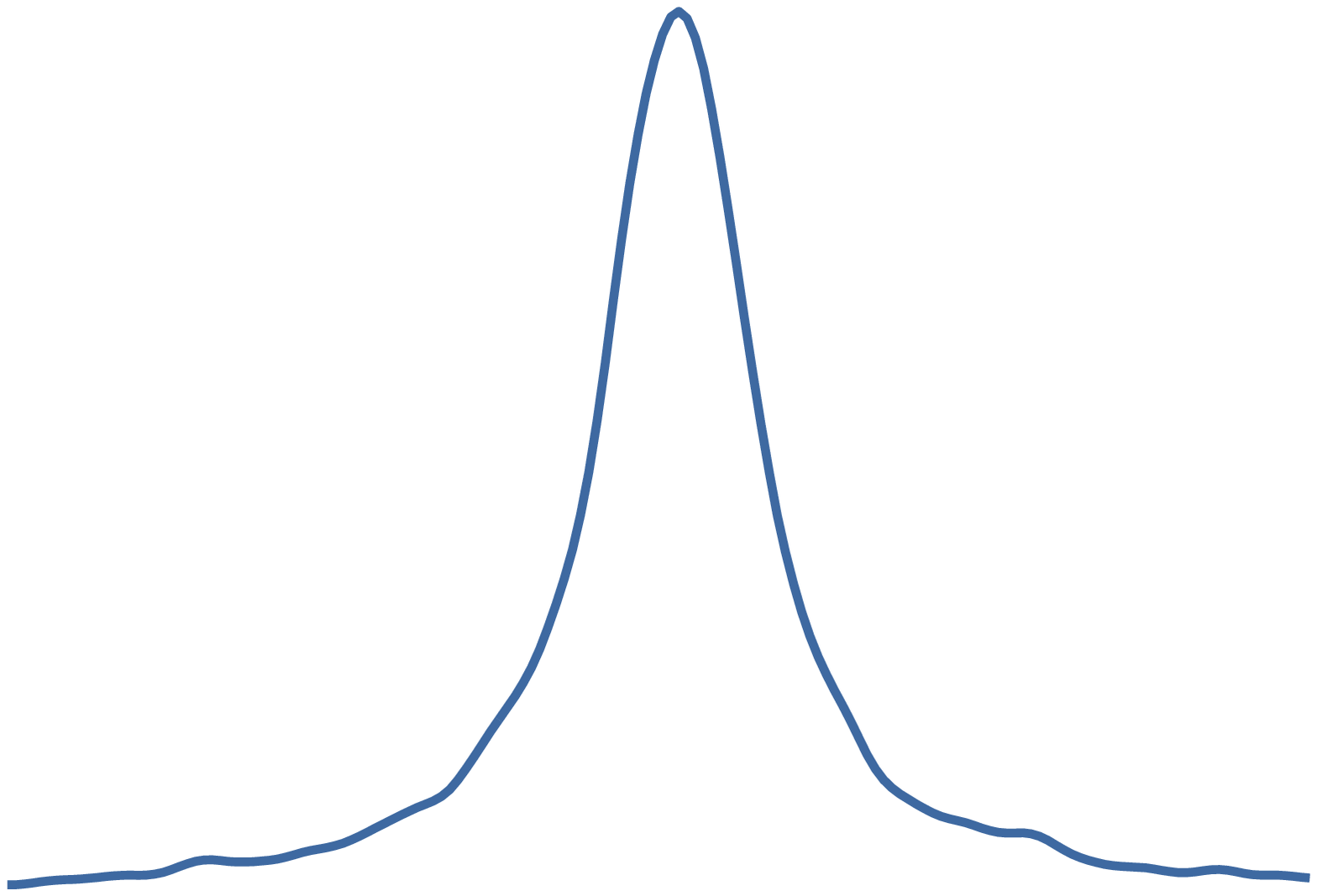};

\end{groupplot}
\end{tikzpicture}
}

\caption[Verification of coarse map accuracy in two dimensional example.]{Comparison of the true coarse prior density and the coarse prior density induced by the map $\ifmap_c$.  A degree-7 Hermite polynomial expansion was used to parameterize $\ifmap_f$. The first coarse parameter on each coarse cell, corresponding to $\crv_1$, is the most difficult for the map to capture because of its log-normal shape.  The color scales, contour levels, and axis bounds are the same for both plots.}
\label{fig:coarsmap2d}
\end{figure}

With confidence in the transport maps, we can move on to posterior sampling.  The preMALA MCMC algorithm, using the Hessian at the MAP as a preconditioner, was again used to sample the coarse posterior. It is relatively simple to compute gradients of the coarse posterior using adjoint methods, allowing us to use the Langevin approach effectively. Even though the coarse sampling problem still has $\pd_{\crv} = 384$ parameters, the coarse map $\ifmap_c$ itself captures much of the problem structure and the coarse MCMC chain mixes remarkably well, achieving a near-optimal acceptance rate of 60\%. 
Each coarse MCMC chain was run for $2 \times 10^5$ steps. Ten independent parallel chains were run, and coarse sampling was completed in 49 minutes.  
After coarse-scale MCMC sampling, the coarse samples were combined with independent samples of $\rrv_f$ through $\ifmap_f$ to generate posterior samples of the fine-scale variable $\frv$.  This coarse-to-fine sampling took 61 minutes.  Figure \ref{fig:post2d} shows the posterior mean and variance as well as two posterior samples.   A single fine-scale sample was generated for each coarse sample.   Notice that the fine-scale realizations have the same rough fine-scale structure as the true $\log(\kappa)$ field. This is an important feature that would not be present in many methods based on \textit{a priori} dimension reduction, such as truncated KL expansions.

\begin{figure}[h!]
\centering
\subfigure[True $\theta=\log(\kappa)$ field.]{
\includegraphics[width=0.46\textwidth]{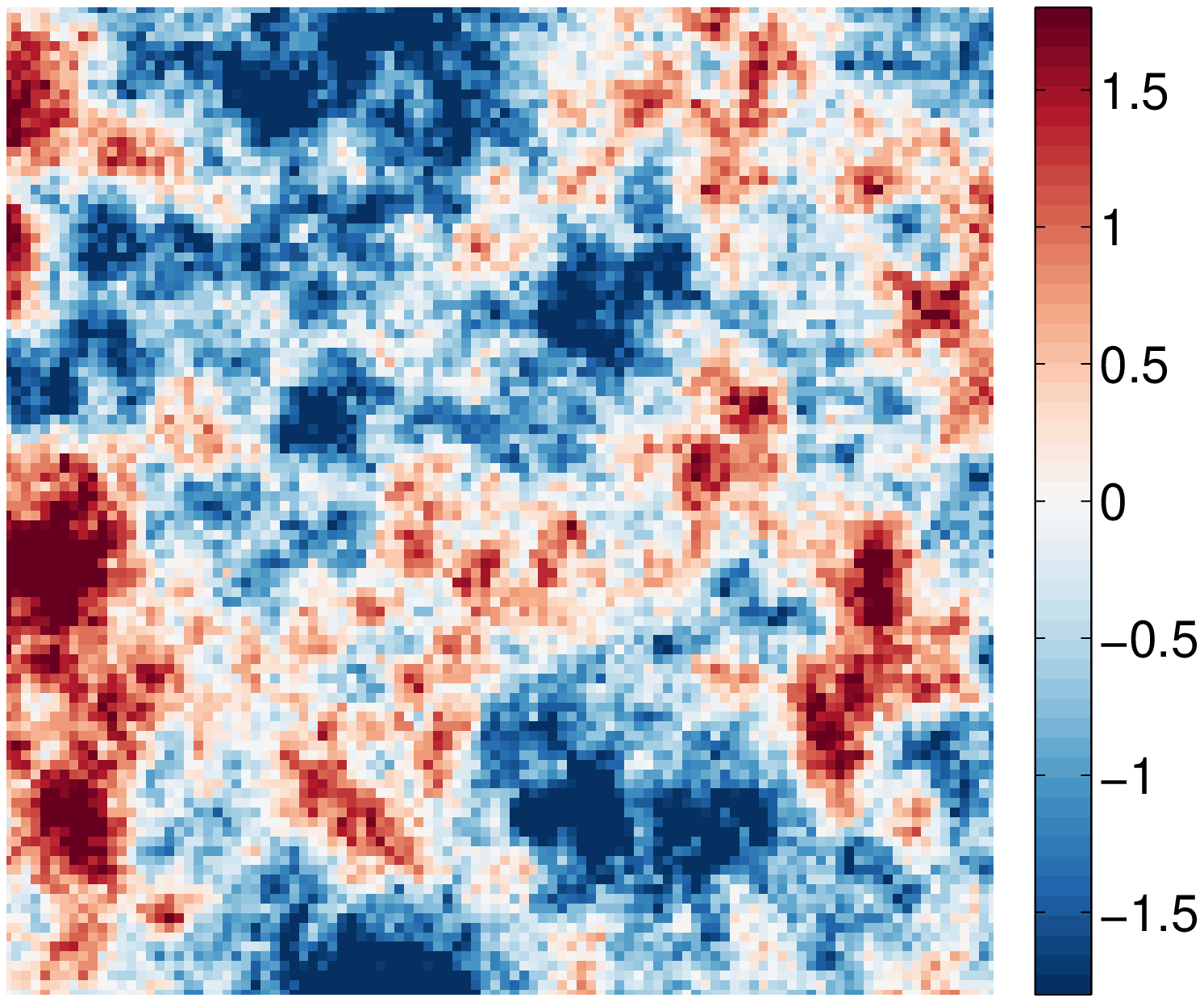}
}

\subfigure[Posterior mean using multiscale approach.]{
\includegraphics[width=0.46\textwidth]{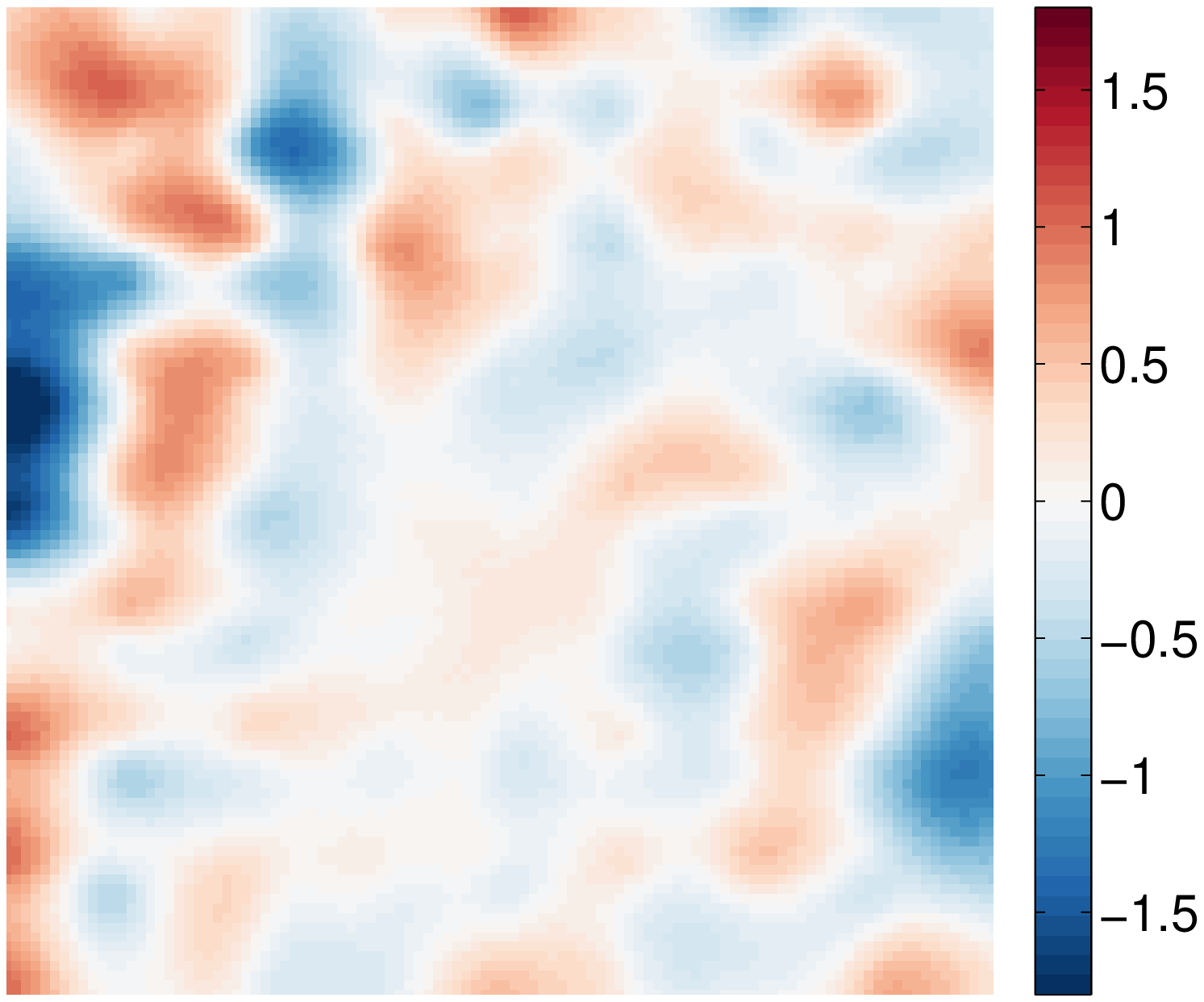}
}
\subfigure[Posterior variance using multiscale approach.]{
\includegraphics[width=0.46\textwidth]{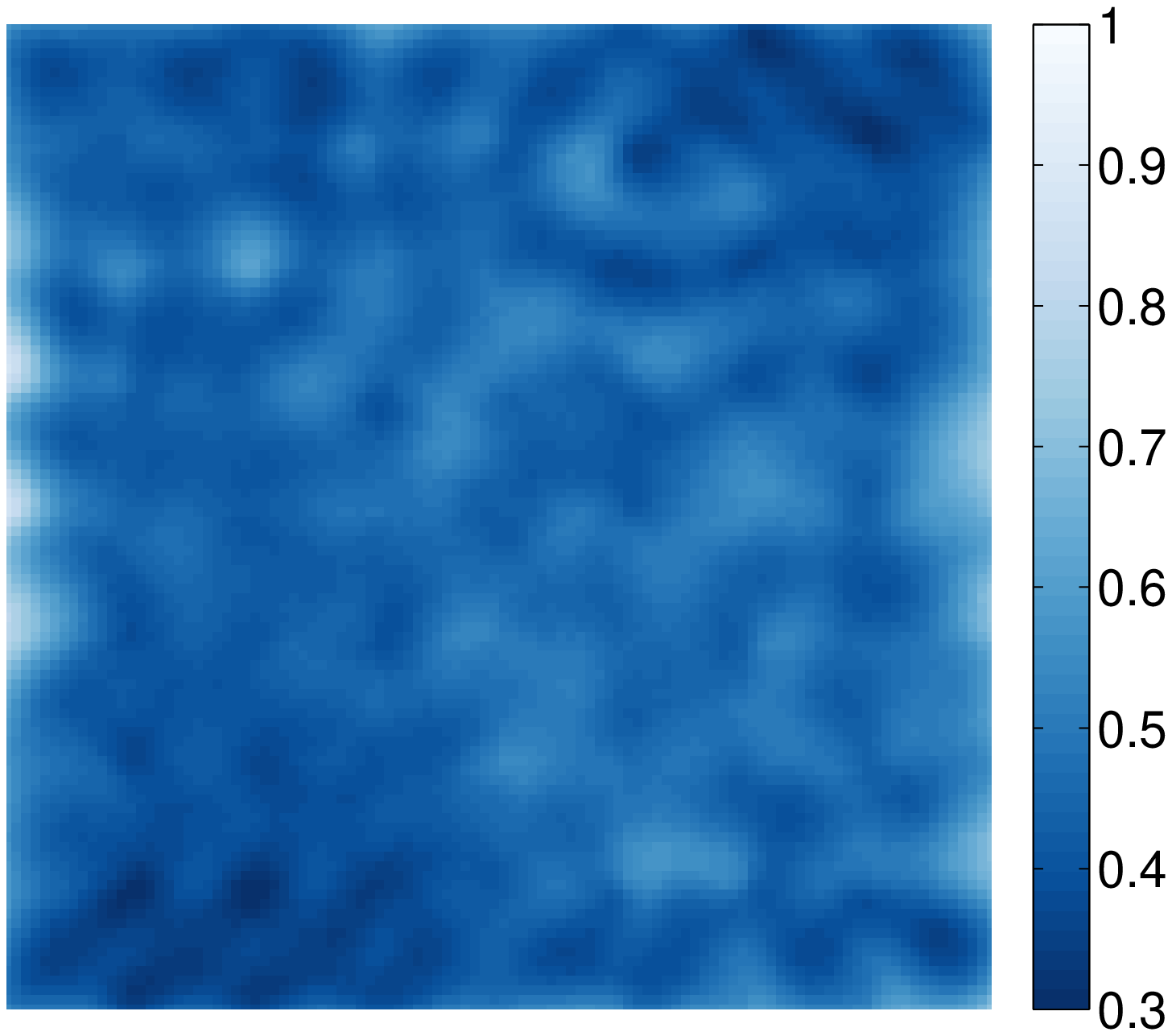}
}

\subfigure[Posterior realization]{
\includegraphics[width=0.46\textwidth]{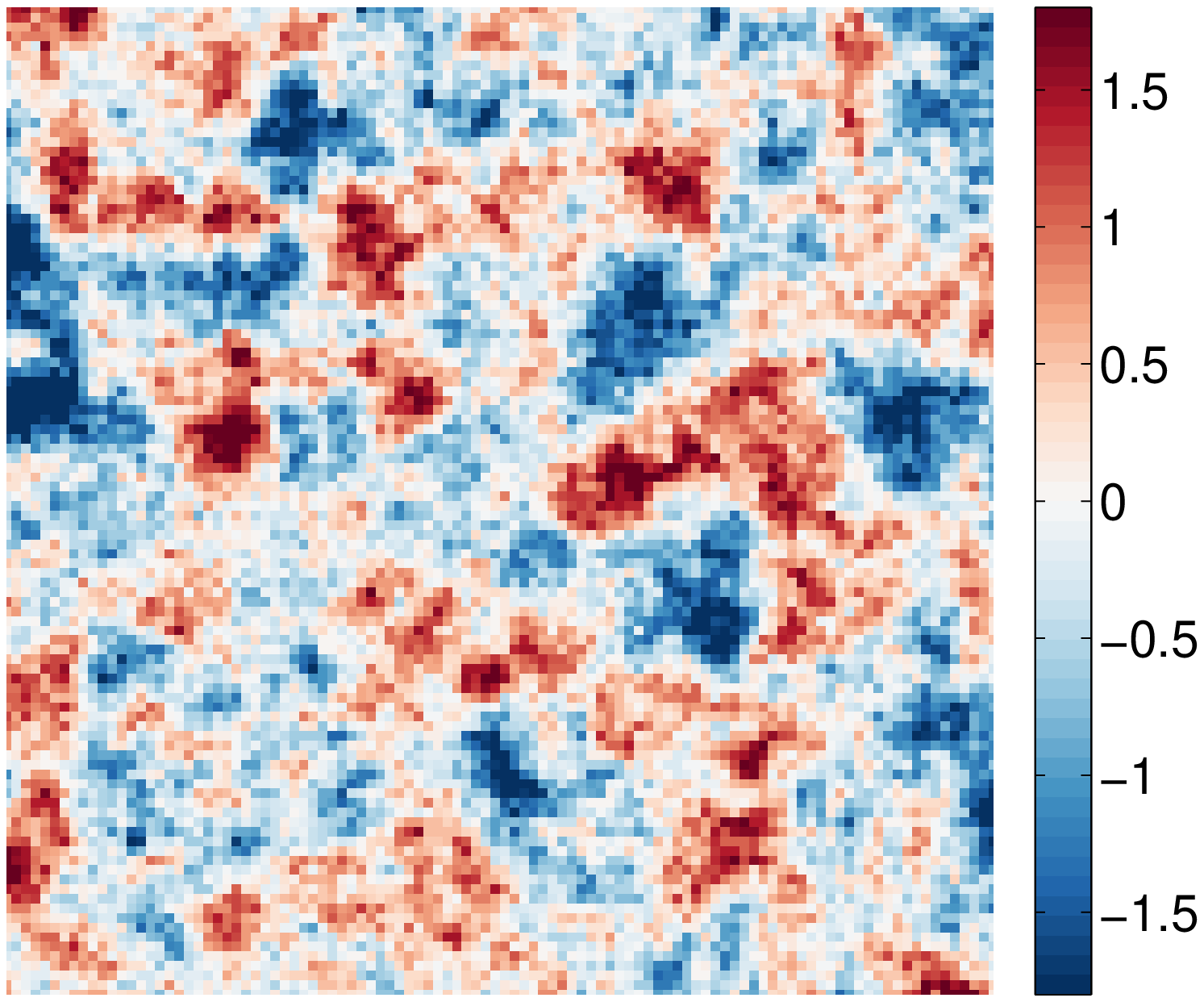}
}
\subfigure[Posterior realization]{
\includegraphics[width=0.46\textwidth]{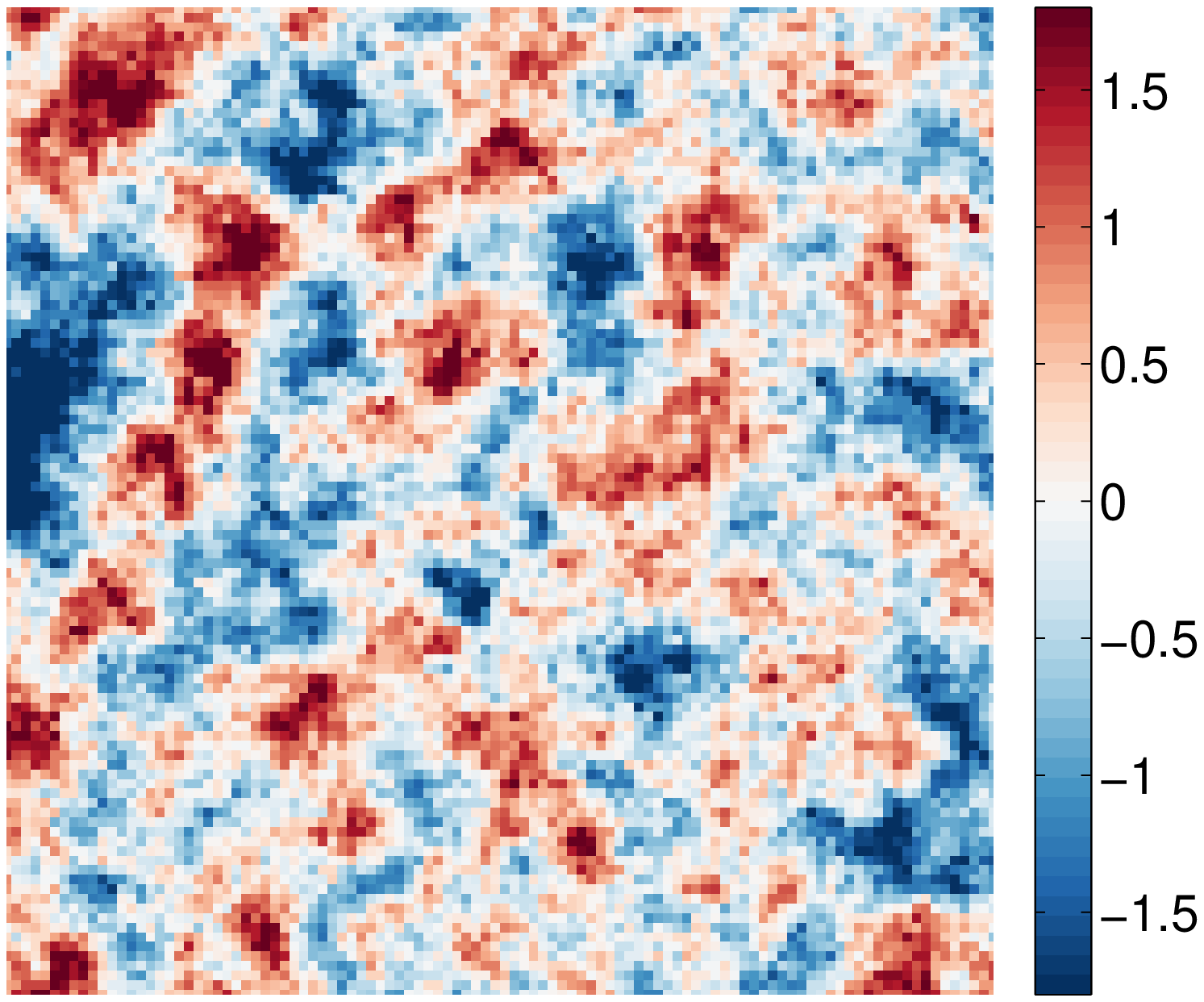}
}
\caption[Posterior summary in two dimensional example.]{Application of the multiscale inference framework to a $10816$-dimensional Bayesian inverse problem.}
\label{fig:post2d}
\end{figure}

\section{Discussion}

We have developed a method for efficiently solving Bayesian inference problems containing multiscale structure, as defined in (\ref{eq:multidef}). The method uses transport maps to decouple the original inference problem into a well-conditioned and lower-dimensional coarse-scale sampling problem (sampling $\pi(\rrv_c|\drv)$), followed by direct coarse-to-fine ``prolongation'' (evaluating $\ifmap_f$ with posterior $\rrv_c$ samples).   By exploiting locality and stationarity, we are able to build these transport maps despite the large dimension of the spatially distributed parameters of interest.  Our method accurately approximates the true posterior and can be applied to very large problems that are essentially intractable with other sampling methods.  We also stress that our approach is not restricted to subsurface flow applications or elliptic PDE forward models. The construction of transport maps relies entirely on prior samples, and therefore is not problem-specific or tied to specific probability distributions.  \EditsText{We should also point out that nothing in our formulation changes when the prior model is hierarchical, e.g.,  $\pi(\theta \vert \zeta)$ with some hyperparameters $\zeta$. If the posterior on the hyperparameters is not of particular interest, then marginal samples of $\pi(\theta) = \int  \pi(\theta \vert \zeta) \pi(\zeta) d \zeta $ can directly be generated and our approach used as described.\footnote{Otherwise, one could construct a larger map using joint prior samples of $(\gamma, \theta, \zeta)$.}}  Indeed, exact or approximate satisfaction of the conditional independence assumption \eqref{eq:multidef} is all that is required to apply our framework. Inverse problems with this structure exist in many areas, ranging from tree physiology \cite{Graf2014} to materials modeling \cite{Miller_Tadmor_2009}, and numerous other problems with scale separation and/or smoothing forward models. 

A typical serial MCMC sampler could take weeks to run on a problem as large as the two-dimensional example from Section \ref{chap:maps:twoD}.  Decoupling the problem using transport maps allowed us to solve it in only two hours.  Part of this improvement lies in the parallelism intrinsic to our approach.  All of the prior sampling, much of the optimization used to build the transport maps, and all of the post-MCMC coarse-to-fine map evaluations can be parallelized.  This level of parallelism is not available in MCMC samplers, even when multiple chains are run, as MCMC is inherently a serial process.  While we used some algorithm-level parallelism (employing MPI for parallel model evaluations and parallel map construction), our approach would lend itself well to more sophisticated distributed-memory or GPU-parallel implementations in the future.  Such an implementation could further reduce the run time of our framework.

Of course, allowing our approach to generate approximate posterior samples enables computational savings as well. As demonstrated in Section \ref{sec:application}, the accuracy of the posterior approximation can be controlled by the representations of $\ifmap_c$ and $\ifmap_f$.  In Section \ref{chap:maps:twoD}, we let $\ifmap_f$ be linear in order to mitigate the required computational effort.  In applications where more exact posterior sampling is required, however, a higher polynomial degree or alternative functional representation could be employed.  Additionally, if problem-specific information is available (such as the locality and stationarity used in Section \ref{sec:twodim}), it can also be incorporated into the map representation to further increase accuracy and reduce computational expense.  This flexibility is important in practice, and should allow our multiscale approach to be applied in diverse areas.  Complementing the use of problem-specific information is the development of more advanced map construction techniques, which could adaptively construct and refine maps within a user-specified form, targeting a specified error in \eqref{eq:klexpr}. This is an important area of future research. We emphasize that the fundamental idea underlying our approach---interpreting multiscale structure as conditional independence, and applying it in a Bayesian setting---is independent of the algorithmic specifics of map construction.  But advances in the latter will enhance the efficiency and applicability of the inference strategy developed here.

\bibliographystyle{mysiam}
\bibliography{main}

\end{document}